\documentclass[aps,prx,superscriptaddress,twocolumn]{revtex4-2}
\usepackage{bm}
\usepackage{graphicx}
\usepackage{color}
\usepackage{braket}
\usepackage{amsmath,amssymb,amsfonts,amsthm,mathtools}
\usepackage{enumerate}
\usepackage{soul}
\usepackage{enumitem}
\usepackage{subfigure}
\usepackage[colorlinks=true,linkcolor=black,anchorcolor=black,citecolor=black,urlcolor=black]{hyperref}
\usepackage{titlesec}
\usepackage{tikz-cd} 
\usepackage{float}
\usepackage{multirow}
\usepackage[title]{appendix} 
\usepackage{comment}
\usepackage{booktabs}

\begin{document}

\author{Zheng Zhang}
\affiliation{Department of Physics, School of Science, Lanzhou University of Technology, Lanzhou 730050, China}

\author{Peiyuan Wang}
\affiliation{Department of Physics and HK
 Institute of Quantum Science \& Technology, The University of Hong Kong, Pokfulam Road, Hong Kong, China}
 
\author{Y. X. Zhao}
\email{yuxinphy@hku.hk}
\affiliation{Department of Physics and HK
 Institute of Quantum Science \& Technology, The University of Hong Kong, Pokfulam Road, Hong Kong, China}

\title{Projective representation theory of spin space groups}

\begin{abstract}
Spin space groups provide the natural symmetry framework for magnetic crystals with weak spin-orbit coupling, but extracting their physical consequences requires a general theory of irreducible representations. The central difficulty is that spin space groups are represented projectively: their factor systems can render lattice translations noncommuting, induce nonsymmorphic actions in momentum space, and modify the projective structure of little cogroups. These effects lie beyond conventional space-group and double-group representation theory. Here, using Mackey's theory of group extensions, we develop a unified constructive framework for all collinear, coplanar, and noncoplanar spin space groups, including antiunitary symmetries and general factor systems. A central technical result is a canonical decomposition of the relevant factor system $\nu$ into a translational factor $\sigma$, a mixed factor $\gamma$ coupling translations to point-group operations, and a point-group factor $\alpha$. This decomposition makes transparent how ordinary representation theory is modified: $\sigma$ determines the projective translation algebra and the appropriate Brillouin zone, $\gamma$ controls the momentum-space group action and can make it nonsymmorphic, and $\alpha$ contributes to the factor systems of little cogroups. On this basis, we construct all projective irreducible corepresentations by induction over momentum-space orbits. The framework identifies which nonsymmorphic momentum-space symmetries can be realized by spin space groups and {how they reconstruct high-symmetry momenta and band degeneracies.} It reveals Brillouin spaces that are compact flat manifolds rather than tori, {Weyl points of the same chirality locked by projective translation symmetry,} and spin space group symmetry-enforced Zak phases as well as Chern insulators. Our results establish the representation-theoretic foundation for systematic studies of weak-spin-orbit-coupled magnetic materials.
\end{abstract}

\maketitle

\section{Introduction}
Symmetry provides the organizing principle for crystalline quantum matter. The 230 space groups classify the spatial symmetries of crystals and 1651 magnetic space groups classify magnetic crystals in which spin and spatial transformations are locked by spin-orbit coupling (SOC) \cite{BradleyMathematical2009}. When SOC is weak, however, spin rotations can be independent of spatial operations, and the appropriate symmetries are spin space groups (SSGs) \cite{BrinkmanTheory1966,LitvinSpin1974,SandratskiiNoncollinear1998,LiuSpinGroup2022,SmejkalConventional2022,SmejkalEmerging2022}. This enlarged symmetry framework has become central to the description of altermagnets and noncollinear magnetic order. The recent enumeration of SSGs \cite{XiaoSpin2024,ChenEnumeration2024,JiangEnumeration2024} has therefore stimulated applications ranging from electronic and magnon band structures to superconductivity and magnetic response \cite{CorticelliSpinspace2022,GuoEightfold2021,YangSymmetry2024,ZengDescription2024,
DevarajInterplay2024,RadaelliTensorial2024,ChenUnconventional2025,FengSuperconducting2025,
EtxebarriaCrystal2025,LiTopological2025,YangThreedimensional2025,YuNeel2025,TakahashiSymmetry2025,
BiniskosSystematic2025,WeissenhoferMagnon2026,ElcoroAutomatic2026,HuangPrediction2026,SongUnified2026}.

A space-group classification acquires much of its physical content through the irreducible representations of the symmetry groups \cite{WeylTheory1950,WignerGroup1959}. SSGs are no exception. Their irreducible representations determine the symmetry and degeneracy of quasiparticles \cite{DresselhausGroup2008,AltmannBand1991,YangSymmetry2024,SongConstructions2025}, constrain the connectivity of numerically calculated band structures \cite{ZakBand1982,MichelElementary2001,CorticelliSpinspace2022}, provide selection rules for spectroscopic probes \cite{BirmanSpace1962}, classify collective excitations \cite{DresselhausGroup2008,CorticelliSpinspace2022,ChenUnconventional2025}, and enable symmetry-based diagnoses of topological phases \cite{BradlynTopological2017,PoSymmetrybased2017,CanoTopology2018,VergnioryComplete2019,ZhangCatalogue2019,TangComprehensive2019}.

{However, there is an intrinsic difference between representations of SSGs and ordinary representations of space groups. That is, when applying to electronic systems (spin-1/2 systems), SSGs are projectively represented, i.e., for two elements $g_1,g_2$ of an SSG $\mathcal{G}$, the multiplication rule of their representation matrices $\rho(g_1),\rho(g_2)$ is modified by a phase factor:}
\begin{equation}
\rho(g_1)\rho(g_2)=\nu(g_1,g_2)\rho(g_1g_2).
\end{equation}
{$\nu$ is called a factor system and $\rho$ is called a $\nu$-representation of $\mathcal{G}$. The factor system arises from the lift of $O(3)$ spin operations into $SU(2)\rtimes Z_2^{\mathcal{T}}$ in spin-1/2 systems \cite{XiaoSpin2024}. Nontrivial factor systems can largely change the structure of representations. For example, the high symmetry points can be shifted, the degeneracies can be changed, and even the Brillouin zone can become non-commuting. Previous works have shown many physical consequences caused by nontrivial factor systems of crystalline groups, including the modification of $k\cdot p$ models, the change of topological classifications, new types of quasiparticles, and nonsymmorphic symmetries in momentum space. Thus, to capture correct physical properties of electronic bands in magnetic systems with weak SOC, one must use projective representations  rather than ordinary representations of SSGs.}

{Despite their fundamental role, SSGs still lack a general theory of projective representations. This gap has two aspects. First, although the factor system of an SSG can in principle be derived, a systematic interpretation of its representation-theoretic significance and physical meaning is still missing. Second, a general constructive procedure for obtaining all irreducible projective representations of a given SSG with its induced factor system has yet to be established, despite its importance for applications. In this work, we address both issues by developing a unified representation theory for all SSGs, including those associated with collinear, coplanar, and noncoplanar magnetic orders. Our approach builds on Mackey's theory of group extensions \cite{MackeyUnitary1958} and accommodates both general factor systems and antiunitary symmetries. We decompose the factor systems of SSGs into physically meaningful components in a manner compatible with Mackey's construction of representations, and then use this decomposition to establish a general method for constructing all irreducible projective representations of any SSG.}

{Let us introduce our theory here briefly.} Recall that an SSG $\mathcal{G}$ can be expressed as a semidirect product $\mathcal{G}\cong \mathcal{S}\rtimes G$, where $\mathcal{S}$ is the spin-only group and $G$ is the quotient spin space group, which is isomorphic to a magnetic space group or a space group. The principal step in constructing the $\nu$-representations of $\mathcal{G}$ is to construct the $\nu_q$-representations of $G$, where $\nu_q$ is the restriction of $\nu$ to $G$.

The factor system $\nu_q$ can be specified as follows. In the semidirect product $\mathcal{G}\cong \mathcal{S}\rtimes G$, the spin operations of elements in $G$ encode the spin-space locking and is specified by a graded homomorphism $\varphi:G\to O(3)$, whose grading distinguishes unitary from antiunitary elements. The spinor double cover $SU(2)\rtimes Z_2^T\to O(3)$ defines a cohomology class $[\alpha]\in H^{2,c}(O(3),U(1))$, where $c$ specifies the complex conjugation action of antiunitary elements on $U(1)$. Pulling this class back gives
\begin{equation}
	[\nu_q]=\varphi^*[\alpha]\in H^{2,c}(G,U(1)).
\end{equation}
The core technical problem is therefore to determine how $\nu_q$ modifies the representation theory of $G$. A central result of this work is a canonical decomposition of the magnetic-group factor system $\nu_q$ into three components: a factor system $\sigma$ of the translational subgroup, a mixed factor $\gamma$ relating translations to point-group operations, and a point-group factor $\alpha$. This form does more than simplify the calculation: it identifies the precise origin of each departure from ordinary representation theory {and provides clear physical interpretation}. The factor $\sigma$ determines whether translations commute and thereby fixes the projective translation algebra and the appropriate Brillouin zone; $\gamma$ determines the point-group action on that zone and can render it nonsymmorphic; and $\alpha$, together with contributions inherited from $\sigma$ and $\gamma$, determines the factor systems of the little cogroups. The canonical decomposition thus provides the organizing principle for the construction below.

To state the modifications in constructing projective irreducible representations, we first recall the Mackey construction for ordinary representations of a space group $G$. Let $\mathrm{T}$ be its translational subgroup and $P=G/\mathrm{T}$ its point group. The Brillouin zone is the space of irreducible representations of $\mathrm{T}$, conventionally labelled by crystal momentum $\boldsymbol{k}$. The point group $P$ acts on this space, and its orbits are the $k$ stars \cite{CornwellGroup1997}. Choosing a representative $\boldsymbol{k}$ from each star, one defines the little group
$
	G_{\boldsymbol{k}}=\{g\in G\,|\,g\boldsymbol{k}\equiv\boldsymbol{k}\},
$
where equivalence is understood modulo reciprocal lattice vectors. The corresponding little cogroup is $G_{\boldsymbol{k}}/\mathrm{T}$. An irreducible representation of $G_{\boldsymbol{k}}$ is obtained by combining the representation of $\mathrm{T}$ at $\boldsymbol{k}$ with an appropriate projective irreducible representation of $G_{\boldsymbol{k}}/\mathrm{T}$. Inducing this little-group representation over the $k$ star then gives an irreducible representation of $G$. 

Through its canonical components, the factor system $\nu_q$ requires five fundamental modifications of this construction. (i) Lattice translations can become noncommuting and, for SSG factor systems, may generate a Clifford algebra. Their irreducible representations can therefore be multidimensional, and the Brillouin zone must be defined as the space of inequivalent irreducible representations of this projective translation algebra. (ii) The point group can act nonsymmorphically in momentum space, so that some point-group elements shift even the conventional $\Gamma$ point. (iii) For each $\boldsymbol{k}$ star, $\nu_q$ modifies the factor system of the little cogroup $G_{\boldsymbol{k}}/\mathrm{T}$. (iv) Correspondingly, a projective irreducible (co)representation of the little cogroup must be combined with the generally multidimensional irreducible representation of $\mathrm{T}$ labelled by $\boldsymbol{k}$ to form a projective irreducible representation of $G_{\boldsymbol{k}}$. (v) Finally, inducing over the $\boldsymbol{k}$ star requires additional $\nu_q$-dependent phases to assemble a projective irreducible representation of $G$.

Once all $\nu_q$-irreducible (co)representations of $G$ have been obtained, a second application of the Mackey construction then combines these (co)representations with the appropriate projective representations of the spin-only group $\mathcal{S}$ to obtain all $\nu$-irreducible corepresentations of the full SSG $\mathcal{G}$.

{Here, we compare our theory with previous approaches to highlight its key advances. The established theory of double-valued representations of ordinary space groups emerges as a particularly simple special case of our framework, in which the underlying structures simplify considerably and the full complexity of the general theory does not arise.} Their factor system is the pullback of the point-group factor system associated with the double cover $SU(2)\to SO(3)$. Consequently, only modification (iii) is essential: translations remain commuting, the point group acts symmorphically in momentum space, and the rest of the Mackey construction proceeds as in the ordinary case. In Reference~\cite{ChenEnumeration2024,SongConstructions2025,ChenUnconventional2025}, projective representation theory of SSGs {is} discussed. Reference~\cite{ChenUnconventional2025} covers the collinear case, in which the projective structure affects the little cogroups but not the translation algebra or its momentum-space action. {Refs.~\cite{ChenEnumeration2024,SongConstructions2025} construct little-cogroup representations for general SSGs. However, they use the magnetic unit cell convention, which mixes all physical information into an undecomposed factor system. The physical consequences such as multidimensional Brillouin zone, nonsymmorphic actions in momentum space are hidden in their framework. Also, their construction relies on the existence of magnetic unit cell, which only applies to commensurate SSGs but not incommensurate SSGs. In contrast, our theory applies to all SSGs.}

Beyond completing the representation-theoretic construction, our framework revises several basic notions of band theory. We determine the appropriate Brillouin zone when translations are represented projectively, derive the resulting real-space--momentum-space correspondence, and identify the nonsymmorphic momentum-space symmetries realizable by SSGs. These structures shift high-symmetry points and {give rise to} Brillouin spaces that are compact flat manifolds rather than tori, same-chirality Weyl points locked by projective translation symmetry, symmetry-enforced Zak phases and Chern insulators. They connect SSG representation theory to the broader physics of projective crystalline symmetry in magnetic materials and artificial crystals \cite{ChenBrillouin2022,ZhangGeneral2023,XiaoSpin2024,HuHigherOrder2024,FonsecaWeyl2024,PuAcoustic2023,TaoHigherorder2024,LaiRealprojectiveplane2024,LiuTopological2024}.

The remainder of the paper is organized as follows. Section~\ref{SSGstr} reviews the group structure of SSGs, introduces the spin-only group and quotient SSG, and defines the convention of spin operations for the quotient SSG. Section~\ref{Rep} develops the representation theory in two stages. After reviewing the ordinary $k$-star construction, we decompose the factor system into its translation, point-group, and mixed components and use the extended Mackey construction to obtain projective corepresentations of general magnetic groups. We then lift these results to the four classes of SSGs and illustrate the procedure with three examples exhibiting, respectively, collinear symmetry, a nonsymmorphic momentum-space action, and noncommuting translations. Section~\ref{Cons} derives physical consequences of this framework: it constrains which nonsymmorphic momentum-space magnetic groups can be realized by SSGs, explains the reconstruction of high-symmetry momenta and little cogroups, identifies the realizable Brillouin platycosms and their topological implications, derives same-chirality Weyl points locked by projective translation symmetry, spin space group symmetry-enforced Zak phases and Chern insulators. Section~\ref{Conclu} discusses the scope of the theory and the representation databases required for future applications. The appendices provide the cohomological classification and decomposition of factor systems, the extension of Mackey theory to antiunitary groups, explicit projective translation irreducible representations, constraints on realizable nonsymmorphic momentum-space actions, and the projective induction and corepresentation formulas used in the main text.

\section{Structure of spin space groups}\label{SSGstr}
In this section, we briefly review the group structure of SSGs \cite{XiaoSpin2024,SongConstructions2025}. A spin space group contains operations in the following general form:
\begin{equation}
g=\{X_gU_g|R_g|\boldsymbol{t}_g\},
\end{equation} 
where $\{R_g|\boldsymbol{t}_g\}$ is an ordinary space group operation on the lattice, $R_g$ is the point group operation and $\boldsymbol{t}_g$ is the translation (may be fractional). $U_g$ is {an} $SO(3)$ rotation operation on the magnetic moment $\boldsymbol{M}$ but not on the lattice. $X_g$ is either time reversal operation $\mathcal{T}$ or trivial operation $I$. Since $\boldsymbol{M}$ is a pseudovector, $\mathcal{T}$ will reverse its direction. For a magnetic {moment} field $\boldsymbol{M}(\boldsymbol{r})$, the action of $g$ on it is
\begin{equation}
g\boldsymbol{M}(\boldsymbol{r})=s_g U_g\boldsymbol{M}(\{R_g|\boldsymbol{t}_g\}^{-1}\boldsymbol{r})
\end{equation}
where $s_g:=-1$ if $X_g=\mathcal{T}$ and $s_g:=1$ if $X_g=I$. For any magnetic {moment} field $\boldsymbol{M}(\boldsymbol{r})$, its spin space group $\mathcal{G}$ is defined as the group which {leaves it invariant}.

{An} SSG operation in the form $\{XU|I|\boldsymbol{0}\}$ is called a pure-spin symmetry operation (this name is somewhat misleading, since the time reversal operation also acts on lattice in general). All pure-spin symmetry {operations} in $\mathcal{G}$ form a normal subgroup $\mathcal{S}$ of $\mathcal{G}$, which is called the spin-only group.

SSGs can be classified into four classes according to their spin-only group:

(1) Non-magnetic. $\mathcal{S}$ is the full spin rotation symmetry group $SO(3)\rtimes Z_2^T\cong O(3)$, i.e.,
\begin{equation}
\mathcal{S}=\{\{XU|1|\boldsymbol{0}\}|U\in SO(3), X\in Z_2^{T}\}.
\end{equation}
For this class of SSGs, $\mathcal{G}=\mathcal{S}\times G$, where $G$ consists of space group operations in $\mathcal{G}$.

(2) Collinear magnetic order. The directions of $\boldsymbol{M}(\boldsymbol{r})$ are collinear. We can set $\boldsymbol{M}(\boldsymbol{r})=(0,0,M_z(\boldsymbol{r}))$ without loss of generality. $\mathcal{S}$ is given by
\begin{multline}
\mathcal{S}=\{\{U_z(\theta)|1|\boldsymbol{0}\}|\theta \in [0,2\pi)\}\\ \cup {\{\{\mathcal{T}U_{\boldsymbol{n}_\theta}(\pi)|1|\boldsymbol{0}\}|\theta \in [0,\pi)\}},
\end{multline}
where $U_z(\theta)$ is a rotation of the magnetic moment around the $z$ axis with angle $\theta$, $U_{\boldsymbol{n}_\theta}(\pi)$ is a rotation around the axis $\boldsymbol{n}_\theta=(\cos\theta,\sin\theta,0)$ in the $x-y$ plane by angle $\pi$. Mathematically, $\mathcal{S}\cong SO(2) \rtimes Z_2^T \cong O(2)$.

(3) Coplanar magnetic order. The directions of $\boldsymbol{M}(\boldsymbol{r})$ are in the same plane. We can set this plane to be the $x-y$ plane without loss of generality. $\mathcal{S}$ is given by
\begin{equation}
\mathcal{S}=\{\{I|1|\boldsymbol{0}\},\{\mathcal{T}U_z(\pi)|1|\boldsymbol{0}\}\}\cong Z_2^T.
\end{equation}

(4) Non-coplanar magnetic order. $\mathcal{S}$ is trivial.

Since $\mathcal{S}$ is a normal subgroup of $\mathcal{G}$, we can define a quotient group $G=\mathcal{G}/\mathcal{S}$, which is called quotient spin space group (qSSG) of $\mathcal{G}$. Except the identity element, an element in $G$ is a spatial operation followed by a spin operation. Abstractly, $G$ is isomorphic to a space group and there is a homomorphism from $G$ to $O(3)$. Thus, given a space group $G$, inequivalent SSGs can be classified by $H^1(G,O(3))$ \cite{XiaoSpin2024}.

In general, $\mathcal{G}$ is a group extension of $\mathcal{S}$ by $G$, but we can choose a section $s$ of $G$ in $\mathcal{G}$ such that $s(G)$ is a subgroup of $\mathcal{G}$, so we can write $\mathcal{G}=\mathcal{S}\rtimes G$. Furthermore, we can choose the section such that $s(G)$ commute with $\mathcal{S}$, then $\mathcal{G}$ is isomorphic to $\mathcal{S}\times G$. In the following, we set the convention of spin operations in $G$ to make $\mathcal{G}=\mathcal{S}\times G$.

For non-magnetic order, the spin operation part of elements in $G$ can be taken to be trivial, and $G$ is exactly a space group. For collinear magnetic order, if the spin operation part is nontrivial, it flips the magnetic moment field. There are different choices of the nontrivial spin operation for elements in $G$. In this work, we choose it to be the time reversal $\mathcal{T}$. With this choice, $G$ is a normal subgroup of $\mathcal{G}$ and $\mathcal{G}=\mathcal{S}\times G$. More explicitly, we have
\begin{equation}
\mathcal{G}=(SO(2)\times G)\rtimes {Z}_2^T,
\end{equation}
where $SO(2)$ is the unitary subgroup of $\mathcal{S}$, and $Z_2^T=\{I, \mathcal{T}U_x(\pi)\}$. $Z_2^T$ acts trivially on $G$ but {nontrivially} on $SO(2)$. Note for collinear magnetic order, the effective time-reversal $T=\mathcal{T}U_x(\pi)$ satisfies ${T}^2=1$. For coplanar magnetic order, the spin operation part of an element in $G$ rotates or reflects the magnetic moment field  in the $x-y$ plane, which can be described by an $O(3)$ {matrix} $\mathrm{diag}(O_{xy},O_z)$, in which $O_{xy}$ is an $O(2)$ {matrix} and $O_z=\pm 1$. Here, we follow Ref. \cite{XiaoSpin2024} to choose $O_z=\det O_{xy}$, which ensures the spin operation to be unitary. With this choice, $G$ is isomorphic to a space group and $\mathcal{G}=G\times Z_2^T$. Therefore, a coplanar SSG is isomorphic to a type-II MSG with effective time reversal {$T^2=1$}. For non-coplanar magnetic order, since $\mathcal{S}$ is trivial, $\mathcal{G}=G$, which is isomorphic to a magnetic space group.


To enumerate all SSGs in this framework, see Ref. \cite{XiaoSpin2024} for details.

\section{Representation theory of spin space groups}\label{Rep}
In this section, we develop the representation theory of SSGs. This task is challenging, since the $SO(3)$ spin operation must be projectively represented by $SU(2)$ matrices, which results in that the whole SSG being projectively represented. Fortunately, Mackey's representation theory for group extensions (also known as Mackey machine) provides a mathematical method for this task. Formally, if a group $G$ has a normal subgroup $N$, and the quotient group $P=G/N$, then $G$ is a group extension from $N$ by $P$:
\begin{equation}\label{extension1}
1\to N \to G \to P \to 1.
\end{equation}
Mackey's theory allows one to construct all projective irreps of $G$ with factor system $\nu$ from projective irreps of $N$ and projective irreps of subgroup of $P$. First, one constructs all irreps of the normal subgroup $N$ with factor system $\nu$. Then one considers the action of $G$ on the irrep space of $N$, which divides the irrep space of $N$ into orbits. For each orbit, one selects a representative irrep of $N$ and calculates its little group and little cogroup. The little cogroup is always isomorphic to a subgroup of $P$. Next, one constructs all irreps of the little cogroup with a certain factor system, from which the irreps of the little group can be constructed. Finally, one gets irreps of $G$ by induced representations of irreps of little groups. 


Mackey's original theory only treats unitary groups, but its extension to antiunitary groups is straightforward \cite{ParthasarathyProjective1969}. The extended Mackey's theory allows $P$ to be an antiunitary group, while $N$ must be unitary. Since SSGs are antiunitary groups in general, we have to use the extended Mackey's theory. We give an introduction to the extended Mackey's theory in Appendix \ref{appcorep} in a language friendly to physicists.


To apply Mackey's theory for SSGs, we write the four classes of SSGs as group extensions. In the following, we still denote the full spin space group as $\mathcal{G}$ and the quotient spin space group as $G$.

(1) Non-magnetic. Since $\mathcal{G}=G\times (SO(3)\rtimes Z_2^T)$, we have 
\begin{equation}
1 \to G \to \mathcal{G} \to SO(3)\rtimes Z_2^T \to 1,
\end{equation}
where $G$ is isomorphic to a space group in general.

(2) Collinear magnetic order. Since $\mathcal{G}=(SO(2)\times G)\rtimes {Z}_2^T$, we have 
\begin{equation}\label{exten2}
1 \to SO(2)\times G \to \mathcal{G} \to  Z_2^T \to 1.
\end{equation}
If the qSSG $G$ is unitary, the extended Mackey's theory is applicable to Eq. (\ref{exten2}). However, for collinear magnetic SSGs, $G$ contains antiunitary elements in general. In order to apply Mackey's theory, we rewrite the extension as
\begin{equation}\label{2z2}
1 \to SO(2)\times G_0 \to \mathcal{G} \to  Z_2^T\times Z_2^a \to 1.
\end{equation}
where $G_0$ is the unitary subgroup of $G$, and $Z_2^a=G/G_0$. $G_0$ is isomorphic to a space group in general. 

(3) Coplanar magnetic order. Since $\mathcal{G}=G\times Z_2^T$, we have 
\begin{equation}
1 \to  G \to \mathcal{G} \to  Z_2^T \to 1,
\end{equation}
where $G$ is isomorphic to a space group in general.

(4) Noncoplanar magnetic order. $\mathcal{G}=G$, where $G$ is {isomorphic} to a magnetic space group in general.

According to Mackey's theory, to construct projective coirreps of $\mathcal{G}$ with factor system $\nu$ (which we call $\nu$-coirreps for short), we have to first construct $\nu$-(co)irreps of the normal subgroup, i.e., {construct} $\nu$-(co)irreps of $G$ for non-magnetic, coplanar and noncoplanar SSGs, but to construct $\nu$-irreps of $SO(2)\times G_0$ for collinear SSGs. After obtaining $\nu$-(co)irreps of $G$ or $SO(2)\times G_0$, the remaining procedures in Mackey machine are relatively simple, because $SO(3)\times Z_2^T, Z_2^T$ and $Z_2^T\times Z_2^a$ are relatively easy to deal with. Thus, the most challenging part in constructing projective coirreps of SSGs resides in the construction of projective (co)irreps of the qSSG $G$ or the unitary subgroup $G_0$ of $G$. Since $G$ or $G_0$ is isomorphic to a magnetic space group in general, so we have to solve the following general problem: How to construct all $\nu$-(co)irreps for a magnetic space group?

The answer is also Mackey's theory. Here we use a general $G$ to illustrate. In general, $G$ can also be written as a group extension
\begin{equation}
1 \to  \mathrm{T} \to G \to P \to 1,
\end{equation}
where $\mathrm{T}$ is the unitary translational subgroup of $G$, and $P\cong G/\mathrm{T}$ is a magnetic point group. Then we can apply Mackey's theory to construct $\nu$-(co)irreps of $G$. 

In fact, as a special case, ordinary irreps of (magnetic) space groups are also constructed by Mackey machine.  Compared to ordinary irreps, projective irreps are much more complicated. To apply Mackey machine to projective irreps, we must have a proper description of factor system first. Without it, the actual calculation is impossible.  For this purpose, we develop the decomposition of factor systems for group extensions, see Appendix \ref{appfactor}. This decomposition is especially important when we construct $\nu$-coirreps of the qSSG $G$. Briefly speaking, any factor system $\nu$ of $G$ can be decomposed into three parts: a factor system $\sigma$ of the translational subgroup $\mathrm{T}$,  a factor $\alpha$ of the (magnetic) point group $P$, and a mixing factor $\gamma$ relating $\mathrm{T}$ and $P$. $\sigma, \gamma, \alpha$ have explicit qualitative impact on projective irreps of $G$. Nontrivial $\sigma$ leads to high-dimensional irreps of $\mathrm{T}$. Nontrivial $\gamma$ may lead to nonsymmorphic action on the Brillouin zone, while nontrivial $\alpha$ changes factor systems of little cogroups. 

In summary, to construct projective (co)irreps of an SSG $\mathcal{G}$, we can take two steps. The first step is to construct the projective (co)irreps of the quotient SSG $G$ (or the unitary subgroup $G_0$ of $G$) by Mackey machine. The second step is to construct projective (co)irreps of $\mathcal{G}$ from (co)irreps of $G$ (or $G_0$) by Mackey machine. 

This section consists of five subsections. In subsection \ref{ordirep}, we first review ordinary representation theory of space groups, and clarify the challenges when considering projective representations. For later convenience, we also introduce the concept of momentum space group. In subsection \ref{factorsec}, we analyze the factor system of SSGs. We first introduce the origin of factor systems of SSGs. Then we discuss the decomposition of factor system for SSGs.
The decomposition also has two steps, providing basis for the two-step construction of (co)irreps by Mackey machine respectively. Subsection \ref{construct} is the central part of this work. We finish the first step of construction by Mackey machine, i.e., we give the method to construct general projective coirreps of magnetic space group, which covers the cases of qSSGs $G$ (and unitary subgroup $G_0$ of $G$). In the construction, important topics such as the definition of the Brillouin zone and nonsymmorphic actions on BZ are discussed. In subsection \ref{corepSSG}, we finish the second step of construction by Mackey machine, i.e., we show how to construct projective (co)irreps of an SSG $\mathcal{G}$ from (co)irreps of its qSSG $G$ (or $G_0$). Finally, we give three examples in subsection \ref{example} to demonstrate how to apply our method.

\subsection{Review of ordinary irreps of space groups}\label{ordirep}

\subsubsection{Ordinary irreps of space groups}

We first review how ordinary irreps of a space group $G$ are constructed. A space group $G$ can be seen as a group extension of $\mathrm{T}$ by $P$:
\begin{equation}
1\to \mathrm{T}\to G \to P \to 1,
\end{equation}
where $\mathrm{T}$ is the translational subgroup, and $P=G/\mathrm{T}$ is a point group. For any element $g\in G$, we can write it as $g=(t,p)$ with $t\in \mathrm{T}, p\in P$. If we take a section $s(p)\in G$ for $p$, we can write $g=ts(p)$. For symmorphic space groups, $s(p)$ can always be chosen as $s(p)=p$, while for nonsymmorphic groups, $s(p)$ may contain fractional translations. The multiplication law is given as
\begin{equation}
(t_1,p_1)\cdot (t_2,p_2)=(t_1\chi_{p_1}(t_2)\omega(p_1,p_2), p_1p_2),
\end{equation}
where $\chi_{p_1}(t_2)=s(p_1)t_2s(p_1)^{-1}$ is the action of $s(p_1)$ on $t_2$, and $\omega(t_1,t_2)$ is defined as 
\begin{equation}\label{nonomega}
s(p_1)s(p_2)=\omega(p_1,p_2)s(p_1p_2).
\end{equation}

To construct irreps of a space group $G$, we first consider the translational subgroup $\mathrm{T}$, which has one-dimensional irreps 
\begin{equation}\label{transrep}
T^{\boldsymbol{k}}({t})=e^{i\boldsymbol{k}\cdot \boldsymbol{t}},\ \ \ \ \forall  {t}\in \mathrm{T}, \ \ \boldsymbol{k}\in \mathrm{BZ},
\end{equation}
where BZ means the Brillouin  zone. The momentum $\boldsymbol{k}$ resides in the BZ because two momenta differing by a reciprocal vector $\boldsymbol{K}$ result in the same irrep, i.e., $e^{i\boldsymbol{k}\cdot\boldsymbol{t}}=e^{i\boldsymbol{(k+\boldsymbol{K}})\cdot\boldsymbol{t}}$. All the inequivalent irreps of $\mathrm{T}$ form an irrep space $\widehat{\mathrm{T}}$. If we write the lattice defined by $\mathrm{T}$ as $\mathcal{L}$ and the reciprocal lattice as $\widehat{\mathcal{L}}$, then $\widehat{\mathrm{T}}=\mathbb{R}^d/\widehat{\mathcal{L}}$, where $d$ is the dimension of the space.

Next, we consider the action of $G$ on the irrep space $\widehat{\mathrm{T}}$, which is defined by 
\begin{equation}\label{action}
\begin{aligned}
&\mathrm{Act}_g(T^{\boldsymbol{k}})({t})=T^{\boldsymbol{k}}(g^{-1}tg)\\
&\ \ \ \ \ =e^{i\boldsymbol{k}\cdot p_g^{-1}\boldsymbol{t}}=e^{ip_g\boldsymbol{k}\cdot \boldsymbol{t}}=T^{p_g\boldsymbol{k}}(t), \forall  g\in G, t\in \mathrm{T}.
\end{aligned}
\end{equation}
This means $g=(t_g,p_g)$ takes the irrep labelled by $\boldsymbol{k}$ to the irrep labelled by $p_g\boldsymbol{k}$ (if $p_g\boldsymbol{k}$ is out of the BZ, $T^{p_g\boldsymbol{k}}$ means the irrep equivalent to it in BZ). Since irreps of $\mathrm{T}$ are labelled by $\boldsymbol{k}$, we can also say the action of $g$ on the Brillouin  zone is $\mathrm{Act}_g: \boldsymbol{k} \to p_g\boldsymbol{k}$. The action of $G$ on $\widehat{\mathrm{T}}$ divides $\widehat{\mathrm{T}}$ into orbits. Two irreps are in the same orbit if some $g\in G$ relate them by Eq. (\ref{action}). All corresponding $\boldsymbol{k}$ in an orbit form a $k$-star.  

For an irrep $T^{\boldsymbol{k}}$, its little group $G_{\boldsymbol{k}}$ is defined as the subgroup of $G$ that keep it invariant, i.e.,
\begin{equation}
G_{\boldsymbol{k}}=\{g|\mathrm{Act}_g(T^{\boldsymbol{k}})\cong T^{\boldsymbol{k}}\},
\end{equation}
or equivalently,
\begin{equation}
G_{\boldsymbol{k}}=\{g|p_g\boldsymbol{k}-\boldsymbol{k} \in \widehat{\mathcal{L}}\}.
\end{equation}
For irreps of $\mathrm{T}$ on the same orbit, their little groups are isomorphic. Since the translational subgroup $\mathrm{T}$ is always a normal subgroup of $G_{\boldsymbol{k}}$, we can define the little cogroup of $T^{\boldsymbol{k}}$ as 
\begin{equation}
\widetilde{G}_{\boldsymbol{k}}=G_{\boldsymbol{k}}/\mathrm{T}.
\end{equation}
Here we note that $\widetilde{G}_{\boldsymbol{k}}$ is always a point group, but not a subgroup of $G$ when $G$ is nonsymmorphic. 

The irreps of $G$ are constructed as follows.

We first construct projective irreps of $\widetilde{G}_{\boldsymbol{k}}$ with the factor system
\begin{equation}\label{cofactor}
\tilde{\nu}_{\boldsymbol{k}}(p_1,p_2)=e^{i\boldsymbol{k}\cdot \boldsymbol{\omega}(p_1,p_2)},\ \ \forall  p_1,p_2\in \widetilde{G}_{\boldsymbol{k}},
\end{equation}
where $\boldsymbol{\omega}(p_1,p_2)$ is defined in Eq. (\ref{nonomega}).
Then for each projective irrep $\widetilde{D}_{\boldsymbol{k}}$ of $\widetilde{G}_{\boldsymbol{k}}$,  we can construct an irrep $D_{\boldsymbol{k}}$ of $G_{\boldsymbol{k}}$ by lifting $\widetilde{D}_{\boldsymbol{k}}$:
\begin{equation}
D_{\boldsymbol{k}}((t,p))=\widetilde{D}_{\boldsymbol{k}}(p)e^{i\boldsymbol{k}\cdot \boldsymbol{t}}, \ \forall (t,p)\in G_{\boldsymbol{k}},
\end{equation}
Finally, for each  $D_{\boldsymbol{k}}$, we construct the induced representation $\mathrm{Ind}_{G_{\boldsymbol{k}}}^G D_{\boldsymbol{k}}$, which is an irrep of $G$.

For each orbit, we only need to select one representative point $\boldsymbol{k}$ to induce irreps of $G$ from irreps of $G_{\boldsymbol{k}}$, since different choices of $\boldsymbol{k}$ lead to equivalent irreps of $G$. All irreps of $G$ are exhausted by running over all orbits and all projective irreps of little cogroups with factor system $\tilde{\nu}_{\boldsymbol{k}}$. This is the ordinary representation theory of space groups.

For projective representations of (magnetic) space groups, the above construction meets challenges. First, if the factor system restricted on the subgroup $\mathrm{T}$ is nontrivial, i.e., translations do not commute, projective irreps of $\mathrm{T}$ are not one-dimensional anymore. In this situation, how to define the Brillouin  zone? Can we still have the concept that $G$ acts on the BZ? Second, when $G$ has a nontrivial factor system, the action of $G$ on BZ can be nonsymmorphic \cite{ZhangGeneral2023,XiaoSpin2024}. What is the relationship between the nonsymmorphic action and the factor system, and what kinds of nonsymmorphic actions are possible in SSGs? Third, when $G$ has a nontrivial factor system $\nu$, the factor system $\tilde{\nu}_{\boldsymbol{k}}$ of the little cogroup is not in the form in Eq. (\ref{cofactor}). How to determine $\tilde{\nu}_{\boldsymbol{k}}$ is very challenging, especially for the situation when translations do not commute. In the following, we resolve all these difficulties by developing a complete projective corepresentation theory of magnetic space groups based on Mackey's theory.


\subsubsection{Momentum space groups}\label{msg}

For later discussions, here we review the concept of momentum space group or reciprocal space group. In Eq. (\ref{action}), we define the action of $G$ on irreps of $\mathrm{T}$. If we do not restrict momentum $\boldsymbol{k}$ in the Brillouin zone, but consider all $\boldsymbol{k}\in \mathbb{R}_F^3$, where $\mathbb{R}_F^3$ is the whole momentum space, then we can define a momentum space group $G_F$ that acts on $\mathbb{R}_F^3$.

First, since $\boldsymbol{k}$ is equivalent to $\boldsymbol{k}+\boldsymbol{K}, \boldsymbol{K}\in \widehat{\mathcal{L}}$, we can define a translation group $\mathrm{T}_F$ in $\mathbb{R}_F^3$ according to the reciprocal lattice $\widehat{\mathcal{L}}$. Then for every point group operation $p$, we can define its corresponding momentum space point group operation $\tilde{p}$ by considering the action of $p$ on irreps of $\mathrm{T}$. Combining $\mathrm{T}_F$ and all $\tilde{p}$, we get a momentum space group $G_F$.

For ordinary representation theory, no matter whether a point group operation $p$ is symmorphic or nonsymmorphic, its momentum dual operation $\tilde{p}$ is symmorphic. Thus, one can only get symmorphic $G_F$ in $k$-space. For projective representation theory, the factor system  of $G$ will influence the action of $p$ on irreps of $\mathrm{T}$, which may lead to nonsymmorphic $\tilde{p}$ and hence a nonsymmorphic $G_F$.

For magnetic space groups, the time reversal operation $\mathcal{T}$ acts on $k$-space as an inversion which is antiunitary. Thus, $G_F$ is a magnetic momentum space group in general for an MSG.

\subsection{Factor systems of SSGs}\label{factorsec}

\subsubsection{Origin of factor systems of SSGs}
When we consider how an SSG operation $g=\{X_gU_g|R_g|\boldsymbol{v}_g\}$ acts on electronic states, spin rotation $U_g$ should be represented by an $SU(2)$ matrix, and $X_g$ should be represented as $i\sigma_2\mathcal{K}$ if $X_g=\mathcal{T}$. However, in the definition of SSG, $U_g$ only takes values in $SO(3)$. Since $SU(2)$ is a double cover of $SO(3)$, the lifting from $SO(3)$ to $SU(2)$ will lead to an ambiguity of $\pm$ sign, and then the multiplication of two group elements $g_1,g_2\in \mathcal{G}$ will also have a $\pm$ sign ambiguity:
\begin{equation}
X_{g_1}U_{g_1}X_{g_2}U_{g_2}=\nu(g_1,g_2) X_{g_1g_2}U_{g_1g_2},
\end{equation}
where $\nu(g_1,g_2)\in \mathbb{Z}_2=\{\pm 1\}$ is called a factor system. $\nu(g_1,g_2)$ can be determined by the $SU(2)$ matrix of $U_g$:
\begin{equation}\label{defact}
\nu(g_1,g_2)=X_{g_1}U_{g_1}X_{g_2}U_{g_2}U_{g_1g_2}^{-1}X_{g_1g_2}^{-1}.
\end{equation}
One can change the sign of $U_g$ when lifting from $SO(3)$ to $SU(2)$, which will lead to a coboundary transformation of $\nu$, 
\begin{equation}
\nu(g_1,g_2)\to \nu(g_1,g_2)\frac{f(g_1)f(g_2)}{f(g_1g_2)},
\end{equation}
where $f(g)\in \mathbb{Z}_2$. Factor systems that differ by a coboundary transformation are said to be equivalent. Mathematically, all inequivalent factor systems of an SSG $\mathcal{G}$ are classified by $H^{2,c}(\mathcal{G},U(1))$ (see Appendix \ref{cohomocl}). A general factor system of $\mathcal{G}$ can take values in $U(1)$. But by lifting $SO(3)$ to $SU(2)$, factor systems can only take values in $\mathbb{Z}_2$, so this mechanism only results in a subset of the full classes of factor systems of $\mathcal{G}$ in general.

\subsubsection{Factor systems decomposition of SSGs}

If a group is a product or semidirect product of two groups, then its factor system can be decomposed into three parts \cite{MackeyUnitary1958}. Formally, if $G=K\rtimes H$, then an element $g=G$ can be written as $g=kh, k\in K, h\in H$. A factor system $\nu$ of $G$ can be decomposed as
\begin{equation}\label{semideco}
\begin{aligned}
\nu(k_1h_1,k_2h_2)=\nu_K(k_1,\chi_{h_1}(k_2))\eta(k_2,h_1)\nu_{H}(h_1,h_2),\\
 \ \forall k_1,k_2\in K, \forall h_1,h_2\in H,
\end{aligned}
\end{equation}
where $\nu_K, \nu_H$ is the factor system $\nu$ restricted on $K$ and $H$ respectively. $\eta$ is a factor relating $K$ and $H$, $\chi_h(k)=hkh^{-1}$. With arbitrary $\nu$-corep $\rho$ of $G$, $\eta$ can be written as $\eta(k,h)=\rho(h)\rho(k)\rho^{-1}(h)\rho^{-1}(hkh^{-1})$ (for more details, see Appendix \ref{appfactor}). This allows us to decompose factor systems of SSGs.

For an SSG $\mathcal{G}$ of non-magnetic order, the factor system $\nu_q$ of the qSSG $G$ is trivial. Since elements in $\mathcal{S}=SO(3)\rtimes Z_2^T$ all commute with elements in $G$, the factor $\eta$ is also trivial. Thus, $\nu=\nu_s$, where $\nu_s$ the factor system of $\mathcal{S}$ caused by lifting $SO(3)$ to $SU(2)$ as well as $T^2=-1$.

For an SSG $\mathcal{G}$ of collinear magnetic order, since $\mathcal{G}=(SO(2)\times G)\rtimes {Z}_2^T$, we can apply the canonical decomposition Eq. (\ref{semideco}) twice and get the following decomposition:
\begin{equation}
\begin{aligned}
\nu_{ssg}&((s_1,g_1,k_1),(s_2,g_2,k_2))\\
&=\nu_s(s_1,s_2)\nu_s(k_1,k_2)\nu_q(g_1,g_2)\eta_1(s_2,g_1)\eta_2(s_2g_2,k_1),\\
&\forall s_1,s_2\in SO(2), \forall k_1,k_2\in Z_2^T, \forall g_1,g_2\in G,
\end{aligned}
\end{equation}
where $\nu_s$ is the factor system of $\mathcal{S}$, $\nu_q$ is the factor system of $G$, $\eta_1$ relates $G$ and $SO(2)$ and $\eta_2$ relates $Z_2^T$ and $SO(2)\rtimes G$. However, under the convention we set in Sec. \ref{SSGstr}, $\eta_1, \eta_2$ are both trivial, and $\nu_s(k_1,k_2)=1$, so we have a simple form of decomposition 
\begin{equation}
\begin{aligned}
\nu_{ssg}((s_1,g_1,k_1),&(s_2,g_2,k_2))=\nu_s(s_1,s_2)\nu_q(g_1,g_2).
\end{aligned}
\end{equation}
$\nu_s$ is still caused by lifting $SO(3)$ matrices to $SU(2)$ matrices, while the form of $\nu_q$ is also simple, that is, $\nu_q(g_1,g_2)=-1$ if $g_1,g_2$ both have nontrivial spin operation, otherwise $\nu_q(g_1,g_2)=1$.

For an SSG $\mathcal{G}$ of coplanar magnetic order, since $\mathcal{G}=G\times \mathcal{S}=G\times Z_2^T$, so we can decompose a factor system $\nu_{ssg}$ of $\mathcal{G}$ as
\begin{equation}\label{copdeco}
\begin{aligned}
\nu_{ssg}((g_1,k_1),(g_2,k_2))=\nu_q&(g_1,g_2)\eta(g_2,k_1),\\
&\forall g_1,g_2\in G, \forall k_1,k_2\in \mathcal{S},
\end{aligned}
\end{equation}
where $\nu_q$ is the factor system of $G$. Note in our setting $\nu_s(k_1,k_2)=1$, so only $\nu_q$ and $\eta$ appear in Eq. (\ref{copdeco}). $\eta(g,k)=-1$ if $k$ is the nontrivial element in $Z_2^T$ and anticommutes with the spin operation part of $g$, otherwise $\eta(g,k)=1$.

For an SSG $\mathcal{G}$ of non-coplanar magnetic order, since $\mathcal{G}=G$, the factor system $\nu_{ssg}=\nu_q$. 

\subsubsection{Factor systems decomposition of MSGs}\label{decoqssg}

In the following, we will analyze the decomposition of the factor system $\nu_q$ of the qSSG $G$. Since $G$ is isomorphic to an MSG in general, we discuss factor systems of a general MSG. 

For a general MSG $G$, its unitary translations form a normal subgroup $\mathrm{T}$, thus we can write $G$ as a group extension
\begin{equation}
1\to \mathrm{T} \to G \to P \to 1,
\end{equation}
where $P\cong G/\mathrm{T}$ is the magnetic point group. With respect to this group extension structure of $G$, an element $g\in G$ can be written as a pair $(t,p), t\in \mathrm{T}, p \in P$. Note $p$ could be a time-reversal operation. We can choose a section $s(p)\in G$ for each $p$ and write $(t,p)=ts(p)$ explicitly. In general, $s(p)$ is the point group operation $p$ followed by a fractional translation.  If $p$ itself is a group element of $G$, we choose $s(p)=p$. Then the multiplication rule reads:
\begin{equation}
(t_1,p_1)\cdot (t_2,p_2) =(t_1\chi_{p_1}(t_2)\omega(p_1,p_2),p_1p_2),
\end{equation}
where $\chi_{p}(t)=s(p)ts(p)^{-1}=ptp^{-1} \in \mathrm{T}$ is the action of $s(p)$ on $t$ and $\omega(p_1,p_2)$ is defined as 
\begin{equation}
s(p_1)s(p_2)=\omega(p_1,p_2)s(p_1p_2).
\end{equation}

By adopting the decomposition $g=(t,p)=ts(p)$, we can decompose a factor system $\nu'$ of $G$ into three parts in the sense of equivalence class. According to the decomposition theorem of factor systems of group extensions in Appendix \ref{appfactor}, formally, $\nu'$ is equivalent to a factor system $\nu$, which can be decomposed as 
\begin{equation}\label{decom}
\begin{aligned}
\nu(g_1,g_2)&=\nu((t_1,p_1),(t_2,p_2))\\
&=\sigma(t_1\chi_{p_1}(t_2),\omega(p_1,p_2))\sigma(t_1,\chi_{p_1}(t_2))\\
&\ \ \ \ \gamma(t_2,p_1)\alpha(p_1,p_2),
\end{aligned}
\end{equation}
where 
\begin{equation}
\begin{aligned}
&\sigma(t_1,t_2):=\nu(t_1,t_2), \\
&\alpha(p_1,p_2):=\nu(s(p_1),s(p_2)), \\
&\gamma(t,p):=\frac{\nu(s(p),t)\nu(s(p)t,s(p)^{-1})}{\nu(s(p),s(p)^{-1})}.
\end{aligned}
\end{equation}
$\sigma$ is the factor system restricted on the translational subgroup $\mathrm{T}$, $\alpha$ is a factor about the point group $P$ (but not a factor system of $P$ in general), and $\gamma$ is a mixing factor relating $\mathrm{T}$ and $P$, which can also be written as $\gamma(t,p)=\rho(s(p))\rho(t)\rho^{-1}(s(p))\rho^{-1}(\chi_{p}(t))$ by an arbitrary $\nu$-corep $\rho$. We call Eq. (\ref{decom}) the canonical decomposition of factor systems for MSGs.

$\sigma,\alpha,\gamma$ satisfy the following self-consistency equations
\begin{equation}
\sigma(t_1,t_2)\sigma(t_1t_2,t_3) = \sigma(t_2,t_3)\sigma(t_1,t_2t_3),
\end{equation}
\begin{equation}\label{alpham}
\begin{aligned}
&\frac{\alpha(p_1,p_2)\alpha(p_1p_2,p_3)}{c_{p_1}[\alpha(p_2,p_3)]\alpha(p_1,p_2p_3)}=\\&\ \ \ \ \ \ \gamma(\omega(p_2,p_3),p_1)\frac{\sigma(\chi_{p_1}(\omega(p_2,p_3)),\omega(p_1,p_2p_3))}{\sigma(\omega(p_1,p_2),\omega(p_1p_2,p_3))},
\end{aligned}
\end{equation}

\begin{equation}\label{sigmam}
\frac{\sigma(\chi_{p}(t_1),\chi_{p}(t_2))}{c_{p}[\sigma(t_1,t_2)]}=\frac{\gamma(t_1t_2,p)}{\gamma(t_1,p)\gamma(t_2,p)},
\end{equation}

\begin{equation}\label{gm}
\frac{c_{p_1}[\gamma(t,p_2)]\gamma(\chi_{p_2}(t),p_1)}{\gamma(t,p_1p_2)}=\frac{\sigma(\omega(p_1,p_2),\chi_{p_1p_2}(t))}{\sigma(\chi_{p_1p_2}(t),\omega(p_1,p_2))}.
\end{equation}
where $c_p:=\mathcal{K}$ if $p$ is antiunitary. 

The decomposition is meaningful. If $\sigma$ is nontrivial, the generators of the translational subgroup $\mathrm{T}$ do not commute, which will lead to two-dimensional $\sigma$-irreps of $\mathrm{T}$. The $\gamma$ factor is related to nonsymmorphic action of the point group on the Brillouin  zone. We will discuss these in detail when constructing coirreps of $G$.

The factor system $\nu$ contains redundancy since it can be changed by coboundary transformations. To capture the essential algebraic relation for a $\nu$-corep, we can use relations of generators and cohomology invariants to describe the projective symmetry algebra \cite{ChenClassification2023}. The values of a complete set of cohomology invariants uniquely determine the equivalence class of the factor system. 
The equivalence class of the factor system is important, but the explicit form is not. In practical calculations and applications, rather than dealing with the whole factor system, we only deal with the modified algebra of generators. But when establishing the representation theory, we still use the full factor system, but only keep it in the sense of equivalence class. 

\subsection{Construction of $\nu$-(co)irreps of MSGs}\label{construct}

Here we introduce the procedures to construct $\nu$-(co)irreps of a general MSG. Since a qSSG $G$ (or its unitary subgroup $G_0$) is always isomorphic to an MSG in general, this section provides general method to construct $\nu_q$-(co)irreps for qSSG $G$ (and $G_0$). In this section, we use $G$ to denote a general MSG and use $\nu$ to denote a general factor system of $G$.

\subsubsection{Projective irreps of the translational subgroup}\label{Ptran}

First, we construct all inequivalent $\nu$-irreps of the translational subgroup $\mathrm{T}$. We can write the constraint of $\nu$ on $\mathrm{T}$ as $\sigma$, i.e., $\sigma(t_1,t_2)=\nu(t_1,t_2)$. If $\sigma$ is trivial, the $\sigma$-irreps of $\mathrm{T}$ are just that given in Eq. (\ref{transrep}). If $\sigma$ is nontrivial, translations do not commute and the $\sigma$-irreps are not one-dimensional.

The equivalence class of $\sigma$ can be determined by three independent cohomological invariants:
\begin{equation}
\begin{aligned}
\sigma_{ab}=\frac{\sigma(t_a,t_b)}{\sigma(t_b,t_a)},\sigma_{bc}=\frac{\sigma(t_b,t_c)}{\sigma(t_c,t_b)},\sigma_{ca}=\frac{\sigma(t_c,t_a)}{\sigma(t_a,t_c)},
\end{aligned}
\end{equation}
where $t_a,t_b,t_c$ are three generators of $\mathrm{T}$.  For SSGs, since $\nu$ can only take $\mathbb{Z}_2=\{\pm1\}$ values, nontrivial $\sigma_{ij}$ can only take $\mathbb{Z}_2$ values. But we note that the possible values of $\sigma_{ij}, i,j=a,b,c$ are constrained by point group symmetries since $\sigma(t_1,t_2)$ should satisfy the consistency equations (\ref{alpham}-\ref{gm}), thus not all $\mathbb{Z}_2$ values of $\sigma_{ij}$ are possible for a certain $G$ in general.

For convenience of discussion, we define the projective translation algebra $\mathrm{T}^\sigma$ for the translational subgroup $\mathrm{T}$. The elements in $\mathrm{T}^\sigma$ are the same as those in $\mathrm{T}$, but their algebraic relations are modified by $\sigma$. For example, 
\begin{equation}
\mathsf{t}_a\mathsf{t}_b=\sigma_{ab}\mathsf{t}_b\mathsf{t}_a, 
\end{equation}
where $\mathsf{t}_a,\mathsf{t}_b$ are the corresponding elements of $t_a,t_b$ in $\mathrm{T}^\sigma$ respectively. One can see that a $\sigma$-irrep of $\mathrm{T}$ is just a rep of the projective translation algebra $\mathrm{T}^\sigma$.

To construct $\sigma$-irreps of $\mathrm{T}$, we first consider the center $Z(\mathrm{T}^\sigma)$ of the projective translation algebra $\mathrm{T}^\sigma$, which consists of elements that commute with all elements in $\mathrm{T}^\sigma$, i.e.,
\begin{equation}
Z(\mathrm{T}^\sigma)=\{\mathsf{t}|\mathsf{t}\mathsf{t}'=\mathsf{t}'\mathsf{t}, \forall \mathsf{t}'\in \mathrm{T}^\sigma\}.
\end{equation}
Elements in $Z(\mathrm{T}^\sigma)$ define a sublattice $\mathcal{Z}_{\mathcal{L}}\subset \mathcal{L}$, which has an enlarged unit cell. 

For the contexts we considered, if there is nontrivial $\sigma_{ij}$, the quotient algebra $\mathrm{T}^\sigma/Z(\mathrm{T}^\sigma)$ is always a Heisenberg algebra (see Appendix \ref{irreptrans})
\begin{equation}\label{Heis}
\mathrm{Heis}(\mathbb{Z}_2\times \mathbb{Z}_2)=\langle \mathcal{P},\mathcal{Q}| \mathcal{P}^2=1, \mathcal{Q}^2=1, \mathcal{P} \mathcal{Q}=- \mathcal{Q} \mathcal{P}\rangle,
\end{equation}
where $ \mathcal{P}, \mathcal{Q}$ are two generators. Thus, we have a short exact sequence 
\begin{equation}\label{tranex}
1\to Z(\mathrm{T}^\sigma)\to \mathrm{T}^\sigma \to \mathrm{Heis}(\mathbb{Z}_2\times \mathbb{Z}_2)\to 1,
\end{equation}
which is a central extension, but not split. 

To construct $\sigma$-irreps of $\mathrm{T}$, we first construct irreps of $Z(\mathrm{T}^\sigma)$, which are given by ordinary irreps
\begin{equation}\label{tsirrep}
T^{\boldsymbol{k}}(t)=e^{i\boldsymbol{k}\cdot \boldsymbol{t}},\ \ \ t\in Z(\mathrm{T}^\sigma), \ \ \boldsymbol{k}\in \mathbb{R}^d/\widehat{\mathcal{Z}}_{\mathcal{L}}.
\end{equation}
Note here the Brillouin  zone $\mathrm{BZ}=\mathbb{R}^d/\widehat{\mathcal{Z}}_{\mathcal{L}}$ is defined according to the reciprocal lattice $\widehat{\mathcal{Z}}_{\mathcal{L}}$ rather than the original reciprocal lattice $\widehat{\mathcal{L}}$. 

Then we can construct $\sigma$-irreps of $\mathrm{T}$ according to Mackey's theory (see Appendix \ref{irreptrans}). There are only three essentially different sets of nontrivial values for cohomological invariants $\sigma_{ij}$, and their $\sigma$-irreps are listed in the following. We use $t_a,t_b,t_c$ to denote the three generators of $\mathrm{T}$.

(1) Only one $\sigma_{ij}=-1$. Without loss of generality, we assume $\sigma_{ab}=-1, \sigma_{bc}=\sigma_{ca}=1$. The $\sigma$-irreps can be taken as
\begin{equation}
\begin{aligned}
&T^{\boldsymbol{k}}(t_a)=e^{i\boldsymbol{k}\cdot \boldsymbol{t}_a}\sigma_1,\ \  T^{\boldsymbol{k}}(t_b)=e^{i\boldsymbol{k}\cdot \boldsymbol{t}_b}\sigma_3,\\
&T^{\boldsymbol{k}}(t_c)=e^{i\boldsymbol{k}\cdot \boldsymbol{t}_c}\sigma_0,\ \ \boldsymbol{k}\in \mathrm{BZ}.
\end{aligned}
\end{equation}

(2) Two $\sigma_{ij}=-1$,  Without loss of generality, we assume $\sigma_{ab}=\sigma_{ca}=-1, \sigma_{bc}=1$. The $\sigma$-irreps can be taken as
\begin{equation}
\begin{aligned}
&T^{\boldsymbol{k}}(t_a)=e^{i\boldsymbol{k}\cdot \boldsymbol{t}_a}\sigma_1,\ \ T^{\boldsymbol{k}}(t_b)=e^{i\boldsymbol{k}\cdot \boldsymbol{t}_b}\sigma_3,\\
&T^{\boldsymbol{k}}(t_c)=e^{i\boldsymbol{k}\cdot \boldsymbol{t}_c}\sigma_3,\ \ \boldsymbol{k}\in \mathrm{BZ}.
\end{aligned}
\end{equation}

(3) Three $\sigma_{ij}=-1$. The $\sigma$-irreps can be taken as
\begin{equation}
\begin{aligned}
&T^{\boldsymbol{k}}(t_a)=-ie^{i\boldsymbol{k}\cdot \boldsymbol{t}_a}\sigma_1,\ \  T^{\boldsymbol{k}}(t_b)=-ie^{i\boldsymbol{k}\cdot \boldsymbol{t}_b}\sigma_2,\\
&T^{\boldsymbol{k}}(t_c)=-ie^{i\boldsymbol{k}\cdot \boldsymbol{t}_c}\sigma_3,\ \ \boldsymbol{k}\in \mathrm{BZ}.
\end{aligned}
\end{equation}

Here we note for each case, the $\sigma$-irreps of $\mathrm{T}$ can still be labelled by momentum $\boldsymbol{k}$, and the Brillouin  zone is defined with respect to the center of $\mathrm{T}^\sigma$. {In the literature}, there is another way to define the Brillouin zone according to the minimal abelian subgroup of $\mathrm{T}^\sigma$ \cite{XiaoSpin2024}. We do not adopt this way here, because when considering action of point group on the Brillouin  zone, only the definition we adopted supports a good picture of orbits, which is necessary for constructing irreps by Mackey machine.

\subsubsection{Actions on the Brillouin  zone}\label{actonBZ}
To construct coirreps of $G$, we need to consider the action of $G$ on the irrep space $\widehat{\mathrm{T}}$ of the translational subgroup $\mathrm{T}$. Since inequivalent irreps in $\widehat{\mathrm{T}}$ are labelled by $\boldsymbol{k}$ in the Brillouin  zone, so we also refer to actions on $\widehat{\mathrm{T}}$ as actions on the Brillouin  zone. Actions on the Brillouin  zone are important, because it defines the symmetry in the $\boldsymbol{k}$ space. Recent studies reveal that projective reps of (magnetic) space groups in real space can lead to nonsymmorphic symmetries in $\boldsymbol{k}$ space, which goes beyond ordinary theories \cite{ChenBrillouin2022,ZhangGeneral2023,XiaoSpin2024}. Here, we will show how nonsymmorphic symmetries in $\boldsymbol{k}$ space are related to factor systems of $G$, especially related to factor $\gamma$.

An element $g\in G$ acts on the irrep space $\widehat{\mathrm{T}}$ by 
\begin{equation}\label{actdefine}
\mathrm{Act}_g(T^{\boldsymbol{k}})(t)=c_g[\xi(g|t)T^{\boldsymbol{k}}(g^{-1}tg)],
\end{equation}
where 
\begin{equation}
\xi(g|t)=\frac{\nu\left(g^{-1}, t\right) \nu(g^{-1} t, g)}{\nu(g^{-1}, g)}.
\end{equation}
This definition can be justified as follows. Suppose $|\psi_{i,\boldsymbol{k}}\rangle$ is a basis of the $\sigma$-irrep $T^{\boldsymbol{k}}$ of the translational subgroup, i.e.,
\begin{equation}
\hat{t}|\psi_{i,\boldsymbol{k}}\rangle=T^{\boldsymbol{k}}_{ji}(t)|\psi_{j,\boldsymbol{k}}\rangle.
\end{equation}
where $\hat{t}$ is the operator of translation $\boldsymbol{t}$.
To see how a group operation $g$ acts on BZ, we consider the action of $\hat{t}$ on $\hat{g}|\psi_{i,\boldsymbol{k}}\rangle$,
\begin{equation}\label{gaction}
\begin{aligned}
\hat{t}\hat{g}|\psi_{i,\boldsymbol{k}}\rangle&=\nu^{-1}(g,g^{-1})\hat{g}\widehat{g^{-1}}\hat{t}\hat{g}|\psi_{i,\boldsymbol{k}}\rangle\\
&=\frac{c_g[\nu(g^{-1},t)\nu(g^{-1}t,g)]}{\nu(g,g^{-1})}\hat{g} T^{\boldsymbol{k}}_{ji}(g^{-1}tg)|\psi_{j,\boldsymbol{k}}\rangle\\
&=c_g[\xi(g|t)T^{\boldsymbol{k}}_{ji}(g^{-1}tg)]\hat{g}|\psi_{j,\boldsymbol{k}}\rangle.
\end{aligned}
\end{equation}
Note in the last step we use $c_g[\nu(g,g^{-1})]= \nu(g^{-1}, g)$.  So $\hat{g}|\psi_{i,\boldsymbol{k}}\rangle$ is exactly the basis of the rep $\mathrm{Act}_g(T^{\boldsymbol{k}})$.

To further proceed, we first discuss the case with trivial factor system of the translational subgroup, i.e., $\sigma_{ij}=1$, in which case the irrep is $T^{\boldsymbol{k}}(t)=e^{i\boldsymbol{k}\cdot \boldsymbol{t}}$. Since actions of translations on $T^{\boldsymbol{k}}$ are trivial, we only need to consider actions of (magnetic) point group operations $g=s(p)$ on $T^{\boldsymbol{k}}$. In this setting, {$\xi(g|t)=\gamma(\chi_{p^{-1}}(t),p)$}, and
{
\begin{equation}\label{s1act1}
\mathrm{Act}_p(T^{\boldsymbol{k}})(t)=c_p[\gamma(\chi_{p^{-1}}(t),p)e^{ip\boldsymbol{k}\cdot \boldsymbol{t}}],
\end{equation}}
where $p\boldsymbol{k}$ is the consequence of the action of group point operation $p$ on vector $\boldsymbol{k}$. Considering that antiunitary operations invert $\boldsymbol{k}$, we define 
\begin{equation}
\begin{aligned}
\tilde{p}\boldsymbol{k}=\left\{\begin{array}{ll} p\boldsymbol{k}, \ \ \ \ \ c_p=1\\
-p\boldsymbol{k}, \ \ \ c_p= \mathcal{K}\end{array}\right.
\end{aligned}.
\end{equation}
For a magnetic point group $P$, all corresponding $\tilde{p}$ for $p \in P$ also consist of a magnetic point group $\tilde{P}$, which we call the dual magnetic point group of $P$. With the definition of $\tilde{p}$, we have 
{
\begin{equation}\label{s1act}
\mathrm{Act}_p(T^{\boldsymbol{k}})(t)=c_p[\gamma(\chi_{p^{-1}}(t),p)]e^{i\tilde{p}\boldsymbol{k}\cdot \boldsymbol{t}}.
\end{equation}}
If the factor $\gamma$ is trivial, then we come back to traditional theory, i.e., 
$\mathrm{Act}_p(T^{\boldsymbol{k}})=T^{\tilde{p}\boldsymbol{k}}$, or simply written, $\mathrm{Act}_p: \boldsymbol{k}\to \tilde{p}\boldsymbol{k}$. In traditional theory, no matter whether $s(p)$ is symmorphic or nonsymmorphic, its action on BZ is always symmorphic. 

In general, the factor $\gamma$ is nontrivial, which may lead to nonsymmorphic actions on BZ. When the factor system $\sigma$ of the translational subgroup is trivial, {Eq. (\ref{sigmam}) and Eq. (\ref{gm}) become}
\begin{equation}\label{s2}
\gamma(t_1,p)\gamma(t_2,p)=\gamma(t_1t_2,p),
\end{equation}
\begin{equation}\label{g2}
c_{p_1}[\gamma(t,p_2)]\gamma(\chi_{p_2}(t),p_1)=\gamma(t,p_1p_2).
\end{equation}
The first equation has solutions in the form
\begin{equation}\label{ept}
\gamma(t,p)=e^{-i\boldsymbol{\kappa}_{\tilde{p}}\cdot p\boldsymbol{t}},
\end{equation}
Here the $p$ in the exponent is for later convenience. Plug Eq. (\ref{ept}) into Eq. (\ref{g2}), we have
{
\begin{equation}\label{frac1}
\boldsymbol{\kappa}_{\tilde{p}_1}+\tilde{p}_1\boldsymbol{\kappa}_{\tilde{p}_2}-\boldsymbol{\kappa}_{\tilde{p}_1\tilde{p}_2} \in \widehat{\mathcal{L}},
\end{equation}}
which is a 1-cocycle equation for $\boldsymbol{\kappa}_{\tilde{p}}$ and is exactly the equation satisfied by fractional translations for a magnetic point group $\tilde{P}$ with lattice $\widehat{\mathcal{L}}$ in $k$-space. With solution Eq. (\ref{ept}), Eq. (\ref{s1act}) gives 
\begin{equation}
\mathrm{Act}_p: \boldsymbol{k} \to \tilde{p}\boldsymbol{k}+\boldsymbol{\kappa}_{\tilde{p}}.
\end{equation}
Thus, $p$ may induce a nonsymmorphic action on BZ. Of course, not all $\boldsymbol{\kappa}_{\tilde{p}}$ means an essential nonsymmorphic action. For a point group $P$, if its corresponding $\boldsymbol{\kappa}_{\tilde{p}}=\boldsymbol{k}-\tilde{p}\boldsymbol{k}$ for some $\boldsymbol{k}$, then fractional translations $\boldsymbol{\kappa}_{\tilde{p}}$ are not essential and can be cancelled by resetting the fixed point of $\tilde{P}$. $\boldsymbol{k}-\tilde{p}\boldsymbol{k}$ is called a 1-coboundary. Two 1-cocycle differing by a 1-coboundary are equivalent. The class of nonsymmorphic actions with (magnetic) point group $\tilde{P}$ and lattice $\widehat{\mathcal{L}}$ are classified by inequivalent $\boldsymbol{\kappa}_{\tilde{p}}$. In fact, a 1-coboundary translation of $\boldsymbol{\kappa}_{\tilde{p}}$ corresponds to a coboundary transformation of $\gamma$. For symmorphic group $G$, since Eq. (\ref{s2}) and Eq. (\ref{g2}) are the full equations $\gamma$ should obey, so we have the following correspondence:

\textit{When $\sigma=1$, the equivalence classes of $\gamma(t,p)$ for a symmorphic MSG $G=\mathcal{L}\rtimes P$ are in one-to-one correspondence to the equivalence classes of k-space MSGs in the arithmetic class $(\tilde{P},\widehat{\mathcal{L}})$.}

For nonsymmorphic MSGs, since $\gamma(t,p)$ should satisfy another constraint Eq. (\ref{alpha}), only a part of fractional translations $\boldsymbol{\kappa}_{\tilde{p}}$ are allowed. 

According to the correspondence between $\gamma$ and fractional translation, all $k$-space (magnetic) groups can be realized by proper choice of factor systems of $r$-space (magnetic) groups. However, in the context of SSGs, some $k$-space nonsymmorphic MSGs cannot be realized, because some factor systems of $r$-space MSGs cannot be realized by SSGs. In general, if the $k$-orbits of a $k$-space nonsymmorphic MSG all have dimensions larger than two, then this $k$-space MSG cannot be realized by SSGs. We arrive at this conclusion by analysing dimensions of coirreps of SSGs, see Appendix. \ref{nonsymapp}. Realizable $k$-space MSGs by SSGs are listed in Sec. \ref{actiond2}.

When the factor system of the translational subgroup is nontrivial, the irreps of $\mathrm{T}$ are two-dimensional but still labelled by $\boldsymbol{k}$, so we can still discuss the action of $G$ on BZ. The action is still defined by Eq. (\ref{actdefine}).  For general factor systems which are nontrivial on the subgroup, there can also be nonsymmorphic actions on BZ (see Ref. \cite{ChenFluxMediated2026} for detailed analysis).  But for factor systems arising from spin space groups, no essential nonsymmorphic action on BZ is possible when factor $\sigma$ is nontrivial. So for noncommuting BZ, group actions on BZ are all symmorphic. This conclusion is also proved by analysing dimensions of coirreps of SSGs, see Appendix. \ref{nonsymapp}. 

\subsubsection{{High-symmetry} points and little groups}\label{hsplg}

The action of $G$ on the irrep space $\widehat{\mathrm{T}}$ divides $\widehat{\mathrm{T}}$ into disjoint orbits. In each orbit, irreps can be related by $G$ actions. To construct irreps of $G$, the next procedure is to find the little group and little cogroup for each orbit. For an irrep $T^{\boldsymbol{k}}$, its little group $G_{\boldsymbol{k}}$ consists of all elements keeping it invariant, i.e.,
\begin{equation}
G_{\boldsymbol{k}}=\{g|\mathrm{Act}_g(T^{\boldsymbol{k}})\cong T^{\boldsymbol{k}}\}.
\end{equation}
Since $T^{\boldsymbol{k}}$ is determined by $\boldsymbol{k}$, we can also write 
\begin{equation}
G_{\boldsymbol{k}}=\{g|\tilde{p}_g\boldsymbol{k}+\boldsymbol{\kappa}_{\tilde{p}_g}-\boldsymbol{k} \in \widehat{\mathcal{L}}\},
\end{equation}
where $g=(t_g,p_g)$ and $\tilde{p}_g$ is the dual point group operation of $p_g$ in $k$-space. For irreps in the same orbit, their little groups are isomorphic to each other. 

If $G_{\boldsymbol{k}}$ is nontrivial, we say $\boldsymbol{k}$ is a high-symmetry point. Compared to ordinary representations of $G$, projective representations can have nonsymmorphic actions on BZ, which will change high-symmetry points. 

Another difference here from ordinary representation theory is that, when $T^{\boldsymbol{k}}$ is two-dimensional, for $h\in G_{\boldsymbol{k}}$, $\mathrm{Act}_h(T^{\boldsymbol{k}})$ and   $T^{\boldsymbol{k}}$ can differ by a unitary transformation in general, i.e.,
\begin{equation}\label{Utrf}
\mathrm{Act}_h(T^{\boldsymbol{k}})(t)=U_{\boldsymbol{k}}(h)^\dagger T^{\boldsymbol{k}}(t) U_{\boldsymbol{k}}(h),\ \forall h \in G_{\boldsymbol{k}}.
\end{equation}
In fact, $U_{\boldsymbol{k}}$ forms a matrix  projective corep of $G_{\boldsymbol{k}}$:
\begin{equation}\label{Ufac}
U_{\boldsymbol{k}}(h_1)c_{h_1}[U_{\boldsymbol{k}}(h_2)]=\tau_{\boldsymbol{k}}(h_1,h_2)U_{\boldsymbol{k}}(h_1h_2),
\end{equation}
where $\tau_{\boldsymbol{k}}$ is a factor system of $G_{\boldsymbol{k}}$, which is determined by $U_{\boldsymbol{k}}$. $U_{\boldsymbol{k}}(h)$ has a $U(1)$ phase degree of freedom. Changing the phase of $U_{\boldsymbol{k}}(h)$ only leads to a coboundary transformation of $\tau_{\boldsymbol{k}}$.

Since the translational subgroup $\mathrm{T}$ is always a normal subgroup of $G_{\boldsymbol{k}}$, we can define the little cogroup of $T^{\boldsymbol{k}}$ as 
\begin{equation}
\widetilde{G}_{\boldsymbol{k}}=G_{\boldsymbol{k}}/\mathrm{T}.
\end{equation}

In constructing ordinary corepresentations of $G$, one procedure is to identify the factor system of $\widetilde{G}_{\boldsymbol{k}}$ as Eq. (\ref{cofactor}). Since now $G$ has a  nontrivial factor system $\nu$, the factor system of $\widetilde{G}_{\boldsymbol{k}}$ will {no longer} be in the simple form in Eq. (\ref{cofactor}). We address this question as follows. Note that $\tau_{\boldsymbol{k}}$ and $\nu$ (constrained on $G_{\boldsymbol{k}}$) are two factor systems for $G_{\boldsymbol{k}}$. If we always choose transformation matrix $U_{\boldsymbol{k}}(t)=T^{\boldsymbol{k}}(t)$ for $\forall t\in \mathrm{T}$, then $\tau_{\boldsymbol{k}}(t_1,t_2)=\nu(t_1,t_2)$ on $\mathrm{T}$. This motivates us to define 
\begin{equation}\label{taufac}
\omega'_{\boldsymbol{k}}(h_1,h_2)=\frac{\nu(h_1,h_2)}{\tau_{\boldsymbol{k}}(h_1,h_2)}, \ \ \ \forall h_1,h_2\in G_{\boldsymbol{k}},
\end{equation}
which is also a factor system of  $G_{\boldsymbol{k}}$, but trivial on $\mathrm{T}$. In fact, by properly choosing the phase of $U(h)$, $\omega'_{\boldsymbol{k}}$ can be made into the form (see Appendix \ref{appcorep})
\begin{equation}\label{omegafac}
\omega'_{\boldsymbol{k}}(h_1,h_2)=\omega_{\boldsymbol{k}}([h_1],[h_2]), 
\end{equation}
where $\omega_{\boldsymbol{k}}$ is a factor system of the little cogroup $\widetilde{G}_{\boldsymbol{k}}$, and $[h]$ means the equivalence class corresponding to $h$. $\omega_{\boldsymbol{k}}$ is just the factor system we need for $\widetilde{G}_{\boldsymbol{k}}$. Since factor system of $\widetilde{G}_{\boldsymbol{k}}$ influences dimensions of irreps of $\widetilde{G}_{\boldsymbol{k}}$, projective coreps can lead to change of degeneracies on high-symmetry points compared to ordinary coreps.

\subsubsection{Constructing $\nu$-coirreps of $G$}

Knowing the factor system $\omega_{\boldsymbol{k}}$ for little cogroup $\widetilde{G}_{\boldsymbol{k}}$, we can construct $\nu$-coirreps of $G$ as follows. First, find $\omega_{\boldsymbol{k}}$-coirreps $\widetilde{D}_{\boldsymbol{k}}^{(i)}$ of $\widetilde{G}_{\boldsymbol{k}}$, then for each $\widetilde{D}_{\boldsymbol{k}}^{(i)}$, we can construct a $\nu$-coirrep ${D}_{\boldsymbol{k}}^{(i)}$ for the little group ${G}_{\boldsymbol{k}}$:
\begin{equation}\label{cotol}
{D}_{\boldsymbol{k}}^{(i)}(h)=U_{\boldsymbol{k}}(h)\otimes \widetilde{D}_{\boldsymbol{k}}^{(i)}([h]),\ \  \forall  h \in {G}_{\boldsymbol{k}}.
\end{equation}
Note that $U_{\boldsymbol{k}}$ has factor system $\tau_{\boldsymbol{k}}$ and $\widetilde{D}_{\boldsymbol{k}}^{(i)}$ has factor system $\omega_{\boldsymbol{k}}$, thus ${D}_{\boldsymbol{k}}^{(i)}$ has factor system $\nu$. 

For each ${D}_{\boldsymbol{k}}^{(i)}$, the induced $\nu$-corep $\mathrm{Ind}_{G_{\boldsymbol{k}}}^{G} {D}_{\boldsymbol{k}}^{(i)}$ is a $\nu$-coirrep of $G$. All inequivalent $\nu$-coirreps of $G$ can be obtained by running over all orbits (only need to select one representative point for each orbit) and all ${D}_{\boldsymbol{k}}^{(i)}$. Compared to ordinary induced representations, induced projective representations need additional modification by factor systems. An introduction to induced projective representation is presented in Appendix \ref{Induce}.

\subsubsection{Procedures to construct $\nu$-coirreps of $G$}

Here we summarize the procedures to construct general $\nu$-coirreps of an MSG $G$.

(1) Determine the irreps of the translational subgroup $\mathrm{T}$ according to $\sigma_{ij}$.

(2) Determine action of $G$ (especially point group operations) on irreps of $\mathrm{T}$ by Eq. (\ref{actdefine}).

(3) Determine high-symmetry points ($k$-stars) and their little cogroups.

(4) For each $k$-star, select one representative point $\boldsymbol{k}$, determine the unitary transformation matrices $U_{\boldsymbol{k}}(h)$ for $h\in G_{\boldsymbol{k}}$ according to Eq. (\ref{Utrf}). Then determine the factor system $\tau_{\boldsymbol{k}}$ of $U_{\boldsymbol{k}}$ 
according to Eq. (\ref{Ufac}).

(5) Determine the factor system $\omega_{\boldsymbol{k}}$ of the little cogroup $\widetilde{G}_{\boldsymbol{k}}$ according to Eq. (\ref{taufac}) and Eq. (\ref{omegafac}).

(6) Construct all $\omega_{\boldsymbol{k}}$-coirreps $\widetilde{D}_{\boldsymbol{k}}^{(i)}$ of the little cogroup $\widetilde{G}_{\boldsymbol{k}}$.

(7) Construct $\nu$-coirreps ${D}_{\boldsymbol{k}}^{(i)}$ of the little group $G_{\boldsymbol{k}}$ by Eq. (\ref{cotol}).

(8) Induce $\nu$-coirreps of $G$ by ${D}_{\boldsymbol{k}}^{(i)}$. 

All $\nu$-coirreps of $G$ are constructed by repeating procedures (5) to (8) for all $k$-stars.

\subsection{Construction of $\nu$-coirreps of SSGs}\label{corepSSG}

After constructing $\nu_q$-(co)irreps of the qSSG $G$ (or its unitary subgroup $G_0$), we can construct $\nu_{ssg}$-coirreps of the SSG $\mathcal{G}$. This can be done by applying Mackey machine once more. Compared to Sec. \ref{construct}, construction in this step is relatively simple. Since non-magnetic, collinear, coplanar and noncoplanar SSGs have different group extension structure, we discuss the construction of their projective (co)irreps separately.

\subsubsection{Non-magnetic} 

For non-magnetic order, an SSG $\mathcal{G}$ is a group extension
\begin{equation}
1 \to G \to \mathcal{G} \to SO(3)\rtimes Z_2^T \to 1,
\end{equation}
which is split. The qSSG $G$ is unitary and is isomorphic to a space group. The factor system of $G$ is trivial, while the factor system of $SO(3)\rtimes Z_2^T$ is given by lifting $SO(3)$ to $SU(2)$ and $T^2=-1$.

Suppose irreps of $G$ are denoted as $\rho_{\boldsymbol{k}}^{(i)}$, which means the irrep induced from the $i$th $\nu$-irrep $D_{\boldsymbol{k}}^{(i)}$ of the little group $G_{\boldsymbol{k}}$. To apply Mackey machine, we consider the action of $SO(3)\rtimes Z_2^T$ on the irrep space of $G$. Since operations in $SO(3)\rtimes Z_2^T$ commute with operations in $G$, the action of unitary operations in $SO(3)\rtimes Z_2^T$ on irreps of $G$ is trivial, so we only need to consider the action of the time-reversal operation $T=\mathcal{T}$, which is given as
\begin{equation}
\mathrm{Act}_T(\rho_{\boldsymbol{k}}^{(i)})=\rho_{\boldsymbol{k}}^{(i)*},
\end{equation}
where $\rho_{\boldsymbol{k}}^{(i)*}$ is the complex conjugate rep of $\rho_{\boldsymbol{k}}^{(i)}$. There are three possibilities of the relation between $\rho_{\boldsymbol{k}}^{(i)}$ and $\rho_{\boldsymbol{k}}^{(i)*}$.

{Case (i)} $\rho_{\boldsymbol{k}}^{(i)*}=U^\dagger(T)\rho_{\boldsymbol{k}}^{(i)}U(T)$, and the transformation matrix satisfies $U(T)U(T)^*=1$.

In this case, the little cogroup is $SO(3)\rtimes Z_2^T$, whose factor system still satisfies $T^2=-1$. Then the $\nu_{ssg}$-coirrep of $\mathcal{G}$ is given as
\begin{equation}
\begin{aligned}
&\rho((g,s))=\rho_{\boldsymbol{k}}^{(i)}(g)\otimes D_s^{(1/2)}(s),\ \ \ \forall g\in G, s\in SO(3),\\
&\rho(T)=U(T)\otimes i\sigma_2 \mathcal{K}.
\end{aligned}
\end{equation}

{Case (ii)} $\rho_{\boldsymbol{k}}^{(i)*}=U^\dagger(T)\rho_{\boldsymbol{k}}^{(i)}U(T)$, and the transformation matrix satisfies $U(T)U(T)^*=-1$.

In this case, the little cogroup is $SO(3)\rtimes Z_2^T$, and the factor system of the little cogroup is changed to $T^2=1$ due to $U(T)$. If the pure spin rotations are represented by the $j=1/2$ projective irrep $D_s^{(1/2)}$ of $SO(3)$, the irrep of $SO(3)\rtimes Z_2^T$ is four-dimensional. And the $\nu_{ssg}$-coirrep of $\mathcal{G}$ is given as
\begin{equation}
\begin{aligned}
&\rho((g,s))=\rho_{\boldsymbol{k}}^{(i)}(g)\otimes D_s^{(1/2)}(s)\otimes \sigma_0,\ \ \forall g\in G, s\in SO(3),\\
&\rho(T)=U(T)\otimes i\sigma_2 \otimes i\sigma_2 \mathcal{K}.
\end{aligned}
\end{equation}
Since a nontrivial $U(T)$ requires $\rho_{\boldsymbol{k}}$ to be at least two-dimensional, a coirrep in case (i) is at least eight-dimensional.

Case (iii) $\rho_{\boldsymbol{k}}^{(i)*}$ is not equivalent to $\rho_{\boldsymbol{k}}^{(i)}$.

In this case, the little cogroup is just $SO(3)$. The $\nu_{ssg}$-coirrep of $\mathcal{G}$ is given as
\begin{equation}
\begin{aligned}
&\rho((g,s))=\left(\begin{array}{cc} \rho_{\boldsymbol{k}}^{(i)}(g) & 0 \\ 0 &   \rho_{\boldsymbol{k}}^{(i)*}(g)\end{array}\right) \otimes D_s^{(1/2)}(s),\\
&\rho(T)=\left(\begin{array}{cc} 0 & I \\ I & 0 \end{array}\right) \otimes i\sigma_2 \mathcal{K},\ \ \ \ \forall g\in G, s\in SO(3).\\
\end{aligned}
\end{equation}
In this case, $\rho_{\boldsymbol{k}}^{(i)}$ and $\rho_{\boldsymbol{k}}^{(i)*}$ form a pair to consist of a coirrep of $\mathcal{G}$.

The three cases can be determined by the following criteria \cite{BradleyMagnetic1968}
\begin{equation}
\sum_{g\in G}\chi(g^2)=
\left\{
\begin{aligned}
&+|G|,\ \ \ \mathrm{case (i)}\\
&-|G|,\ \ \ \mathrm{case (ii)}\\
&0,\ \ \ \ \ \  \ \ \ \mathrm{case (iii)}
\end{aligned}
\right.
\end{equation}
$\chi(g^2)$ is the character of $g^2$ in irreps of $G$. This criterion is a special case of the general criteria in Appendix \ref{Criteria}. The general procedures to construct projective coirreps of an antiunitary group from its projective irreps of its unitary subgroup are also introduced in Appendix \ref{Criteria}.


\subsubsection{Collinear magnetic order}\label{collinearrep}

For collinear magnetic order,  according to Eq. (\ref{2z2}), we can write an SSG $\mathcal{G}$ as the following group extension:
\begin{equation}\label{coex}
1\to SO(2)\times G_0 \to \mathcal{G} \to Z_2^T\times Z_2^a \to 1,
\end{equation}
where $G_0$ is the unitary subgroup of the qSSG $G$ and $Z_2^a\cong G/G_0$. This extension is not split in general since $G$ is not a (semi)direct product of $G_0$ and $Z_2^a$ in general. Writing $\mathcal{G}$ as a group extension in Eq. (\ref{coex}) makes the application of Mackey machine possible, since $SO(2)\times G_0$ is unitary.

To apply Mackey machine, the first step is to construct projective irreps of $SO(2)\times G_0$, which are just tensor products of projective irreps of $SO(2)$ and irreps of $G_0$. For spin-1/2 fermions, there are two nonequivalent projective irreps of $SO(2)$, 
\begin{equation}
R_s(\theta)=e^{-is\theta/2},\ \ \ \forall \theta\in [0,2\pi),\ \ \ s=\pm1.
\end{equation}
$s=+1$ and $s=-1$ correspond to spin-up and spin-down states respectively. Since $G_0$ is isomorphic to a space group, we can label its irreps as $\rho_{\boldsymbol{k}}^{(i)}$. Note the factor system restricted on $G_0$ is trivial, so $\rho_{\boldsymbol{k}}^{(i)}$ are ordinary irreps of a space group. The projective irreps of $SO(2)\times G_0$ can be written as
\begin{equation}
\rho_{\boldsymbol{k},s}^{(i)}((\theta,g))=e^{-is\theta/2}\rho_{\boldsymbol{k}}^{(i)}(g),\ \ \forall \theta\in [0,2\pi), \ g\in G_0.
\end{equation}
This means, at every $\boldsymbol{k}$, there are spin-up and spin-down states. In the following, we will see that elements in $G$ with nontrivial spin operations will relate spin-up and spin-down states at different $\boldsymbol{k}$ points.

After knowing projective irreps of $SO(2)\times G_0$, the next step is to consider the action of $Z_2^T\times Z_2^a$ on them. Since the nontrivial element in $Z_2^a$ interchanges $R_+$ and $R_-$, $Z_2^a$ is not in the little cogroup $\tilde{G}_{\rho_{\boldsymbol{k},s}^{(i)}}$. The effective time reversal operation $T \in Z_2^T$ keeps $R_{s}$ invariant, so we only need to consider the action of $T$ on $\rho_{\boldsymbol{k}}^{(i)}$, which simply results in $\rho_{\boldsymbol{k}}^{(i)*}$. Then according to the relation between $\rho_{\boldsymbol{k}}^{(i)*}$ and $\rho_{\boldsymbol{k}}^{(i)}$, there are three cases.

Case (i) $\rho_{\boldsymbol{k}}^{(i)*}=U^\dagger(T)\rho_{\boldsymbol{k}}^{(i)}U(T)$ with $U(T)U(T)^*=1$. 

In this case, the little cogroup is $Z_2^T$, whose factor system is trivial. The $\nu_{ssg}$-coirrep of the little group ${G}_{\rho_{\boldsymbol{k},s}^{(i)}}=(SO(2)\times G_0)\rtimes Z_2^T$  is given by:
\begin{equation}
\begin{aligned}
&{\rho'}_{\boldsymbol{k},s}^{(i)}((\theta,g))=e^{-is\theta/2}\rho_{\boldsymbol{k}}^{(i)}(g),\ \ \ \forall \theta\in [0,2\pi), \  g\in G_0,\\
&{\rho'}_{\boldsymbol{k},s}^{(i)}(T)=U(T)\mathcal{K}.
\end{aligned}
\end{equation}
Then the $\nu_{ssg}$-coirrep of $\mathcal{G}$ is induced by ${\rho'}_{\boldsymbol{k},s}^{(i)}$, which can be written as
\begin{equation}
\begin{aligned}
&\rho(\theta)=\left(\begin{array}{cc} e^{-is\theta/2} I & 0 \\ 0 &  e^{is\theta/2}I \end{array}\right),\ \ \forall \theta\in [0,2\pi),\\
&\rho(h)=\left(\begin{array}{cc} \rho_{\boldsymbol{k}}^{(i)}(h) & 0 \\ 0 &  \rho_{\boldsymbol{k}}^{(i)*}(a^{-1}ha) \end{array}\right),\ \ \forall h\in G_0,\\
&\rho(a)=\left(\begin{array}{cc} 0 & -\rho_{\boldsymbol{k}}^{(i)}(a^2)\\ I  & 0 \end{array}\right)  \mathcal{K},\\
&\rho(T)=\left(\begin{array}{cc} U(T) &\ 0 \\ \ 0 & U(T)^{*} \end{array}\right)  \mathcal{K},
\end{aligned}
\end{equation}
where the minus sign in $\rho(a)$ is due to the factor system $\nu(a,a)=-1$. Since $R_{\pm}$ is in the same orbit under the action of $a$, $s=\pm 1$ result in equivalent $\nu_{ssg}$-coirreps of $\mathcal{G}$.

Case (ii) $\rho_{\boldsymbol{k}}^{(i)*}=U^\dagger(T)\rho_{\boldsymbol{k}}^{(i)}U(T)$ with $U(T)U(T)^*=-1$. 

In this case, the little cogroup is $Z_2^T$, whose factor system satisfies $T^2=-1$. The $\nu_{ssg}$-coirrep of the little group ${G}_{\rho_{\boldsymbol{k},s}^{(i)}}=(SO(2)\times G_0)\rtimes Z_2^T$ is given by:
\begin{equation}
\begin{aligned}
&{\rho'}_{\boldsymbol{k},s}^{(i)}((\theta,g))=\sigma_0 \otimes e^{-is\theta/2}\rho_{\boldsymbol{k}}^{(i)}(g),\ \ \forall \theta\in [0,2\pi), \ g\in G_0,\\
&{\rho'}_{\boldsymbol{k},s}^{(i)}(T)=i\sigma_2\otimes U(T)\otimes\mathcal{K}.
\end{aligned}
\end{equation}
Then the $\nu_{ssg}$-coirrep of $\mathcal{G}$ is induced by ${\rho'}_{\boldsymbol{k},s}^{(i)}$, which can be written as
\begin{equation}
\begin{aligned}
&\rho(\theta)=\left(\begin{array}{cc} e^{-is\theta/2} \sigma_0 \otimes I & 0 \\ 0 & e^{is\theta/2} \sigma_0\otimes I  \end{array}\right),\ \ \forall\theta\in [0,2\pi),\\
&\rho(h)=\left(\begin{array}{cc} \sigma_0 \otimes \rho_{\boldsymbol{k}}^{(i)}(h)& 0 \\ 0 & \sigma_0 \otimes \rho_{\boldsymbol{k}}^{(i)*}(a^{-1}ha) \end{array}\right),\ \ \forall h\in G_0,\\
&\rho(a)=\left(\begin{array}{cc} 0 & -\sigma_0 \otimes \rho_{\boldsymbol{k}}^{(i)}(a^2)\\ \sigma_0 \otimes I & 0 \end{array}\right) \mathcal{K},\\
&\rho(T)=\left(\begin{array}{cc} i\sigma_2 \otimes U(T)&\ 0 \\ \ 0 & i\sigma_2 \otimes U(T)^{*}\end{array}\right) \mathcal{K},
\end{aligned}
\end{equation}
Again, $s=\pm 1$ result in equivalent $\nu_{ssg}$-coirreps of $\mathcal{G}$.

Case (iii) $\rho_{\boldsymbol{k}}^{(i)*}$ is not equivalent to $\rho_{\boldsymbol{k}}^{(i)}$. 

In this case, the little cogroup is trivial, and the $\nu_{ssg}$-coirrep of $\mathcal{G}$ is induced from $\rho_{\boldsymbol{k}}^{(i)}$, which can be written as
\begin{equation}
\begin{aligned}
&\rho(\theta)=\left(\begin{array}{cc} e^{-is\theta/2} \sigma_0\otimes I & 0 \\ 0 &  e^{is\theta/2}\sigma_0 \otimes I \end{array}\right),\ \ \forall \theta\in [0,2\pi),\\
&\rho(h)=\left(\begin{array}{cccc} \rho_{\boldsymbol{k}}^{(i)}(h) & 0 & 0 & 0 \\ 0 & \rho_{\boldsymbol{k}}^{(i)*}(h) & 0 & 0 \\ 0& 0 & \rho_{\boldsymbol{k}}^{(i)*}(a^{-1}ha) & 0 \\0 & 0& 0 & \rho_{\boldsymbol{k}}^{(i)}(a^{-1}ha)\end{array}\right),\\
&\ \ \ \ \ \ \ \ \ \ \ \ \ \ \ \ \ \ \  \  \ \ \ \ \ \ \ \ \ \ \ \ \ \ \ \ \ \   \ \ \ \ \ \ \ \ \ \ \ \ \ \ \ \ \ \  \forall h\in G_0,\\
&\rho(a)=\left(\begin{array}{cccc} 0 & 0 & -\rho_{\boldsymbol{k}}^{(i)}(a^2)  & 0 \\ 0 & 0 & 0 & -\rho_{\boldsymbol{k}}^{(i)*}(a^2) \\ I & 0 &0 & 0 \\0 &  I & 0 & 0 \end{array}\right)\mathcal{K},\\
&\rho(T)=\left(\begin{array}{cc} \sigma_1 \otimes I & 0 \\ 0 & \sigma_1 \otimes I \end{array}\right)  \mathcal{K}.
\end{aligned}
\end{equation}
In this case, $s=\pm 1$ lead to equivalent coirreps, $\rho_{\boldsymbol{k}}^{(i)}$ and $\rho_{\boldsymbol{k}}^{(i)*}$ also lead to equivalent coirreps.

The three cases can be determined by the following criteria \cite{BradleyMagnetic1968}
\begin{equation}
\sum_{g\in G_0}\chi(g^2)=
\left\{
\begin{aligned}
&+|G_0|,\ \ \ \mathrm{case (i)}\\
&-|G_0|,\ \ \ \mathrm{case (ii)}\\
&0,\ \ \ \ \ \  \ \ \ \ \mathrm{case (iii)}
\end{aligned}
\right.
\end{equation}

{From the above} construction, we notice that $\nu_{ssg}$-coirreps of $\mathcal{G}$ can be seen as induced from coirreps of its subgroup $SO(2)\rtimes (G_0\times Z_2^T)$. Thus, we can take another simpler way to construct $\nu_{ssg}$-coirreps of $\mathcal{G}$. First, since $G_0\times Z_2^T$ is isomorphic to a type-II MSG with $T^2=1$ and trivial factor system, its coirreps are already known and labelled by $\boldsymbol{k}$, which we denote as $M_{\boldsymbol{k}}^{(i)}$. Next, since $G_0\times Z_2^T$ acts trivially on irreps $R_s$ of $SO(2)$, the coirreps of $SO(2)\rtimes (G_0\times Z_2^T)$ are
\begin{equation}
M_{\boldsymbol{k},s}^{(i)}((\theta,g))=e^{-is\theta/2}M_{\boldsymbol{k}}^{(i)}(g),\ \ \forall \theta\in [0,2\pi),\ \ g\in G_0\times Z_2^T.
\end{equation}
Finally, the $\nu_{ssg}$-coirreps of $\mathcal{G}$ can be induced from $M_{\boldsymbol{k},s}^{(i)}$, whose matrices can be written as
\begin{equation}\label{collimagcon}
\begin{aligned}
&\rho(\theta)=\left(\begin{array}{cc} e^{-is\theta/2} I & 0 \\ 0 &  e^{is\theta/2}I \end{array}\right),\ \ \forall \theta\in [0,2\pi),\\
&\rho(h)=\left(\begin{array}{cc} M_{\boldsymbol{k}}^{(i)}(h) & 0 \\ 0 &  M_{\boldsymbol{k}}^{(i)*}(a^{-1}ha) \end{array}\right),\ \ \forall h\in G_0\times Z_2^T,\\
&\rho(a)=\left(\begin{array}{cc} 0 & -M_{\boldsymbol{k}}^{(i)}(a^2)\\ I  & 0 \end{array}\right)  \mathcal{K}.
\end{aligned}
\end{equation}

Here we note all the above discussions in this section assume $Z_2^a$ is nontrivial. If $Z_2^a$ is trivial, i.e., $G$ is itself unitary, the $\nu_{ssg}$-coirreps of $\mathcal{G}$ are just $M_{\boldsymbol{k},s}^{(i)}$, where $s=\pm 1$ are inequivalent coirreps. This means, if $G$ is unitary, i.e., spatial operations are not modified by spin operations, spin up and spin down states are not related by spatial operations. But for general cases when $G$ contains antiunitary elements, $s=\pm1$ states are related by spatial operations.

\subsubsection{Coplanar magnetic order} 

For coplanar magnetic orders, $\mathcal{G}=G\times {Z}_2^T$. The qSSG $G$ is isomorphic to a space group and has factor system $\nu_q$. The effective time-reversal operation {$T^2=1$}. The mixing factor between $G$ and $Z_2^T$ is $\eta$, where $\eta(g,T)=-1$ if $g$ anticommutes with $T$. 

We denote the $\nu_q$-irreps of $G$ as  $\rho_{\boldsymbol{k}}^{(i)}$, which can be constructed by the method in Sec. \ref{construct}. To apply Mackey machine to construct $\nu_{ssg}$-coirreps of $\mathcal{G}$, we consider the action of $T$ on $\nu_q$-irreps of $G$, which is given by:
\begin{equation}
\mathrm{Act}_{T}(\rho_{\boldsymbol{k}}^{(i)})(g)=\eta(g,T)\rho_{\boldsymbol{k}}^{(i)*}(g),\ \ \forall g\in G.
\end{equation}

There are three cases.

Case (i) $\mathrm{Act}_{T}(\rho_{\boldsymbol{k}}^{(i)})=U^\dagger(T)\rho_{\boldsymbol{k}}^{(i)}U(T)$ with {$U(T)U(T)^*=1$}.

In this case, the little cogroup is $Z_2^T$ with factor system $T^2=1$, and the $\nu_{ssg}$-coirrep of $\mathcal{G}$ is given as
\begin{equation}
\begin{aligned}
&\rho(g)=\rho_{\boldsymbol{k}}^{(i)}(g) ,\ \ \forall g\in G, \\
&\rho(T)=U(T) \mathcal{K}.
\end{aligned}
\end{equation}

Case (ii) $\mathrm{Act}_{T}(\rho_{\boldsymbol{k}}^{(i)})=U^\dagger(T)\rho_{\boldsymbol{k}}^{(i)}U(T)$ with {$U(T)U(T)^*=-1$}.

In this case, the little cogroup is $Z_2^T$ with factor system {$T^2=-1$}, and the $\nu_{ssg}$-coirrep of $\mathcal{G}$ is given as
\begin{equation}
\begin{aligned}
&\rho(g)=\sigma_0\otimes \rho_{\boldsymbol{k}}^{(i)}(g),\ \ \forall g\in G, \\
&\rho(T)= i\sigma_2\otimes U(T) \mathcal{K}.
\end{aligned}
\end{equation}

Case (iii) $\mathrm{Act}_{T}(\rho_{\boldsymbol{k}}^{(i)})$ is not equivalent to $\rho_{\boldsymbol{k}}^{(i)}$.

In this case, the little cogroup is trivial, and the $\nu_{ssg}$-coirrep of $\mathcal{G}$ is given as
\begin{equation}
\begin{aligned}
&\rho(g)=\left(\begin{array}{cc} \rho_{\boldsymbol{k}}^{(i)}(g) & 0 \\ 0 &   \eta(g,T)\rho_{\boldsymbol{k}}^{(i)*}(g)\end{array}\right),\\
&\rho(T)=\left(\begin{array}{cc} 0 & I \\ I & 0 \end{array}\right)  \mathcal{K},\ \ \ \ \forall g\in G.\\
\end{aligned}
\end{equation}
Note in this case, $\rho_{\boldsymbol{k}}^{(i)}$ and $\rho_{\boldsymbol{k}}^{(i)*}$ come into a pair to form a $\nu_{ssg}$-coirrep of $\mathcal{G}$.

The three cases can be determined by the following criteria
\begin{equation}
\sum_{g\in G}\eta(g,T)\nu_q(g,g)\chi(g^2)=
\left\{
\begin{aligned}
&+|G|,\ \ \ \mathrm{case (i)}\\
&-|G|,\ \ \ \mathrm{case (ii)}\\
&0,\ \ \ \ \ \  \ \  \ \mathrm{case (iii)}
\end{aligned}
\right.
\end{equation}
This criterion is obtained from the general criteria in Appendix \ref{Criteria}.

We can also take another way to construct $\nu_{ssg}$-coirreps of $\mathcal{G}$. Notice that $\mathcal{G}$ is isomorphic to a type-II MSG with nontrivial factor system, so we can apply the method in Sec. \ref{construct} directly to construct $\nu_{ssg}$-coirreps of it. The group extension can be written as
\begin{equation}
1\to \mathrm{T} \to \mathcal{G} \to P \to 1,
\end{equation}
where the magnetic point group $P$ contains the effective time reversal $T$.

\subsubsection{Noncoplanar magnetic order}
For noncoplanar magnetic orders, the qSSG $G$ is just the whole SSG $\mathcal{G}$, i.e., $\mathcal{G}=G$, so $\nu_{ssg}$-coirreps of $\mathcal{G}$ are just $\nu_q$-coirreps of $G$, which can be constructed by the method in Sec. \ref{construct}. 

\subsection{Examples}\label{example}

Here we give three examples to demonstrate how to construct projective (co)irreps of spin space groups by our method. We adopt the notations of SSGs from Ref. \cite{XiaoSpin2024}. 

\subsubsection{$\mathrm{L140.2.8\ M1^+}$}\label{example1}
The first example we consider is a collinear spin space group $\mathcal{G}=\mathrm{L140.2.8\ M1^+}$, which was treated in Ref. \cite{XiaoSpin2024} and is considered to be the symmetry group of $\mathrm{FeGe_2}$. Here, we construct $\nu_{ssg}$-coirreps of $\mathcal{G}$ by our method. The parent group of $\mathcal{G}$ is $I4/mcm$. The generators of $\mathcal{G}/SO(2)=G\times Z_2^T$ can be taken to be $t_i, i=1,2,3$, $C_{4z}=\{I|C_{4z}|0\}$,  $s_y=\{I|C_{2y}|0,0,1/2\}$, $I=\{I|\bar{1}|0\}$, $\tau=\{\mathcal{T}|1|1/2,1/2,1/2\}$, $T=\mathcal{T}U_{x}(\pi)$ \cite{XiaoSpin2024}, where $t_i$ are three translations corresponding to three lattice vectors of the conventional unit cell, $C_{4z}$ is {$\pi$/2}-rotation around $z$-axis, $s_y$ is a screw rotation, $I$ is the inversion, and $\tau$ is a half translation with a nontrivial spin operation, and $T$ is the effective time-reversal.

According to Sec. \ref{collinearrep}, to construct projective coirreps of $\mathcal{G}$, we can first construct ordinary coirreps of $G_0\times Z_2^T$, where $G_0$ is the unitary subgroup of $G$. $G_0\times Z_2^T$ is isomorphic to the type-II MSG $P4/mcc1'$. The ordinary coirreps of $P4/mcc1'$ is known and labelled by high-symmetry points, which we denote as $M_{\boldsymbol{k}}^{(i)}$. For demonstration, here we only discuss coirreps at the $\Gamma$ point as an example. 

At the $\Gamma$ point, the irrep of the translational subgroup is trivial. The little group is $G_0\times Z_2^T$ itself and the little cogroup is the magnetic point group 4/mmm1'. 4/mmm1' has eight 1D coirreps and two 2D coirreps. We denote them as $\Gamma^{(i)}$. Then the coirreps of the little group $G_0\times Z_2^T$ are just
\begin{equation}
\begin{aligned}
&M_{\Gamma}^{(i)}((t,R))=\Gamma^{(i)}(R),\\
&M_{\Gamma}^{(i)}(T)=\Gamma^{(i)}(T).
\end{aligned}
\end{equation}
where $R$ is the point group operation part of an element. Then according to Eq. (\ref{collimagcon}), the $\nu_{ssg}$-coirreps of $\mathcal{G}$ are induced by $M_{\Gamma}^{(i)}$:
\begin{equation}
\begin{aligned}
&\rho(\theta)=\left(\begin{array}{cc} e^{-is\theta/2} I & 0 \\ 0 &  e^{is\theta/2}I \end{array}\right),\ \ \forall \theta\in [0,2\pi),\\
&\rho(h)=\left(\begin{array}{cc} M_{\Gamma}^{(i)}(h) & 0 \\ 0 &  M_{\Gamma}^{(i)*}(\tau^{-1}h\tau) \end{array}\right),\ \ \forall h\in G_0,\\
&\rho(\tau)=\left(\begin{array}{cc} 0 & -I\\ I  & 0 \end{array}\right)  \mathcal{K},\\
&\rho(T)=\left(\begin{array}{cc} \Gamma^{(i)}(T) &\ 0 \\ \ 0 & \Gamma^{(i)}(T)^{*} \end{array}\right)  \mathcal{K},
\end{aligned}
\end{equation}
Since $M_{\Gamma}^{(i)*}(\tau^{-1}h\tau) $ is also a coirrep at $\Gamma$, the degeneracy at $\Gamma$ is doubled by $\tau$.

\subsubsection{$\mathrm{P6.3.2\ GM2\oplus A1}$}\label{example2}
The next example we consider is a coplanar spin space group $\mathcal{G}=\mathrm{P6.3.2\ GM2\oplus A1}$, which has nonsymmorphic actions on BZ. $\mathcal{G}$ is isomorphic to a type-II MSG Pm1'. The parent group of $\mathcal{G}$ is Pm, whose presentation can be written as

\begin{equation}
\begin{aligned}
&\mathrm{Pm}=\langle t_a,t_b,t_c,\sigma_z|t_at_bt_a^{-1}t_b^{-1}=1,\\ &t_bt_ct_b^{-1}t_c^{-1}=1,t_ct_at_c^{-1}t_a^{-1}=1, \sigma_zt_a\sigma_z^{-1}t_a^{-1}=1, \\
 &\sigma_zt_b\sigma_z^{-1}t_b^{-1}=1,\sigma_zt_c\sigma_z^{-1}t_c=1,\sigma_z^2=1\rangle.
\end{aligned}
\end{equation}
The quotient spin space group $G$ is isomorphic to Pm. The $SU(2)$ spin rotation matrices for generators are 
\begin{equation}\label{Ug}
\begin{aligned}
&U(t_a)=-i\sigma_2,\ \  U(t_b)=-i\sigma_2, \\
&U(t_c)=\sigma_0, \ \ U(\sigma_z)=-i\sigma_1.
\end{aligned}
\end{equation}
This leads to a nontrivial factor system $\nu_q$ for $G$. We can characterize the factor system (in the sense of the equivalence class) by the following modified relations of generators:
{
\begin{equation}
\begin{aligned}
&\mathsf{t}_a\mathsf{t}_b\mathsf{t}_a^{-1}\mathsf{t}_b^{-1}=1,\\
&\mathsf{t}_b\mathsf{t}_c\mathsf{t}_b^{-1}\mathsf{t}_c^{-1}=1,\\
&\mathsf{t}_c\mathsf{t}_a\mathsf{t}_c^{-1}\mathsf{t}_a^{-1}=1,\\
&\sigma_z\textsf{t}_a\sigma_z^{-1}\textsf{t}_a^{-1}=-1, \\
&\sigma_z\textsf{t}_b\sigma_z^{-1}\textsf{t}_b^{-1}=-1,\\
&\sigma_z\textsf{t}_c\sigma_z^{-1}\textsf{t}_c=1,\\
&\sigma_z^2=-1.
\end{aligned}
\end{equation}}

Now we construct $\nu_q$-irreps of $G$. First, since $t_a,t_b,t_c$ commute with each other, the irreps of the translational subgroup are given by 
\begin{equation}
T^{\boldsymbol{k}}(t)=e^{i\boldsymbol{k}\cdot \boldsymbol{t}},\ \ \ \boldsymbol{k}\in \mathrm{BZ}.
\end{equation}
Then we consider the action of $\sigma_z$ on the Brillouin  zone. The $\gamma$ factors between three translational generators and $\sigma_z$ are
\begin{equation}
\begin{aligned}
&\gamma(t_a,\sigma_z)=\mathsf{\sigma}_z \mathsf{t}_a \mathsf{\sigma}_z^{-1}\mathsf{t}_a^{-1}=-1, \\
&\gamma(t_b,\sigma_z)=\mathsf{\sigma}_z \mathsf{t}_b \mathsf{\sigma}_z^{-1}\mathsf{t}_b^{-1}=-1, \\
&\gamma(t_c,\sigma_z)=\mathsf{\sigma}_z \mathsf{t}_c \mathsf{\sigma}_z^{-1}\mathsf{t}_c=1.
\end{aligned}
\end{equation}
This leads to a nonsymmorphic action of $\sigma_z$ on BZ, with fractional translation $(\pi,\pi,0)$. 

Since the action of $\sigma_z$ is free, there is no fixed point in BZ, so every point in BZ has a trivial little cogroup and the little group is just the translational subgroup. For a momentum $\boldsymbol{k}$, we can induce a $\nu_q$-irrep $\rho_{\boldsymbol{k}}$ of $G$ from $T^{\boldsymbol{k}}$:
\begin{equation}\label{ex2repq}
\begin{aligned}
&\rho_{\boldsymbol{k}}(t_a)=e^{ik_1}\left(\begin{array}{cc} 1 & 0 \\ 0 &  -1 \end{array}\right),\ \ \rho_{\boldsymbol{k}}(t_b)=e^{ik_2}\left(\begin{array}{cc} 1 & 0 \\ 0 &  -1 \end{array}\right),\\
&\rho_{\boldsymbol{k}}(t_c)=\left(\begin{array}{cc} e^{ik_3} & 0 \\ 0 &  e^{-ik_3} \end{array}\right),\ \ \rho_{\boldsymbol{k}}(\sigma_z)=\left(\begin{array}{cc} 0 & 1 \\ -1 &  0 \end{array}\right).
\end{aligned}
\end{equation}
Here we note $\boldsymbol{k}=(k_1,k_2,k_3)$ and $\boldsymbol{k}'=(k_1+\pi,k_2+\pi,-k_3)$ are in the same orbit (note here we write the coordinates of $\boldsymbol{k}$ under the reciprocal basis $\{\boldsymbol{b}_i\})$, so they induce the same $\nu_q$-irrep of $G$.

Finally, we can obtain coirreps of $\mathcal{G}$ from $\nu_q$-irreps of $G$. According to the construction in  Sec. \ref{corepSSG}, we need to consider the action of $T$ on $\rho_{\boldsymbol{k}}$. For coplanar magnetic order, $T=i\sigma_1\mathcal{K}$, which anticommutes with $U(t_a), U(t_b)$ and $U(\sigma_z)$ in Eq. (\ref{Ug}), leading to factor $\eta(t_a,T)=\eta(t_b,T)=\eta(\sigma_z,T)=-1$. The action of $T$ on $\rho_{\boldsymbol{k}}$ is then given by 
{
\begin{equation}
\mathrm{Act}_{T}(\rho_{\boldsymbol{k}})(g)=\eta(g,T)\rho_{\boldsymbol{k}}^*(g),\ \ \forall g\in G.
\end{equation}}
Due to the factor $\eta$, $T$ sends $\boldsymbol{k}=(k_1,k_2,k_3)$ to $\boldsymbol{k}'=(-k_1+\pi, -k_2+\pi,-k_3)$. {At general $\boldsymbol{k}$ points}, $\boldsymbol{k}$ and $\boldsymbol{k}'$ have inequivalent irreps, the coirreps of $\mathcal{G}$ obtained by $\rho_{\boldsymbol{k}}$ are given by case (iii):
\begin{equation}
\begin{aligned}
&\rho(t_a)=\left(\begin{array}{cccc} e^{ik_1} & 0 & 0 & 0\\ 0 &   e^{i(k_1+\pi)} &0 &0\\0 & 0 & e^{-i(k_1-\pi)} & 0\\ 0 &   0&0 &e^{-ik_1}\end{array}\right),\\
&\rho(t_b)=\left(\begin{array}{cccc} e^{ik_2} & 0 & 0 & 0\\ 0 &   e^{i(k_2+\pi)} &0 &0\\0 & 0 & e^{-i(k_2-\pi)} & 0\\ 0 &   0&0 &e^{-ik_2}\end{array}\right),\\
&\rho(t_c)=\left(\begin{array}{cccc} e^{ik_3} & 0 & 0 & 0\\ 0 &   e^{-ik_3} &0 &0\\0 & 0 & e^{-ik_3} & 0\\ 0 &   0&0 &e^{ik_3}\end{array}\right),\\
&\rho(\sigma_z)=\left(\begin{array}{cc} i\sigma_2 &  0\\ 0 &-i\sigma_2 \end{array}\right),\\
&\rho(T)=\left(\begin{array}{cc} 0 & \sigma_0 \\ \sigma_0 & 0 \end{array}\right)  \mathcal{K}.
\end{aligned}
\end{equation}

{Now consider high-symmetry points. Since $\mathrm{Act}_{T}(\rho_{\boldsymbol{k}})$ is the irrep at $T\boldsymbol{k}$, it can be equivalent to $\rho_{\boldsymbol{k}}$ only if $T\boldsymbol{k}$ lies in the orbit $\{\boldsymbol{k},\sigma_z\boldsymbol{k}\}$, and the two possibilities must be treated separately. (A) $T\boldsymbol{k}\equiv\boldsymbol{k}$ requires $k_1,k_2=\pm\pi/2$ together with $k_3=0$ or $\pi$. At these eight points $T$ enters the little group, and the little cogroup is $Z_2^T$ (or $1'$), $\mathrm{Act}_{T}(\rho_{\boldsymbol{k}})$ is equivalent to $\rho_{\boldsymbol{k}}$, and the coirreps of $\mathcal{G}$ are given by Eq.~(\ref{ex2repq}) with $\rho(T)=\sigma_3\mathcal{K}$.  (B) $T\boldsymbol{k}\equiv\sigma_z\boldsymbol{k}$ requires $k_1,k_2\in\{0,\pi\}$ with $k_3$ arbitrary. There $T$ alone does not preserve $\boldsymbol{k}$; what does is the antiunitary combination $\sigma_zT$, which acts on momentum as $(k_1,k_2,k_3)\mapsto(-k_1,-k_2,k_3)$. These are the high-symmetry lines, with little cogroup $2'$ generated by $\sigma_zT$. The coirreps of $\mathcal{G}$ are given by Eq.~(\ref{ex2repq}) with $\rho(T)=\sigma_1\mathcal{K}$.  Counting inequivalent orbits, the above gives 4 high-symmetry points (8 points in the BZ) and 2 high-symmetry lines (4 lines in the BZ). Compared with ordinary representations, the positions of the HSPs are therefore shifted, and the high-symmetry lines appear at different momenta.}

\subsubsection{$\mathrm{N22.9.42\ T1 \oplus Y1 \oplus Z1}$}\label{example3}

The last example we consider is a {noncoplanar} spin space group $\mathcal{G}=\mathrm{N22.9.42\ T1 \oplus Y1 \oplus Z1}$, which has a non-commuting Brillouin  zone. The parent group is F222, whose presentation can be written as
{
\begin{equation}
\begin{aligned}
F222=&\langle t_a,t_b,t_c, C_{2x},C_{2y}| t_at_bt_a^{-1}t_b^{-1}=1,\\&t_bt_ct_b^{-1}t_c^{-1}=1,t_ct_at_c^{-1}t_a^{-1}=1, \\&C_{2x}t_aC_{2x}^{-1}=t_a^{-1},  C_{2x}t_bC_{2x}^{-1}=t_a^{-1}t_c, \\&C_{2x}t_cC_{2x}^{-1}=t_bt_a^{-1},  C_{2y}t_aC_{2y}^{-1}=t_b^{-1}t_c, \\&C_{2y}t_bC_{2y}^{-1}=t_b^{-1}, C_{2y}t_cC_{2y}^{-1}=t_b^{-1}t_a, \\&C_{2x}C_{2y}C_{2x}^{-1}C_{2y}^{-1}=1, C_{2x}^2=1,C_{2y}^2=1 \rangle.
\end{aligned}
\end{equation}}
Here, we choose the primitive vectors of the face-centered lattice to be 
\begin{equation}
\boldsymbol{a}_1=\frac{1}{2}(0,b,c),\ \ \boldsymbol{a}_2=\frac{1}{2}(a,0,c),\ \ \boldsymbol{a}_3=\frac{1}{2}(a,b,0),\ \ 
\end{equation}
where $a,b,c$ are the lattice constants in three directions. 

The quotient spin space group $G$ is $\mathcal{G}$ itself and is isomorphic to F222. The $SU(2)$ spin operations of generators are
\begin{equation}
\begin{aligned}
&U(t_a)=-i\sigma_1, \ \ U(t_b)=-i\sigma_2, \ \ U(t_c)=-i\sigma_3, \\
&U(C_{2x})=\sigma_0,\ \ U(C_{2y})=\sigma_0.
\end{aligned}
\end{equation}
This leads to a nontrivial factor system $\nu$ for $G$. We can characterize the factor system by the following modified relations of generators:
{
\begin{equation}
\begin{aligned}
&\mathsf{t}_a\mathsf{t}_b\mathsf{t}_a^{-1}\mathsf{t}_b^{-1}=-1,\\
&\mathsf{t}_b\mathsf{t}_c\mathsf{t}_b^{-1}\mathsf{t}_c^{-1}=-1,\\
&\mathsf{t}_c\mathsf{t}_a\mathsf{t}_c^{-1}\mathsf{t}_a^{-1}=-1,\\
&\mathsf{C}_{2x}\mathsf{t}_a\mathsf{C}_{2x}^{-1}=-\mathsf{t}_a^{-1}, \\
&\mathsf{C}_{2x}\mathsf{t}_b\mathsf{C}_{2x}^{-1}=\mathsf{t}_a^{-1}\mathsf{t}_c, \\
&\mathsf{C}_{2x}\mathsf{t}_c\mathsf{C}_{2x}^{-1}=\mathsf{t}_b\mathsf{t}_a^{-1},\\
&\mathsf{C}_{2y}\mathsf{t}_a\mathsf{C}_{2y}^{-1}=\mathsf{t}_c\mathsf{t}_b^{-1}, \\
&\mathsf{C}_{2y}\mathsf{t}_b\mathsf{C}_{2y}^{-1}=-\mathsf{t}_b^{-1},\\
&\mathsf{C}_{2y}\mathsf{t}_c\mathsf{C}_{2y}^{-1}=\mathsf{t}_b^{-1}\mathsf{t}_a, \\
&\mathsf{C}_{2x}\mathsf{C}_{2y}\mathsf{C}_{2x}^{-1}\mathsf{C}_{2y}^{-1}=1,\\
&\mathsf{C}_{2x}^2=1,\\
&\mathsf{C}_{2y}^2=1.
\end{aligned}
\end{equation}}
Now we construct $\nu$-irreps of $G$. First, we define the Brillouin  zone by the center $Z(\mathrm{T}^\sigma)$ of the translational subgroup. The generators of $Z(\mathrm{T}^\sigma)$ can be chosen to be 
\begin{equation}\label{newge}
\begin{aligned}
&t_a'=t_bt_ct_a^{-1}, \\
&t_b'=t_ct_at_b^{-1}, \\
&t_c'=t_at_bt_c^{-1}.
\end{aligned}
\end{equation}
$t_a',t_b',t_c'$ generate a simple {orthorhombic} lattice $\mathcal{Z}_\mathcal{L}$, which is a sublattice of the original face-centered {orthorhombic} lattice. The unit cell vectors of $\mathcal{Z}_\mathcal{L}$ are:
\begin{equation}\label{basict}
\begin{aligned}
\boldsymbol{a}_1'=\boldsymbol{a}_2+\boldsymbol{a}_3-\boldsymbol{a}_1,\\ 
\boldsymbol{a}_2'=\boldsymbol{a}_3+\boldsymbol{a}_1-\boldsymbol{a}_2,\\
\boldsymbol{a}_3'=\boldsymbol{a}_1+\boldsymbol{a}_2-\boldsymbol{a}_3.
\end{aligned}
\end{equation}
The Brillouin  zone should be defined according to $\mathcal{Z}_\mathcal{L}$. According to Sec. \ref{Ptran}, the $\nu$-irreps of the translational subgroup can be written as
\begin{equation}
\begin{aligned}
&T^{\boldsymbol{k}}(t_a)=-ie^{i\boldsymbol{k}\cdot \boldsymbol{a}_1}\sigma_1,\\ &T^{\boldsymbol{k}}(t_b)=-ie^{i\boldsymbol{k}\cdot \boldsymbol{a}_2}\sigma_2, \\
&T^{\boldsymbol{k}}(t_c)=-ie^{i\boldsymbol{k}\cdot \boldsymbol{a}_3}\sigma_3, \ \ \ \ \boldsymbol{k}\in \mathrm{BZ}.
\end{aligned}
\end{equation}

Next, we consider the action of point group on BZ. The $\gamma$ factors between $t_a',t_b',t_c'$ and $C_{2x}, C_{2y}$ are evaluated to be trivial, so the action of $C_{2x},C_{2y}$ on BZ is symmorphic. More explicitly,
\begin{equation}
\begin{aligned}
C_{2x}: (k_1,k_2,k_3) \to (k_1,-k_2,-k_3),\\
C_{2y}: (k_1,k_2,k_3) \to (-k_1,k_2,-k_3),
\end{aligned}
\end{equation}
where $k_1,k_2,k_3$ are coefficients of three reciprocal lattice vectors, i.e., $\boldsymbol{k}=k_1\boldsymbol{b}_1'+k_2\boldsymbol{b}_2'+k_3\boldsymbol{b}_3'$, in which $\boldsymbol{b}_i'\cdot \boldsymbol{a}_j'=\delta_{ij}$. {High-symmetry} points in BZ are just those for the simple {orthorhombic} lattice and can be labelled as $\mathrm{\Gamma}$, R, S, T, U, X, Y, Z, etc. 

It is too lengthy to construct all irreps at every high-symmetry point, so we only demonstrate the construction at $\mathrm{Y}=(0,\pi,0)$ and $R=(\pi,\pi,\pi)$ as examples. The little cogroup at Y is $D_2$, generated by $C_{2x}$ and $C_{2y}$. The irrep of the translational subgroup at Y is
\begin{equation}
\begin{aligned}
&T^{Y}(t_a)=\sigma_1, \ T^{Y}(t_b)=-i\sigma_2, \ T^{Y}(t_c)=\sigma_3.
\end{aligned}
\end{equation}
To construct the irrep of the full group, we need to investigate how $C_{2x}$ and $C_{2y}$ act on $T^Y$. According to Eq. (\ref{gaction}), we have
{
\begin{equation}
\begin{aligned}
&\mathrm{Act}_{C_{2x}}(T^Y)(t_a)=-T^Y(t_a)^{-1}=-\sigma_1,\\
&\mathrm{Act}_{C_{2x}}(T^Y)(t_b)=T^Y(t_a)^{-1}T^Y(t_c)=-i\sigma_2,\\
&\mathrm{Act}_{C_{2x}}(T^Y)(t_c)=T^Y(t_b)T^Y(t_a)^{-1}=-\sigma_3,\\
&\mathrm{Act}_{C_{2y}}(T^Y)(t_a)=T^Y(t_c)T^Y(t_b)^{-1}=\sigma_1,\\
&\mathrm{Act}_{C_{2y}}(T^Y)(t_b)=-T^Y(t_b)^{-1}=-i\sigma_2,\\
&\mathrm{Act}_{C_{2y}}(T^Y)(t_c)=T^Y(t_b)^{-1}T^Y(t_a)=\sigma_3.
\end{aligned}
\end{equation}}
where the minus sign after the first equal sign in the first and fifth lines are due to the factor system. Although $\mathrm{Act}_{C_{2x}}(T^Y)$ is not the same as $T^Y$, they are indeed equivalent and can be related by a unitary transformation:
\begin{equation}
\mathrm{Act}_{C_{2x}}(T^Y)(t)=\sigma_2^\dagger T^Y(t)\sigma_2,\ \ \ \forall t \in \mathrm{T}.
\end{equation}
So we can take $U_{Y}(C_{2x})=\sigma_2$. Since $C_{2y}$ does not change $T^Y$, so $U_{Y}(C_{2y})=\sigma_0$. For elements in the translational subgroup, the transformation matrices are taken to be $U_Y(t)=T^Y(t)$. $U_Y$ is a projective representation of the little group $G_Y$ at $Y$ (here $G_Y=G$). The factor system $\tau_Y$ of $U_Y$ is the same as $\nu$, so the quotient factor system $\omega_Y$ of the little cogroup is trivial.

There are four inequivalent irreps of the little cogroup $D_2$, and they are all one-dimensional. We label them as $\Gamma^{(i)}, i=1,2,3,4.$ For an irrep $\Gamma^{(i)}$, we can construct a $\nu$-irrep $\rho_Y^{(i)}$ of the little group $G_Y$ (and also the group $G$) by
\begin{equation}
\rho_Y^{(i)}(g)=U_Y(g)\Gamma^{(i)}(R_g),\ \ \forall  g\in G_Y.
\end{equation}
where $R_g$ is the point operation part of $g$. More explicitly, we can write $\rho_Y^{(i)}$ in terms of generators:
\begin{equation}
\begin{aligned}
&\rho_Y^{(i)}(t_a)=\sigma_1, \ \ \rho_Y^{(i)}(t_b)=-i\sigma_2,\ \ \rho_Y^{(i)}(t_c)=\sigma_3,\\
&\rho_Y^{(i)}(C_{2x})=\sigma_2\Gamma^{(i)}(C_{2x}),\ \ \rho_Y^{(i)}(C_{2y})=\sigma_0\Gamma^{(i)}(C_{2y}).
\end{aligned}
\end{equation}

{At high-symmetry point $R=(\pi,\pi,\pi)$, the little cogroup is also $D_2$, generated by $C_{2x}$ and $C_{2y}$. The irrep of the translational subgroup at R is 
\begin{equation}
T^{R}(t_a)=i\sigma_1, \ T^{R}(t_b)=i\sigma_2, \ T^{R}(t_c)=i\sigma_3.
\end{equation}
Consider the action of $C_{2x}$ and $C_{2y}$ on $T^R$, the transformation {matrices} can be taken as $U_R(C_{2x})=\sigma_1, U_R(C_{2y})=\sigma_2$. Since $U_R(C_{2x})$ anticommutes with $U_R(C_{2y})$, the factor system $\tau_R$ of $U_R$ is {different from} $\nu$, leading to a nontrivial quotient factor system $\omega_R$ of the little cogroup $D_2$, characterized by $\omega_R(C_{2x},C_{2y})/\omega_R(C_{2y},C_{2x})=-1$. There is one $\omega_R$-irrep of $D_2$ that is two dimensional: $\widetilde{D}(C_{2x})=\sigma_1$, $\widetilde{D}(C_{2y})=\sigma_2$. Combining with $U_R$, we obtain the $\nu$-irrep $\rho_R$ of the little group at R, i.e., the $\nu$-irrep of $G$ at R: }
\begin{equation}
\begin{aligned}
&\rho_R(t_a)=i\sigma_1\otimes I, \ \ \rho_R(t_b)=i\sigma_2\otimes I,\ \ \rho_R(t_c)=i\sigma_3\otimes I,\\
&\rho_R(C_{2x})=\sigma_1\otimes \sigma_1,\ \ \rho_R(C_{2y})=\sigma_2\otimes \sigma_2.
\end{aligned}
\end{equation}
{This $\nu$-irrep of $G$ is four-dimensional, meaning there is a fourfold degeneracy at R. This is rather special, because in ordinary representation theory, little cogroup $D_2$ at R in symmorphic space groups only has 1-dimensional irreps. The fourfold degeneracy at R here arises from two distinct origins: one from the noncommuting BZ, and the other from nontrivial factor systems contributed by the transformation matrix $U_R$; both are absent in ordinary representation theory.}


\section{Physical consequences}\label{Cons}

Here we discuss some physical consequences caused by projective coreps of SSGs, which are absent in ordinary representations of MSGs. Consequences in Sec. \ref{Briplaty} and Sec. {\ref{zakphase}} are known from previous studies for projective crystalline symmetry\cite{ZhangBrillouin2025,ChenClassification2023}, but have not been investigated in SSGs. 

\subsection{Realizable $k$-space MSGs by SSGs}\label{actiond2}

Here we list all realizable $k$-space MSGs by SSGs. Since all symmorphic $k$-space MSGs can be realized by SSGs, we only concentrate on nonsymmorphic $k$-space MSGs.  According to Sec. \ref{actonBZ} and Appendix \ref{nonsymapp}, for a nonsymmorphic $k$-space MSG, the necessary condition that it can be realized by SSGs is that it has at least one orbit with dimension two in $k$-space, i.e., $d_{\mathrm{o,min}}= 2$, where $d_{\mathrm{o,min}}$ is the minimal dimension of orbits in $k$-space. For type-I, II, III MSGs in $k$-space, since they are isomorphic to their parent space groups, any operation $g$ can be seen as a $k$-space spatial operation followed by a $k$-space time reversal $\mathcal{T}_k$ (does not change $k$-points) or trivial operation. Thus, an MSG of type-I, II, III has the same orbit dimension as its parent space groups. In Table~\ref{dim2}, we list all nonsymmorphic space groups whose $d_{\mathrm{o,min}}= 2$. Type-I, II, III nonsymmorphic MSGs whose parent groups are not in Table \ref{dim2} cannot be realized as $k$-space symmetry groups by SSGs. {For a type-IV $k$-space MSG $\mathcal{M}$, the {antiunitary} translation is always a free action. To keep $d_{\mathrm{o,min}}= 2$, the parent group (OG notation) of $\mathcal{M}$ must be symmorphic. }

{In reverse, one can show that $d_{\mathrm{o,min}}= 2$ is also a sufficient condition that a $k$-space MSG can be realized by SSGs.}

{To summarize,  whether a $k$-space MSG can be realized by SSGs is determined by its type and its parent space group (OG notation). If its parent space group  is symmorphic, then it can be realized by SSGs. If its parent space group is nonsymmorphic, then only if its parent space group is in Table~\ref{dim2} and it is of type-I, II or III, it can be realized by SSGs. }

\begin{table*}[htbp]
\centering
\small
\begin{tabular}{@{} p{2.8cm} p{\dimexpr\textwidth-2.8cm-4\tabcolsep\relax} @{}}
\toprule
Crystal system & Nonsymmorphic space groups with $d_{\mathrm{o,min}}=2$ \\
\midrule
Triclinic
  & --- \\
Monoclinic (7)
  & 4 $P2_1$, 7 $Pc$, 9 $Cc$, 11 $P2_1/m$, 13 $P2/c$, 14 $P2_1/c$, 15 $C2/c$ \\[4pt]
Orthorhombic (35)
  & 17 $P222_1$, 18 $P2_12_12$, 20 $C222_1$, 24 $I2_12_12_1$, 26 $Pmc2_1$, 27 $Pcc2$, 28 $Pma2$, 30 $Pnc2$, 31 $Pmn2_1$, 32 $Pba2$, 34 $Pnn2$, 36 $Cmc2_1$, 37 $Ccc2$, 39 $Aem2$, 40 $Ama2$, 41 $Aea2$, 43 $Fdd2$, 45 $Iba2$, 46 $Ima2$, 48 $Pnnn$, 49 $Pccm$, 50 $Pban$, 51 $Pmma$, 53 $Pmna$, 55 $Pbam$, 58 $Pnnm$, 59 $Pmmn$, 63 $Cmcm$, 64 $Cmce$, 66 $Cccm$, 67 $Cmme$, 68 $Ccce$, 70 $Fddd$, 72 $Ibam$, 74 $Imma$ \\[4pt]
Tetragonal (39)
  & 77 $P4_2$, 80 $I4_1$, 84 $P4_2/m$, 85 $P4/n$, 86 $P4_2/n$, 88 $I4_1/a$, 90 $P42_12$, 93 $P4_222$, 94 $P4_22_12$, 98 $I4_122$, 100 $P4bm$, 101 $P4_2cm$, 102 $P4_2nm$, 103 $P4cc$, 104 $P4nc$, 105 $P4_2mc$, 108 $I4cm$, 109 $I4_1md$, 112 $P\overline{4}2c$, 113 $P\overline{4}2_1m$, 114 $P\overline{4}2_1c$, 116 $P\overline{4}c2$, 117 $P\overline{4}b2$, 118 $P\overline{4}n2$, 120 $I\overline{4}c2$, 122 $I\overline{4}2d$, 124 $P4/mcc$, 125 $P4/nbm$, 126 $P4/nnc$, 127 $P4/mbm$, 128 $P4/mnc$, 129 $P4/nmm$, 131 $P4_2/mmc$, 132 $P4_2/mcm$, 134 $P4_2/nnm$, 136 $P4_2/mnm$, 137 $P4_2/nmc$, 140 $I4/mcm$, 141 $I4_1/amd$ \\[4pt]
Trigonal (6)
  & 158 $P3c1$, 159 $P31c$, 161 $R3c$, 163 $P\overline{3}1c$, 165 $P\overline{3}c1$, 167 $R\overline{3}c$ \\[4pt]
Hexagonal (11)
  & 173 $P6_3$, 176 $P6_3/m$, 182 $P6_322$, 184 $P6cc$, 185 $P6_3cm$, 186 $P6_3mc$, 188 $P\overline{6}c2$, 190 $P\overline{6}2c$, 192 $P6/mcc$, 193 $P6_3/mcm$, 194 $P6_3/mmc$ \\[4pt]
Cubic (11)
  & 201 $Pn\overline{3}$, 203 $Fd\overline{3}$, 208 $P4_232$, 210 $F4_132$, 218 $P\overline{4}3n$, 219 $F\overline{4}3c$, 222 $Pn\overline{3}n$, 223 $Pm\overline{3}n$, 224 $Pn\overline{3}m$, 226 $Fm\overline{3}c$, 227 $Fd\overline{3}m$ \\
\bottomrule
\end{tabular}
\caption{Nonsymmorphic space groups with minimal orbit dimension $d_{\mathrm{o,min}}=2$, grouped by crystal system. Among 157 nonsymmorphic space groups, 109 have $d_{\mathrm{o,min}}=2$. For a type-I, II or III $k$-space MSG $\mathcal{M}$ whose parent space group $G$ (OG {notation}) is nonsymmorphic, only if $G$ is in this table, $\mathcal{M}$ is realizable by SSGs. }
\label{tab:d2-space-groups}\label{dim2}
\end{table*}

\subsection{Change of high-symmetry points and little cogroups}\label{hspchange}

{High-symmetry} points (HSPs) and little cogroups of (magnetic) space groups in $k$-space are very important data in condensed matter physics. They are used in the analysis of physical properties of energy bands, such as nodal points, nodal lines and nodal surfaces. The topological quantum chemistry, which aims to diagnose nontrivial topological states, also relies on the data of HSPs and little cogroups. For ordinary representation theory of (magnetic) space groups, these data are already thoroughly worked out and accessible in datasets. However, for projective representation theory of (magnetic) space groups or SSGs, one cannot use the data of HSPs and little cogroups in existing datasets directly without checking the applicability, because factor systems can change HSPs and little cogroups, as we mentioned in Sec. \ref{hsplg}.

Projective reps change HSPs via two mechanisms. The first is nonsymmorphic actions in $k$ space. In ordinary reps, a point group operation $p$ always keeps some $k$ points invariant. In projective reps, $p$ can be made into a nonsymmorphic action by nontrivial factor systems. Since nonsymmorphic actions are free, i.e., move all $k$ points, high-symmetry points (lines, surfaces) relating to $p$ have lower symmetry than before, i.e., their little cogroups become smaller. Thus, HSPs and their little cogroups will be changed by nontrivial factor systems.

Formally, ordinary coreps of an MSG $G$ always lead to a unique $k$-space symmorphic MSG $G_F$, then HSPs and their little cogroups are determined by $G_F$. But for projective coreps of $G$, the $k$-space MSG $G_F$ can be nonsymmorphic. HSPs and their little cogroups determined by nonsymmorphic $G_F$ will be essentially different from that determined by symmorphic $G_F$. For example, in Sec. \ref{example2}, $\mathrm{P6.3.2\ GM2\oplus A1}$ is isomorphic to a type-II MSG $\mathrm{Pm1'}$. In ordinary reps, $\mathrm{Pm1'}$ leads to a $k$-space symmetry group $\mathrm{P2'/m}$. There are 8 high-symmetry points with little cogroup m1', four {high-symmetry} lines with little cogroup $\mathrm{m'}$, and two {high-symmetry planes} with little cogroup m; {the counts are numbers of symmetry-inequivalent orbits unless stated otherwise.} However, the factor system of $\mathrm{P6.3.2\ GM2\oplus A1}$ is nontrivial, and the $k$-space symmetry group is changed to a nonsymmorphic group $\mathrm{P2'/c}$, which determines 4 high-symmetry points with little cogroup $\mathrm{1'}$, and 2 {high-symmetry} lines with little cogroup $\mathrm{2'}$.

Non-commuting relation of translations is another mechanism to change HSPs. As we discussed in Sec. \ref{example3}, when translations are not commuting due to nontrivial factor systems, we should define the Brillouin zone by the center of the projective translational algebra $\mathrm{T}^\sigma$, which means defining the Brillouin zone by a sublattice $\mathcal{Z_L}$ of the original lattice $\mathcal{L}$. Thus, the original reciprocal lattice $\widehat{\mathcal{L}}$ is changed to a reciprocal lattice $\widehat{\mathcal{Z}}_\mathcal{L}$. In general, $\widehat{\mathcal{L}}$ and $\widehat{\mathcal{Z}}_\mathcal{L}$ have different HSPs. For example, in Sec. \ref{example3}, 
$\mathrm{N22.9.42\ T1 \oplus Y1 \oplus Z1}$ is isomorphic to a space group F222, which has a face-centered lattice in real space and a body-centered reciprocal lattice. However, the three translational generators in $\mathrm{N22.9.42\ T1 \oplus Y1 \oplus Z1}$ are all anti-commuting with each other, and the center of $\mathrm{T}^\sigma$ defines a simple {orthorhombic} lattice (see Eq. (\ref{basict})). Thus, the reciprocal lattice also becomes a simple {orthorhombic} lattice.

Yet, the HSPs and their little cogroups are not tabulated for general projective representations of (magnetic) space groups. To enable future applications of projective reps of (magnetic) space groups, it is necessary to work out these data.

\subsection{Brillouin  platycosms}\label{Briplaty}
The usual Brillouin  zone manifold is a torus $T^d$. If there is a free action on BZ, then we can further reduce the torus to a smaller domain, in which all information of bands is contained. The reduced Brillouin zone can be another manifold rather than $T^d$. In ordinary rep theory, all actions on BZ are point-group-like, which are not free. Only when nonsymmorphic symmetries on BZ are found \cite{ChenBrillouin2022}, free actions on BZ are possible. In 2D, certain nonsymmorphic actions on $T^2$ can lead to Klein bottle Brillouin zone, while in 3D, certain nonsymmorphic actions on $T^3$ can lead to Brillouin platycosms.

For a $k$-space MSG $G_F$, we can find a subgroup $G_F^{\mathrm{free}}\subseteq G_F$ that contains only free actions (except the identity element). $G_F^{\mathrm{free}}$ is called a Bieberbach group. In our setting, we require the translational subgroup $\mathrm{T}\subset G_F^{\mathrm{free}}$, then the reduced Brillouin zone is defined to be $\mathrm{RBZ}=\mathbb{R}_F/G_F^{\mathrm{free}}$. If there is no nonsymmorphic action in $k$-space except translations, $G_F^{\mathrm{free}}=\mathrm{T}\cong \mathbb{Z}^d$, and $\mathrm{RBZ}=\mathbb{R}_F/\mathbb{Z}^d\cong T^d$, so we get the usual torus Brillouin zone. But when $G_F^{\mathrm{free}}$ is a nontrivial Bieberbach group, the reduced Brillouin zone can become other manifold. In general, $G_F$ can contain different Bieberbach subgroups, so there can be different ways to define the reduced Brillouin zone. In 2D, there are only two different Bieberbach groups i.e., $P1$ and $Pg$. The former leads to the usual torus Brillouin zone, while the latter leads to the Klein bottle Brillouin zone \cite{ChenBrillouin2022}. In 3D, there are 10 Bieberbach groups, and they lead to ten Brillouin  platycosms \cite{ZhangBrillouin2025}. In Table~\ref{tab1}, we list all 10 platycosms, their corresponding $k$-space Bieberbach groups and whether they can be realized by projective reps of SSGs. We note that only four of ten platycosms can be realized by SSGs. The other six platycosms cannot be realized because the dimensions of their $k$-orbits are larger than two.

\begin{table*}[htbp]
  \centering
  \begin{tabular}{cccc}          
    \toprule
   Platycosms & Ori  & $G_{F}^{\mathrm{free}}$ & Realizable by SSG \\
    \midrule
   Cubical torocosm & Y  & P1 & Y \\[4pt]
 First amphicosm & N  & Pc & Y
\\[4pt]
    Second amphicosm & N & Cc & Y\\[4pt]
  First amphidicosm  & N  & $\mathrm{Pca2_1}$ & N\\[4pt]
   Second amphidicosm  & N  & $\mathrm{Pna2_1}$ & N\\[4pt]
    Dicosm  & Y  & $\mathrm{P2_1}$ & Y\\[4pt]
    Tricosm & Y  & $\mathrm{P3_1}$& N\\[4pt]
    Tetracosm & Y  & $\mathrm{P4_1}$ & N\\[4pt]
    Hexacosm  & Y  & $\mathrm{P6_1}$ & N\\[4pt]
    Didicosm  & Y  & $\mathrm{P2_12_12_1}$ & N\\
    \bottomrule
  \end{tabular}
  \caption{Brillouin  platycosms and their realization by SSGs. The first two columns list all ten platycosms. The third column lists their orientability. The fourth column lists their corresponding Bieberbach groups $G_F^\mathrm{free}$ in $k$-space. The last column lists whether some SSGs  can realize $G_F^\mathrm{free}$ in $k$-space. Only four of ten platycosms can be realized by SSGs.}
  \label{tab1}
\end{table*}

Traditional topological phases live on the torus Brillouin zone. For example, the Chern insulator is characterized by the Chern class, which is defined as the second cohomology class of line bundles on the torus for systems with translational symmetry. Since $k$-space nonsymmorphic symmetries can lead to non-torus Brillouin zone, new topological phases can appear. For example, in 2D, the Klein bottle insulator can appear when the Brillouin zone is the Klein bottle, which can be characterized by a new $Z_2$ topological number \cite{ChenBrillouin2022}. In 3D,  Brillouin platycosms allow new topological phases, which can be characterized by new topological numbers \cite{ZhangBrillouin2025}. In Ref. \cite{ZhangBrillouin2025}, these new topological states were demonstrated in artificial systems with flux. Here, since projective reps of SSGs can make the Brillouin zone to be the first amphicosm, the second amphicosm and the Dicosm, the new topological states on these three Brillouin manifolds can possibly be realized in real magnetic materials.

In orientable Brillouin zone, there is Nielsen–Ninomiya (NN) theorem: the total chirality of Weyl points is zero. However, this theorem relies on the condition that the underlying manifold is orientable. When the underlying manifold is non-orientable, the NN theorem should be modified, that is, the total  chirality of Weyl points is even \cite{ZhangBrillouin2025}. Since ordinary Brillouin zones are tori, only the original NN theorem applies. Nonsymmorphic $k$-space actions provide the possibility that the reduced BZ is non-orientable. In 2D, the Klein bottle BZ is non-orientable. In 3D, four of ten Brillouin  platycosms are non-orientable, whose Bieberbach groups are Pc, Cc, $\mathrm{Pca}2_1$ and $\mathrm{Pna}2_1$. Pc and Cc  can be realized by SSGs (see Table~\ref{tab1}), which gives the chance to violate the traditional NN theorem in real weak SOC magnetic systems.

\subsection{Same-chirality Weyl points locked by projective symmetry}\label{weylchiral}

{In Sec.~\ref{example3} the fourfold degeneracy at the $R$ point of the noncoplanar SSG $\mathcal{G}=\mathrm{N22.9.42\ T1\oplus Y1\oplus Z1}$ was traced to two independent sources: the two-dimensional irreducible $\sigma$-irrep of the translational subgroup and the nontrivial factor $\omega_R$ of the little cogroup. The decomposition of the $\nu$-irrep determines more than the degeneracy. It determines that the low-energy excitations are two-fold Weyl points with the same chirality. Such Weyl points are protected by projective translation symmetry and cannot be removed by perturbations that break the point-group symmetry.  In this subsection $\boldsymbol{q}$ is measured from $R$, and $\mu_i$ acts on the two-dimensional multiplicity space of the $\omega_R$-irrep.}

At $R$ the three spin translations are represented by
\begin{equation}\label{Rtrans}
T^{R}(t_a)=i\sigma_1,\qquad T^{R}(t_b)=i\sigma_2,\qquad T^{R}(t_c)=i\sigma_3.
\end{equation}
These three matrices generate the full matrix algebra $M_2(\mathbb{C})$. By Schur's lemma, any Bloch Hamiltonian that preserves the spin translations, $[H,\rho(t)]=0$ for all $t\in\mathrm{T}$, must therefore be a block scalar,
{
\begin{equation}\label{Rblock}
H_R(\boldsymbol{q})=I_2\otimes h(\boldsymbol{q}).
\end{equation}}
{The whole freedom sits in the $2\times2$ matrix $h$. Any Hermitian $h$ can be expanded as $h=h_0+d_i(\boldsymbol{q})\mu_i$ with real $d_i$ that vanish at $R$, so it is the point-group symmetry that constrains the coefficients. The two spatial rotations act as $\rho_R(C_{2x})=\sigma_1\otimes\mu_1$ and $\rho_R(C_{2y})=\sigma_2\otimes\mu_2$ and reverse the momentum, so the condition they impose is the covariance $H_R(\boldsymbol{q})=\rho_R(g)H_R(g\boldsymbol{q})\rho_R(g)^{-1}$ with $g=C_{2x},C_{2y}$. The coefficients transform as $(d_1,d_2,d_3)\to(d_1,-d_2,-d_3)$ and $(-d_1,d_2,-d_3)$ under $C_{2x}$ and $C_{2y}$, respectively, which at linear order leaves}
{
\begin{equation}\label{Rkp}
H_R(\boldsymbol{q})=I_2\otimes\big(v_xq_x\mu_1+v_yq_y\mu_2+v_zq_z\mu_3\big)+O(q^2).
\end{equation}}

To the order written, the spectrum of Eq.~(\ref{Rkp}) is $E=\pm(v_x^2q_x^2+v_y^2q_y^2+v_z^2q_z^2)^{1/2}$, each level doubly degenerate. {The $O(q^2)$ terms permitted in Eq.~(\ref{Rkp}) sit in the scalar $h_0$ and shift the two doublets equally, so they do not affect the node.} The four bands are thus two identical Weyl cones placed at the same momentum, and their charges add. {Writing $V_i=\partial_{q_i}H_R|_{\boldsymbol{q}=0}=I_2\otimes h_i$ and $V_{ij}=\tfrac12\mathrm{tr}_{\mu}(h_i\mu_j)$, where $\mathrm{tr}_{\mu}$ denotes the trace over the two-dimensional multiplicity space, the $3\times3$ velocity matrix $V$ satisfies}
{
\begin{equation}\label{detV}
\det V=v_xv_yv_z=\frac{1}{4i}\mathrm{tr}\big(V_1V_2V_3\big).
\end{equation}}
{The trace in the second equality of Eq.~(\ref{detV}) runs over the full four-dimensional space. A nonvanishing $\det V$ is exactly the statement that the two cones carry the same chirality; for a Dirac-type node, built from two cones of opposite chirality, this trace vanishes identically. Correspondingly the occupied subspace is $I_2\otimes P_{\mathrm{occ}}$, with $P_{\mathrm{occ}}$ the occupied band of $h$, and carries $|C|=2$. More generally, every irrep of $\mathrm{T}^{\sigma}$ in this Brillouin zone is two-dimensional, so every band of this SSG is at least doubly degenerate, and every node is locally a pair of equal-chirality cones.}

Perturbations of the quartet fall into two families. Those preserving the spin translations are spanned by $I_2\otimes\{I,\mu_1,\mu_2,\mu_3\}$; those preserving both spatial rotations are spanned by $\{I,\mathcal{D}_1,\mathcal{D}_2,\mathcal{D}_3\}$, with $\mathcal{D}_i\equiv\sigma_i\otimes\mu_i$ and {$\mathcal{D}_3=-\mathcal{D}_1\mathcal{D}_2$}. Each family is four-dimensional and their intersection contains only the identity, so a perturbation that preserves the full symmetry at $R$ can only shift the energy. Any nontrivial one breaks either the spin translations or a spatial rotation, and the two choices lead to different fates for the node.

For a translation-preserving perturbation, the Hamiltonian has the form
{
\begin{equation}\label{Rmove}
\begin{gathered}
H_R=I_2\otimes\big[(v_iq_i+m_i)\mu_i\big],\\
\boldsymbol{q}_{\mathrm{node}}=-\big(m_1/v_x,\;m_2/v_y,\;m_3/v_z\big).
\end{gathered}
\end{equation}}
{The node stays fourfold with $|C|=2$ and only shifts. No perturbation of this family can split or gap the quartet, and the three masses are in one-to-one correspondence with the three coordinates of the node: within this family the node position is a faithful coordinate. A practical consequence is that four distinct energies at $R$ do not indicate that the charge-two node has been removed; it has merely left $R$. This is different from four-fold nodes protected by point groups, which will be removed by perturbations breaking the point group symmetry.}

For a rotation-preserving perturbation, $\delta H=a\mathcal{D}_1+b\mathcal{D}_2$. On the line $q_y=q_z=0$ the spectrum is
{
\begin{equation}\label{Rsplit}
E=\pm a\pm\sqrt{v_x^2q_x^2+b^2},
\end{equation}}
{which for $|a|>|b|$ has two crossings at $q_x=\pm\sqrt{a^2-b^2}/v_x$ with energies $\{-2|a|,0,0,2|a|\}$, each of which is a two-band Weyl point with $|C|=1$, and both carry the same sign. The parent node therefore splits as two nodes with the same chirality. The two fragments cannot annihilate one another; they can disappear only by meeting a node of opposite charge elsewhere in the Brillouin zone. For $|a|<|b|$ the two branches never cross and the quartet is gapped, so a rotation-preserving perturbation that breaks the spin translations can remove the node altogether: what protects it is the translational symmetry alone. The third direction of the same family, $\mathcal{D}_3$, instead splits the node along $q_z$ into two fragments that again carry the same chirality.}

{Finally, because the node carries $|C|=2$ rather than $|C|=1$, it is expected to terminate two surface Fermi arcs rather than one.}

\subsection{Enforced topological phases by projective symmetries}\label{Enforce}

\subsubsection{Enforced Zak phase by type-IV $k$-MSGs}\label{zakphase}

For type-IV $k$-MSGs, the symmetry group in $k$ space  always contains an antiunitary half translation $\boldsymbol{K}_a/2$. This kind of symmetry operation arises {from} projective $PT$ symmetry in real space. $PT$ is an onsite antiunitary symmetry in $k$-space, i.e., $\widetilde{PT}=I$. However, if the factor $\gamma(t,PT)$ is nontrivial, $PT$ symmetry can lead to fractional translation in $k$-space. According to the 1-cocycle equation (\ref{frac1}), we have 
\begin{equation}
\boldsymbol{\kappa}_{\widetilde{PT}}=\boldsymbol{K}/2,
\end{equation}
where $\boldsymbol{K}$ is a reciprocal translation vector. If $\boldsymbol{K}/2$ is a half reciprocal translation vector, $\boldsymbol{\kappa}_{\widetilde{PT}}$ is nontrivial, and $PT$ leads to an antiunitary half translation in $k$-space.

When the $k$-MSG is type-IV, following the discussion in Ref. \cite{ChenClassification2023},  the Zak phase $\theta_a$ along any $\boldsymbol{K}_a$-periodic path for a single band is given by 
\begin{equation}
\theta_a = i\ln \alpha\   \mathrm{mod}\  2\pi,
\end{equation}
where $\alpha=(\mathsf{PT})^2$ is exactly the factor $\nu(PT,PT)$, which can only take values $\pm1$ according to the consistency condition. So $\alpha=-1$ will enforce a nontrivial Zak phase $\pi$ along $\boldsymbol{K}_a$ and $\alpha=1$ will enforce a trivial Zak phase. 

\subsubsection{Chern insulator enforced by projective translation symmetry}

{The Zak phase above is enforced by the mixed factor $\gamma$, which makes the momentum-space action nonsymmorphic. A different enforced topological datum follows from the translational factor $\sigma$: when the translations do not commute, the parity of the Chern number of an isolated band is fixed. In particular a single isolated doublet cannot carry an even Chern number.}

{Consider a two-dimensional SSG whose translational factor is of the noncommuting type, with two independent spin translations satisfying $\mathsf{t}_a\mathsf{t}_b=-\mathsf{t}_b\mathsf{t}_a$. The Brillouin zone is defined by the center of the projective translation algebra, which is generated by translations of $2\boldsymbol{a}$ and $2\boldsymbol{b}$, so the momenta $k_x,k_y$ are $\pi$-periodic. Every irrep of $\mathrm{T}^{\sigma}$ is two-dimensional and every band is a doublet. }

{The two spin translations can be represented as}

\begin{equation}\label{Slocal}
\rho_{\boldsymbol{k}}(t_a)=-ie^{ik_x}\sigma_1,\ \ 
\rho_{\boldsymbol{k}}(t_b)=-ie^{ik_y}\sigma_3,\ \ 
\end{equation}
{where $k_x,k_y\in [-\pi/2,\pi/2)$.} Advancing $k_x$ by $\pi$ flips the sign of $\rho_{\boldsymbol{k}}(t_a)$ and leaves $\rho_{\boldsymbol{k}}(t_b)$ unchanged, whereas advancing $k_y$ does the opposite. The frame must therefore be changed at the zone boundary by unitary transformations that flip $\sigma_1$ while preserving $\sigma_3$ and vice versa, i.e.,  
\begin{equation}
\begin{aligned}
\rho_{(k_x+\pi,k_y)}=G_x(k_y)^{-1}\rho_{(k_x,k_y)}G_x(k_y),\\
 \rho_{(k_x,k_y+\pi)}=G_y(k_x)^{-1}\rho_{(k_x,k_y)}G_x(k_x),
 \end{aligned}
 \end{equation}
 with
{
\begin{equation}\label{Gtrans}
G_x(p_y)=e^{ik_y}\sigma_3,
\qquad
G_y=\sigma_1 .
\end{equation}}
{These transition functions satisfy the corner compatibility relation}

\begin{equation}\label{cocycleG}
G_x(k_y+\pi)\,G_y=G_y\,G_x(k_y).
\end{equation}
$\rho_{\boldsymbol{k}}$ defines a two-dimensional bundle $E_T$, whose Chern number is 
\begin{equation}
C(E_T)=\frac{1}{2\pi i} \int  \mathrm{tr}(G_x(k_y)^{-1}dG_x(k_y)),
\end{equation}
which is $\pm1$ according to the convention. An isolated group of bands compatible with the translations has rank $2r$ and decomposes as $E=E_T\otimes F$ with $\mathrm{rank}\,F=r$. Using $c_1(E_T\otimes F)=\mathrm{rank}(F)\,c_1(E_T)+\mathrm{rank}(E_T)\,c_1(F)$,
{
\begin{equation}\label{chernparity}
\begin{gathered}
c_1(E)=r\,c_1(E_T)+2c_1(F),\\
C(E)\equiv r\pmod 2
\end{gathered}
\end{equation}}
{In particular, for an isolated doublet, $r=1$, and the Chern number is enforced to be odd. A gapped filling of $r$ doublets carries an anomalous Hall conductance that is an odd multiple of $e^{2}/h$ when $r$ is odd and an even multiple when $r$ is even \cite{LuFilling2020}. }

Here we give a minimal model. Take a two-dimensional Bravais lattice with chemical lattice vectors $\boldsymbol{a}$ and $\boldsymbol{b}$ and a four-site magnetic cell, with local moments
{
\begin{equation}\label{Mmoments}
\mathbf{M}_{\boldsymbol{n}}
=\frac{\big((-1)^{n_y},\,(-1)^{n_x},\,m(-1)^{n_x+n_y}\big)}{\sqrt{2+m^{2}}},
\end{equation}}
{and the Hamiltonian}
{
\begin{equation}\label{Hchern}
H=-\sum_{\boldsymbol{n},\boldsymbol\delta}t_{\boldsymbol\delta}\,
c^{\dagger}_{\boldsymbol{n}+\boldsymbol\delta}c_{\boldsymbol{n}}
+\mathrm{h.c.}
+J\sum_{\boldsymbol{n}}c^{\dagger}_{\boldsymbol{n}}
(\mathbf{M}_{\boldsymbol{n}}\cdot\mathbf{s})\,c_{\boldsymbol{n}},
\end{equation}}
{where $\boldsymbol\delta=\boldsymbol{a},\boldsymbol{b},\boldsymbol{a}+\boldsymbol{b}$, $(t_{\boldsymbol{a}},t_{\boldsymbol{b}},t_{\boldsymbol{a}+\boldsymbol{b}})=(1,0.8,0.7)$ and $J=3$. The hoppings are real and spin independent, and the symmetries include $S_a=-i\sigma_xt_{{a}}$ and $S_b=-i\sigma_yt_{{b}}$, which are anticommuting. The magnetic cell contains four sites and hence eight electronic states, i.e., four doublets. At $m=1$ the Chern numbers of the four doublets, from lowest to highest, are $(C_1,C_2,C_3,C_4)=(-1,+1,+1,-1)$, all odd as required; at $m=-1$ every sign is reversed.}

The parity constraint can also forbid an insulator. At $m=0$ the moments lie in the $xy$ plane and an effective antiunitary symmetry $T=\sigma_x\mathcal{K}$ with $T^{2}=+1$ is restored. It reverses the Berry curvature, so the total Chern number of any $T$-invariant occupied subspace vanishes. Combined with Eq.~(\ref{chernparity}), a vanishing total Chern number requires $r$ to be even, and therefore
{
\begin{equation}\label{nocc}
n_{\mathrm{occ}}\equiv0\pmod 4 .
\end{equation}}
A filling with a half-integer number of electrons per chemical cell can then not be insulating while these symmetries are preserved. The model realizes the obstruction: at $m=0$ the first and second doublets, and likewise the third and fourth, touch at $M=(\pi/2,\pi/2)$ at the energy $E=\pm\sqrt{J^{2}+(2t_{\boldsymbol{a}+\boldsymbol{b}})^{2}}$, 
{whereas the middle gap remains open. The closure occurs precisely at the excluded fillings, which turns the parity constraint into a checkable prediction rather than a statement about degeneracies in general.}

\section{Discussion and conclusion}\label{Conclu}

In this work, we have developed a unified projective representation theory for spin space groups. The construction applies to collinear, coplanar, and noncoplanar magnetic orders and incorporates both general factor systems and antiunitary symmetries. Given the defining data of an SSG, the framework provides a systematic procedure for constructing all of its projective irreducible corepresentations.

A central ingredient is the decomposition of SSG factor systems. Because SSGs are infinite groups, a useful description must retain the full cohomological information while reducing it to tractable data. We first decompose the factor system of an SSG into those of its quotient SSG (qSSG) and spin-only group, together with a factor relating the two sectors, which can be nontrivial for coplanar SSGs. The factor system of the qSSG contains the richer structure. By extending Mackey's factor-system decomposition, we separate it into a factor system of the translational subgroup, a point-group factor, and a mixed factor relating translations to point-group operations. Each component admits a finite description up to equivalence and has a distinct role in the representation theory: the translational factor determines the projective translation algebra and hence the Brillouin zone, the mixed factor controls the momentum-space action of the point group and can make it nonsymmorphic, and the point-group factor contributes to the factor systems of little cogroups.

This decomposition also clarifies the relation between our construction and earlier treatments of SSG representations \cite{XiaoSpin2024,ChenEnumeration2024,SongConstructions2025}. {There are two basic routes to enumerate SSGs, one uses crystallographic unit cell convention and another uses magnetic unit cell convention. Refs.~\cite{ChenEnumeration2024,SongConstructions2025} construct corepresentations of little cogroups in the latter convention, while we construct corepresentations of SSGs in the former convention, which is the remaining task of Ref. \cite{XiaoSpin2024}. Compared with magnetic unit cell convention, our theory in crystallographic unit cell convention reveals more physical information explicitly and is complete for all SSGs.} 

The projective representation theory for magnetic space groups developed in Sec.~\ref{construct} is not restricted to SSGs. It applies whenever a (magnetic) space group is represented with a general factor system. Such projective crystalline symmetries arise in artificial crystals, including acoustic and photonic systems \cite{XueProjectively2022,LiAcoustic2022,MengSpinful2023,ChenClassification2023,LongNonAbelian2024,XiaoRevealing2024}, as well as in moir\'e systems \cite{CalugaruMoire2025} and two-dimensional materials in magnetic fields \cite{Herzog-ArbeitmanHofstadter2023,FangSymmetry2023}. The present framework therefore supplies a general tool for analyzing these systems and for identifying physical phenomena that have no counterpart in ordinary crystalline representation theory.

Practical applications will require a corresponding extension of crystallographic representation databases \cite{ElcoroDouble2017}. In particular, high-symmetry momenta and little cogroups must be tabulated for nonsymmorphic momentum-space groups, and the factor systems of space groups and their little cogroups must be classified. Ultimately, such a database should provide the irreducible representations of every space group for all admissible factor systems, including the modified Brillouin zones and momentum-space actions needed to interpret them.

With this representation-theoretic foundation in place, topological quantum chemistry and symmetry-based indicator theory can be extended to magnetic crystals with weak SOC. This extension would enable systematic, high-throughput diagnoses of topological phases in weak-SOC magnetic materials. We leave its development to future work.

\section*{Acknowledgements}

This work was supported by the General Research Fund of the Research Grants Council of Hong Kong (Grant Nos. 17301224 and 17302525).

\appendix

\section{Classification of factor system of SSGs}\label{cohomocl}
A projective corepresentation $\rho$ for a spin space group $\mathcal{G}$ satisfies a modified multiplication rule 
\begin{equation}
\rho(g_1)\rho(g_2)=\nu(g_1,g_2) \rho(g_1g_2), \ \ \forall g_1,g_2 \in \mathcal{G}.
\end{equation}
 $\rho(g)$ is antiunitary when $g$ reverses time. The extra phase $\nu(g_1,g_2)\in U(1)$ is called a factor system.

Due to the associativity $(g_1g_2)g_3=g_1(g_2g_3)$ for $\forall g_1,g_2,g_3\in \mathcal{G}$, the factor system $\nu$ should satisfy the twisted cocycle equation:
\begin{equation}\label{cocycleeq}
\nu(g_1,g_2)\nu(g_1g_2,g_3)=c_{g_1}[\nu(g_2,g_3)]\nu(g_1,g_2g_3),
\end{equation}
where $c_{g}$ means the complex conjugate and the identity map when $\rho(g)$ is antiunitary and unitary respectively. If a $\nu$ satisfies {Eq. (\ref{cocycleeq})}, it is called a 2-cocycle. All 2-cocycles form an abelian group $Z^{2,c}(\mathcal{G},U(1))$, where $c$ means the complex conjugation twist.

Since modifying each $\rho(g)$ to $f(g)\rho(g)$ by a $U(1)$ phase $f(g)$ leads to an equivalent projective corepresentation, one defines equivalence relation between factor systems (2-cocycles):
\begin{equation}
\nu(g_1,g_2)\sim \nu(g_1,g_2)\frac{f(g_1)c_{g_1}[f(g_2)]}{f(g_1g_2)}.
\end{equation}
$f(g_1)c_{g_1}[f(g_2)]/{f(g_1g_2)}$ is called a 2-coboundary. All 2-coboundaries form an abelian group $B^{2,c}(\mathcal{G},U(1))$. Thus, the equivalence classes of factor systems are classified by the quotient group {$H^{2,c}(\mathcal{G}, U(1))=Z^{2,c}(\mathcal{G},U(1))/B^{2,c}(\mathcal{G},U(1))$}, which is called the second cohomology group of $\mathcal{G}$.

\section{Factor system decomposition for group extensions}\label{appfactor}
Here, we introduce the factor system decomposition for group extensions. It extends Mackey's canonical factor decomposition theorem (theorem 9.4 in \cite{MackeyUnitary1958}) to more general group extensions with antiunitary elements. The result has a similar form to that in \cite{ZhaoUnified2020}, but is more general. The decomposition of factor systems for SSGs and magnetic groups in Sec. \ref{factorsec} can be obtained by applying the theorem here.

We first set up the basic notations. Suppose $N$ is a normal subgroup of $G$, $P\cong G/N$, i.e., $G$ is a group extension from $N$ by $P$:
\begin{equation}\label{extension}
1 \rightarrow N \rightarrow G \rightarrow P \rightarrow 1.
\end{equation}
An element of $G$ can be written as $(x,y), x\in N, y\in P$. For every $y\in P$, we can take a section $s(y)$ in $G$ and define $(x,y)_s=xs(y)$. Then the multiplication rule reads:
\begin{equation}
\begin{aligned}
(x_1,y_1)_s\cdot (x_2,y_2)_s &= x_1s(y_1)x_2s(y_2)\\&=x_1(s(y_1)x_2s(y_1)^{-1})s(y_1)s(y_2)\\
&=x_1\chi_{s,y_1}(x_2)\omega_s(y_1,y_2)s(y_1y_2)\\&=(x_1\chi_{s,y_1}(x_2)\omega_s(y_1,y_2),y_1y_2),
\end{aligned}
\end{equation}
where $\chi_{s,y}(x)=s(y)xs(y)^{-1}$ is the action of $s(y)$ on $x$ and $\omega_s(y_1,y_2)$ is defined as 
\begin{equation}
s(y_1)s(y_2)=\omega_s(y_1,y_2)s(y_1y_2).
\end{equation}
$\omega_s$ satisfies the consistency condition:
\begin{equation}\label{consist}
\omega_s(y_1,y_2)\omega_s(y_1y_2,y_3)=\chi_{s,y_1}(\omega_s(y_2,y_3))\omega_s(y_1,y_2y_3).
\end{equation}

 When $N$ is abelian, $\chi_{s,y}: P \to \mathrm{Aut}(N)$ is a homomorphism and independent of $s$, i.e., $\chi_{y_1}\circ \chi_{y_2}=\chi_{y_1y_2}$. Also, Eq. (\ref{consist}) becomes a twisted cocycle equation since $N$ is abelian. So the group extension $G$ in Eq. (\ref{extension}) is classified by $H^{2+\chi}(P,N)$. But when $N$ is nonabelian, $\chi_{s,y}$ is not a homomorphism and depends on $s$. $\chi_{s,y}$ is a homomorphism only when the extension is split, i.e., $\omega_s(y_1,y_2)$ is trivial and $G=N \rtimes P$.

A factor system $\nu: G\times G \to U(1)$ is defined as: 
\begin{equation}
\mathrm{(a)}\  \nu(g,e)=\nu(e,g)=1. 
\end{equation}
\begin{equation}\label{cocycle}
\mathrm{(b)}\  \nu(g_1,g_2)\nu(g_1g_2,g_3)=c_{g_1}(\nu(g_2,g_3))\nu(g_1,g_2g_3).
\end{equation}
Here, $c_g$ is the complex conjugate if $g$ is an antiunitary element. 

Now, we give the decomposition theorem of the factor systems.

\textbf{Theorem}: If $G$ is a group extension of $N$ by $P$ and $N$ is unitary, a factor system $\nu'$ of $G$ is equivalent to a factor system $\nu$, which can be decomposed as
\begin{equation}\label{deco}
\begin{aligned}
\nu(g_1,g_2)&=\nu((x_1,y_1)_s,(x_2,y_2)_s)\\
&=\sigma(x_1\chi_{s,y_1}(x_2),\omega(y_1,y_2))\sigma(x_1,\chi_{s,y_1}(x_2))\\&\ \ \ \  \gamma_s(x_2,y_1)\alpha_s(y_1,y_2),
\end{aligned}
\end{equation}
where $\sigma$ is the restriction of $\nu$ on $N$, i.e., $\sigma(x_1,x_2)=\nu(x_1,x_2)$, $\forall x_1,x_2\in N$. $\alpha_s(y_1,y_2):=\nu(s(y_1),s(y_2))$ and 
$\gamma_s(x,y):=\nu(s(y),x)\nu(s(y)x,s(y)^{-1})/{\nu(s(y),s(y)^{-1})}$. The decomposed factors satisfy the following self-consistency equations:
\begin{equation}\label{sigmacy}
\sigma(x_1,x_2)\sigma(x_1x_2,x_3)=\sigma(x_2,x_3)\sigma(x_1,x_2x_3)
\end{equation}
\begin{equation}\label{alpha}
\begin{aligned}
&\frac{\alpha(y_1,y_2)\alpha(y_1y_2,y_3)}{c_{y_1}(\alpha(y_2,y_3))\alpha(y_1,y_2y_3)}=\\
&\ \ \ \ \gamma(\omega(y_2,y_3),y_1)\frac{\sigma(\chi_{y_1}(\omega(y_2,y_3)),\omega(y_1,y_2y_3))}{\sigma(\omega(y_1,y_2),\omega(y_1y_2,y_3))},
\end{aligned}
\end{equation}

\begin{equation}\label{sigma}
\frac{\sigma(\chi_{y}(x_1),\chi_{y}(x_2))}{c_{y}(\sigma(x_1,x_2))}=\frac{\gamma(x_1x_2,y)}{\gamma(x_1,y)\gamma(x_2,y)},
\end{equation}

\begin{equation}\label{g}
\frac{c_{y_1}(\gamma(x,y_2))\gamma(\chi_{y_2}(x),y_1)}{\gamma(x,y_1y_2)}=\frac{\sigma(\omega(y_1,y_2),\chi_{y_1y_2}(x))}{\sigma(\chi_{y_1}\chi_{y_2}(x),\omega(y_1,y_2))}.
\end{equation}
 Note the factor $\gamma_s(x_2,y_1)$ and $\alpha_s(y_1,y_2)$ depend on the section $s$, and we omit the subscript $s$ in the self-consistency equations for simplicity. The decomposition of factor systems in Sec.\ref {decoqssg} is obtained by applying this theorem to magnetic space groups, where $\chi_{p_1}\chi_{p_2}(t)=\chi_{p_1p_2}(t)$ since the translational subgroup is abelian.
 
\textit{Proof}: Let $\nu'$ be a factor system of $G$ and let $\rho'$ be any $\nu'$ corepresentation of $G$. Then $\rho'((x,0)_s)\rho'((0,y)_s)$$=\nu'(x,(0,y)_s)\rho'((x,y)_s)$, or $\rho'(x)\rho'(s(y))$$=\nu'(x,s(y))\rho'(xs(y))$. Let 
\begin{equation}
\rho((x,y)_s)=\nu'(x,s(y))\rho'((x,y)_s), 
\end{equation}
then $\rho$ is a $\nu$ corepresentation for a factor system $\nu$ that is equivalent to $\nu'$ and $\rho(x)\rho(s(y))=\rho(xs(y))$. The restriction of $\nu$ on $N$ is $\sigma$, which naturally satisfies Eq. (\ref{sigmacy}). Consider
\begin{equation}
\begin{aligned}
\rho&((x_1,y_1)_s)\rho((x_2,y_2)_s)=\\
&\nu((x_1,y_1)_s, (x_2,y_2)_s)\rho((x_1\chi_{s,y_1}(x_2)\omega_s(y_1,y_2),y_1y_2)),
\end{aligned}
\end{equation}

\begin{equation}
\begin{aligned}
LHS=&\rho((x_1,y_1)_s)\rho((x_2,y_2)_s)=\rho(x_1)\rho(s(y_1))\rho(x_2)\rho(s(y_2)),\\
RHS=&\rho((x_1\chi_{s,y_1}(x_2)\omega_s(y_1,y_2))\rho(s(y_1y_2))\\
=&\sigma^{-1}(x_1\chi_{s,y_1}(x_2),\omega_s(y_1,y_2))\sigma^{-1}(x_1,\chi_{s,y_1}(x_2))\\&\rho(x_1)\rho(\chi_{s,y_1}(x_2))\rho(\omega_s(y_1,y_2)s(y_1y_2))\\
=&\sigma^{-1}(x_1\chi_{s,y_1}(x_2),\omega_s(y_1,y_2))\\
&\sigma^{-1}(x_1,\chi_{s,y_1}(x_2))\nu^{-1}(s(y_1),s(y_2))\\& \rho(x_1)\rho(\chi_{s,y_1}(x_2))\rho(s(y_1))\rho(s(y_2)).
\end{aligned}
\end{equation}

Equating LHS and RHS, we have
\begin{equation}
\begin{aligned}
\rho&(s(y_1))\rho(x_2)\rho^{-1}(s(y_1))\rho^{-1}(\chi_{s,y_1}(x_2))\\
&=\nu((x_1,y_1)_s, (x_2,y_2)_s)\sigma^{-1}(x_1\chi_{s,y_1}(x_2),\omega_s(y_1,y_2))\\&\ \ \ \ \sigma^{-1}(x_1,\chi_{s,y_1}(x_2))\alpha_s^{-1}(y_1,y_2),
\end{aligned}
\end{equation}
where we write $\nu(s(y_1),s(y_2))=\alpha_s(y_1,y_2)$. Since $\rho(s(y_1))\rho(x_2)\rho^{-1}(s(y_1))\rho^{-1}(\chi_{s,y_1}(x_2))$ only depends on $x_2$ and $y_1$, we can define it as $\gamma_s(x_2,y_1)$. By the definition of factor system, we can also write  it equivalently by  
\begin{equation}
\gamma(x_2,y_1):=\frac{\nu(s(y_1),x_2)\nu(s(y_1)x_2,s(y_1)^{-1})}{{\nu(s(y_1),s(y_1)^{-1})}}.
\end{equation}
Thus we obtain the decomposition in Eq. (\ref{deco}). 

Now we prove the consistency equations. For simplicity, we omit the subscript $s$ below.

\textit{2-cocycle equation for $\alpha$}

Consider the equation:
\begin{equation}
[\rho(s(y_1))\rho(s(y_2))]\rho(s(y_3))=\rho(s(y_1))[\rho(s(y_2))\rho(s(y_3))],
\end{equation}
\[
\begin{aligned}
LHS&=\alpha(y_1,y_2)\rho(\omega(y_1,y_2))\rho(s(y_1y_2))\rho(s(y_3))\\
&=\alpha(y_1,y_2)\alpha(y_1y_2,y_3)\rho(\omega(y_1,y_2))\rho(\omega(y_1y_2,y_3))\\&\ \ \ \ \ \ \ \rho(s(y_1y_2y_3))\\
&=\alpha(y_1,y_2)\alpha(y_1y_2,y_3)\sigma(\omega(y_1,y_2),\omega(y_1y_2,y_3))\\&\ \ \ \ \ \rho(\omega(y_1,y_2)\omega(y_1y_2,y_3))\rho(s(y_1y_2y_3)).
\end{aligned}
\]

\[
\begin{aligned}
RHS&=c_{y_1}(\alpha(y_2,y_3))\rho(s(y_1))\rho(\omega(y_2,y_3))\rho(s(y_2y_3))\\
&=c_{y_1}(\alpha(y_2,y_3))\gamma(\omega(y_2,y_3),y_1)\rho(\chi_{y_1}(\omega(y_2,y_3)))\\&\ \ \ \ \rho(s(y_1))\rho(s(y_2y_3))\\
&=c_{y_1}(\alpha(y_2,y_3))\alpha(y_1,y_2y_3)\gamma(\omega(y_2,y_3),y_1)\\
&\ \ \ \ \rho(\chi_{y_1}(\omega(y_2,y_3)))\rho(\omega(y_1,y_2y_3))\rho(s(y_1y_2y_3))\\
&=c_{y_1}(\alpha(y_2,y_3))\alpha(y_1,y_2y_3)\gamma(\omega(y_2,y_3),y_1)\\
&\ \ \ \ \ \sigma(\chi_{y_1}(\omega(y_2,y_3)),\omega(y_1,y_2y_3))\\
&\ \ \ \ \ \ \ \ \rho(\chi_{y_1}(\omega(y_2,y_3))\omega(y_1,y_2y_3))\rho(s(y_1y_2y_3)).
\end{aligned}
\]
Equating $LHS$ and $RHS$ and using Eq. (\ref{consist}), we get the 2-cocycle equation for $\alpha$:
\begin{equation}
\begin{aligned}
&\frac{\alpha(y_1,y_2)\alpha(y_1y_2,y_3)}{c_{y_1}(\alpha(y_2,y_3))\alpha(y_1,y_2y_3)}=\\
&\ \ \ \ \gamma(\omega(y_2,y_3),y_1)\frac{\sigma(\chi_{y_1}(\omega(y_2,y_3)),\omega(y_1,y_2y_3))}{\sigma(\omega(y_1,y_2),\omega(y_1y_2,y_3))}.
\end{aligned}
\end{equation}

\vspace{0.5cm}
\textit{Consistency condition for $\sigma$}

Expressing the cocycle equation (\ref{cocycle}) in terms of the decomposed factor system (\ref{deco}), we have
\begin{equation}\label{decocyc}
\frac{c_{y_1}(\gamma(x_3,y_2))\gamma(x_2\chi_{y_2}(x_3)\omega(y_2,y_3),y_1)}{\gamma(x_2,y_1)\gamma(x_3,y_1y_2)}=\frac{A}{B}C,
\end{equation}
where
\[
\begin{aligned}
A=&\sigma(x_1,\chi_{y_1}(x_2))\sigma(x_1\chi_{y_1}(x_2),\omega(y_1,y_2))\\&\sigma(x_1\chi_{y_1}(x_2)\omega(y_1,y_2),\chi_{y_1y_2}(x_3))\\
&\sigma(x_1\chi_{y_1}(x_2)\omega(y_1,y_2)\chi_{y_1y_2}(x_3),\omega(y_1y_2,y_3)),
\end{aligned}
\]

\[
\begin{aligned}
B=&c_{y_1}(\sigma(x_2,\chi_{y_2}(x_3))) c_{y_1}(\sigma(x_2\chi_{y_2}(x_3),\omega(y_2,y_3)))\\
&\sigma(x_1,\chi_{y_1}(x_2\chi_{y_2}(x_3)\omega(y_2,y_3)))\\
&\sigma(x_1\chi_{y_1}(x_2\chi_{y_2}(x_3)\omega(y_2,y_3)), \omega(y_1,y_2y_3)),
\end{aligned}
\]
\[
\begin{aligned}
C&=\frac{\alpha(y_1,y_2)\alpha(y_1y_2,y_3)}{c_{y_1}(\alpha(y_2,y_3))\alpha(y_1,y_2y_3)}\\&=\gamma(\omega(y_2,y_3),y_1)\frac{\sigma(\chi_{y_1}(\omega(y_2,y_3)),\omega(y_1,y_2y_3))}{\sigma(\omega(y_1,y_2),\omega(y_1y_2,y_3))}.
\end{aligned}
\]

Take $y_2=e$, then $\omega(y_1,y_2)=e$, $C=1$,
\[
LHS=\frac{\gamma(x_2x_3,y_1)}{\gamma(x_2,y_1)\gamma(x_3,y_1)},
\]
\[
\begin{aligned}
RHS&=\frac{\sigma(x_1,\chi_{y_1}(x_2))\sigma(x_1\chi_{y_1}(x_2),\chi_{y_1}(x_3))}{c_{y_1}(\sigma(x_2,x_3))\sigma(x_1,\chi_{y_1}(x_2)\chi_{y_1}(x_3))}\\
&=\frac{\sigma(\chi_{y_1}(x_2),\chi_{y_1}(x_3))}{c_{y_1}(\sigma(x_2,x_3))}.
\end{aligned}
\]
So we have the consistency condition for $\sigma$:
\begin{equation}
\frac{\sigma(\chi_{y}(x_1),\chi_{y}(x_2))}{c_{y}(\sigma(x_1,x_2))}=\frac{\gamma(x_1x_2,y)}{\gamma(x_1,y)\gamma(x_2,y)}.
\end{equation}

\vspace{0.5cm}
\textit{Twisted covariant equation for $\gamma$}

Plugging the consistency condition (\ref{alpha}) and $(\ref{sigma})$ into the general cocycle equation (\ref{decocyc}), it is simplified to 
\[
\frac{c_{y_1}(\gamma(x_3,y_2))\gamma(\chi_{y_2}(x_3),y_1)}{\gamma(x_3,y_1y_2)}=\frac{A}{B}\frac{c_{y_1}(\sigma(x_2,\chi_{y_2}(x_3)))}{\sigma(\chi_{y_1}(x_2),\chi_{y_1}\chi_{y_2}(x_3))}.
\]
Because the LHS is independent of $y_3$, the RHS must be so. So we can take $y_3=e$ to simplify the RHS, we have
\[
\begin{aligned}
\mathrm{RHS}&=\frac{\sigma(x_1,\chi_{y_1}(x_2))\sigma(x_1\chi_{y_1}(x_2)\omega(y_1,y_2))}{\sigma(x_1,\chi_{y_1}(x_2)\chi_{y_1}\chi_{y_2}(x_3))\sigma(\chi_{y_1}(x_2),\chi_{y_1}\chi_{y_2}(x_3))}\\
&\ \ \ \ \frac{\sigma(x_1\chi_{y_1}(x_2)\omega(y_1,y_2),\chi_{y_1y_2}(x_3))}{\sigma(x_1\chi_{y_1}(x_2)\chi_{y_1}\chi_{y_2}(x_3),\omega(y_1,y_2))}\\
&=\frac{\sigma(x_1\chi_{y_1}(x_2),\omega(y_1,y_2))}{\sigma(x_1\chi_{y_1}(x_2),\chi_{y_1}\chi_{y_2}(x_3))}\\
&\ \ \ \ \frac{\sigma(x_1\chi_{y_1}(x_2)\omega(y_1,y_2),\chi_{y_1y_2}(x_3))}{\sigma(x_1\chi_{y_1}(x_2)\chi_{y_1}\chi_{y_2}(x_3),\omega(y_1,y_2))}\\
&=\frac{\sigma(\omega(y_1,y_2),\chi_{y_1y_2}(x_3))}{\sigma(\chi_{y_1}\chi_{y_2}(x_3),\omega(y_1,y_2))}\\
&\ \ \ \ \frac{\sigma(x_1\chi_{y_1}(x_2),\omega(y_1,y_2)\chi_{y_1y_2}(x_3))}{\sigma(x_1\chi_{y_1}(x_2),\chi_{y_1}\chi_{y_2}(x_3)\omega(y_1,y_2))}\\
&=\frac{\sigma(\omega(y_1,y_2),\chi_{y_1y_2}(x_3))}{\sigma(\chi_{y_1}\chi_{y_2}(x_3),\omega(y_1,y_2))}.
\end{aligned}
\]
In the last line, we use the equation:
\[
\begin{aligned}
\chi_{y_1y_2}(x_3)&=s(y_1y_2)x_3s^{-1}(y_1y_2)\\&=\omega^{-1}(y_1,y_2)s(y_1)s(y_2)x_3s^{-1}(y_2)s^{-1}(y_1)\omega(y_1,y_2)\\
&=\omega^{-1}(y_1,y_2)\chi_{y_1}\chi_{y_2}(x_3)\omega(y_1,y_2).
\end{aligned}
\]
Hence, the final twisted covariant equation for $\gamma$ is 
\begin{equation}
\frac{c_{y_1}(\gamma(x,y_2))\gamma(\chi_{y_2}(x),y_1)}{\gamma(x,y_1y_2)}=\frac{\sigma(\omega(y_1,y_2),\chi_{y_1y_2}(x))}{\sigma(\chi_{y_1}\chi_{y_2}(x),\omega(y_1,y_2))}.
\end{equation}

From the procedures above, we see that Eqs. (\ref{sigmacy}-\ref{g}) are equivalent to the cocycle equation of $\nu$.

\section{Corepresentation theory of group extensions}\label{appcorep}
Here, we extend Mackey's representation theory of group extensions to corepresentation theory of group extensions. We follow the notations in Appendix \ref{appfactor}.

We consider a group extension $G$ from a group $N$ by $P$:
\begin{equation}
1 \rightarrow N \rightarrow G \rightarrow P \rightarrow 1,
\end{equation}
where $N$ only contains unitary operations, while $G$ may contain antiunitary operations.
As a normal subgroup of $G, N$ is closed under conjugation by any element $g \in G$, i.e.,
\begin{equation}
g^{-1} N g=N.
\end{equation}
Thus, $G$ canonically acts on $N$. Consider a factor $\nu$ of $G$, whose restriction on $N$ is a factor system $\sigma$ of $N$.  Let $L$ be a $\sigma$-representation of $N$. Then, we have
\begin{equation}
L(x_{1}) L(x_{2})=\sigma(x_{1}, x_{2}) L(x_{1} x_{2}).
\end{equation}
All $\sigma$-irreps of $N$ form a rep space $\mathrm{Rep}^\sigma(N)$. 

\textit{The action of $G$ on the irrep space of $N$}

For any $g \in G$, we define the action of $g$ on $\mathrm{Rep}^{\sigma}(N)$ as
\begin{equation}
\mathrm{Act}_g(L)(x)=c_g[\xi(g | x) L(g^{-1} x g)], \ \ \forall g\in G, x\in N.
\end{equation}
Here, $c_g$ is the complex conjugate if $g$ is antiunitary. The front phase $\xi$ as a function from $G \times G$ to $U(1)$ is defined as
\begin{equation}
\xi(g |h):=\frac{\nu\left(g^{-1}, h\right) \nu(g^{-1} h, g)}{\nu(g^{-1}, g)},\ \ \forall g, h \in G.
\end{equation}
It is easy to check  two facts: i) As desired, $\mathrm{Act}_g(L)$ is a $\sigma$-representation, and is unitary (irreducible) iff $L$ is unitary (irreducible). ii) $\mathrm{Act}$ is indeed a group action on the set $\operatorname{Rep}^{\sigma}(N)$ of all $\sigma$-representations of $N$, i.e.,
\begin{equation}
\mathrm{Act}_{g_{1}} \mathrm{Act}_{g_{2}}=\mathrm{Act}_{g_{1} g_{2}}.
\end{equation}

Let $\widehat{N}$ be the set of equivalence classes of irreducible unitary $\sigma$-representations of $N$. Then, the $G$-action $\mathrm{Act}$ on $\operatorname{Rep}^{\sigma}(N)$ endows a $G$-action $\mathrm{Act}$ on $\widehat{N}$ :
\begin{equation}
\mathrm{Act}_{g}([L]):=\left[\mathrm{Act}_g(L)\right], \ \ \forall g\in G.
\end{equation}
Here, $[L]$ is the equivalence class of $L$. It is clear that two equivalent $\sigma$-representations are mapped to the same equivalence class by $\mathrm{Act}_g$ for any $g \in G$. Hence, $\mathrm{Act}$ is well defined as a $G$-action on $\widehat{N}$.

Given $[L] \in \widehat{N}$, let $G_{[L]}$ be the little group of $[L]$. That is, $G_{[L]}$ is the subgroup of $G$ consisting of group elements that leave $[L]$ invariant under the action. In more concrete terms, for each $s \in G_{[L]}, \mathrm{Act}_{s}(L)$ is equivalent to $L$, i.e., there exists a unitary operator $U_{L}(s)$ such that
\begin{equation}
\mathrm{Act}_s(L)(x)=U_{L}^{\dagger}(s) L(x) U_{L}(s),\ \ \forall x\in N.
\end{equation}
Since $L$ is irreducible, $U_{L}(s)$ is uniquely determined by $L$ and $s$ up to a phase factor $\lambda_{L}(s) \in U(1)$. Because if there is another unitary transformation $U_L'(s)$ relates $\mathrm{Act}_s(L)$ and $L$, then one can easily check that $U_L'^{\dagger}(s)U_L(s)$ commutes with all $L(x)$, thus $U_L'^\dagger(s)U_L(s)=\lambda_L(s)I$ according to the Schur's lemma.

It is significant to observe that once we specify $U_{L}$ for a given $L$, we know how to construct $U_{\tilde{L}}$ for any equivalent irreducible unitary representation $\tilde{L} \in[L]$. Since there exists a unitary operator $V$ such that $\tilde{L}(x)=V^{\dagger} L(x) V$ for all $x \in N$. Then, we define
\begin{equation}
U_{\tilde{L}}(s):=V^{\dagger}U_{L}(s) c_s(V), \ \ \  \forall s \in G_{[L]},
\end{equation}
and it is straightforward to check that
\begin{equation}
\mathrm{Act}_s(\tilde{L})(x)=U_{\tilde{L}}^{\dagger}(s) \tilde{L}(x) U_{\tilde{L}}(s)
\end{equation}
for all $s \in G_{[L]}$ and all $x \in N$.

With the definition of $U_{L}, \mathrm{Act}_{s_1}(\mathrm{Act}_{s_2}(L))$ corresponds to $U_{L}(s_{1}) c_{s_1}(U_{L}(s_{2}))$ for all $s_{1}, s_{2} \in G_{[L]}$, and $\mathrm{Act}_{s_1s_2}(L)$ corresponds to $U_L(s_1s_2)$. Due to the aforementioned $U(1)$ degrees of freedom, $U_{L}(s_{1}) c_{s_1}(U_{L}(s_{2}))$ is equal to $U_L(s_1s_2)$ up to a phase factor $\tau_{L}(s_{1}, s_{2})$, i.e.,
\begin{equation}
U_{L}(s_{1}) c_{s_1}(U_{L}(s_{2}))=\tau_{L}(s_{1}, s_{2}) U_{L}(s_{1} s_{2}).
\end{equation}
This means $U_L$ is a matrix $\tau_L$-corep of the little group. Clearly, $\tau_{L}$ is a twisted 2-cocycle, and is uniquely determined up to coboundaries, and therefore $L$ is associated with a cohomological class $\left[\tau_{L}\right] \in H^{2,c}\left(G_{[L]}, U(1)\right)$. Note that obviously $\left[\tau_{L}\right]$ is independent of the choice $L$ of a representative of $[L]$, and may be denoted as $\left[\tau_{[L]}\right]$.

It is clear that $N$ is a subgroup of $G_{[L]}$. Restricting $\tau_{L}$ on $N \times N$, we can set
\begin{equation}
\tau_{L}(x_{1}, x_{2})=\sigma(x_{1}, x_{2}),\ \ \forall x_{1}, x_{2} \in N,
\end{equation}
This can be realized by setting
\begin{equation}
U_{L}(x)=L(x),\ \ \forall x \in N.
\end{equation}
The equation $\tau_{L}\left(x_{1}, x_{2}\right)=\sigma\left(x_{1}, x_{2}\right)=\nu(x_1,x_2)$ for all $x_{1}, x_{2} \in N$ motivates us to consider $\omega_L'=\nu / \tau_{L} \in$ $Z^{2,c}\left(G_{[L]}, U(1)\right)$, which is specified by
\begin{equation}
\omega_L'(s_{1}, s_{2})=\frac{\nu(s_{1}, s_{2})}{\tau_{L}(s_{1}, s_{2})}, \ \ \forall s_{1}, s_{2} \in G_{[L]}.
\end{equation}
Clearly, $\omega_L'=1$ when restricted on $N \subset G_{[L]}$. We claim that $\omega_L'$ is equivalent to the form $\omega_{L} \circ(\pi \times \pi)$. Here, $\omega_{L}$ is a twisted 2-cocycle over the little cogroup $G_{[L]} / N$, namely $\omega_{L} \in Z^{2,c}\left(G_{[L]} / N, U(1)\right)$, and $\pi$ is the canonical projection $G_{[L]} \rightarrow G_{[L]} / N$. Our claim means that $L$ and $\nu$ uniquely determine a 2-cohomological class $\left[\omega_{L}\right] \in H^{2,c}(G_{[L]} / N, U(1))$.

We now explicitly present $\omega_{L}$. For each coset $N s$, we choose a representative $c(N s)$, or denote it simply as $c(s)$ without ambiguity. We require that $c(x)=1$ for all $x \in N$. We then introduce the function
\begin{equation}
g(s):=\omega_L'(c(s) s^{-1}, s)
\end{equation}
from $G_{[L]}$ to $U(1)$, and transform $\omega_L'$ by it, i.e.,
\begin{equation}
\omega_L''(s_{1}, s_{2})=\omega_L'(s_{1}, s_{2}) \frac{g(s_{1} s_{2})}{g(s_{1}) c_{s_1}(g(s_{2}))}.
\end{equation}
This transformation can be seen as rechoosing the phase of $U_L(s)$. We claim that $\omega_L''$ is of the form $\omega_{L} \circ(\pi \times \pi)$, which is showed in the following.

Since $\omega_L'=1$ on $N \times N$, the trivial rep $I$ is a $\omega_L'$-rep of $N$. Then, let $W$ be the induced $\omega_L'$-corep of $G_{[L]}$ from the trivial rep $I$ of $N$. It is significant to observe that $W(x)=I$ for all $x \in N$. Then, $W(s)=W(x) W(s)=\omega_L'(x,s) W(xs)$ for all $s \in G_{[L]}$ and all $x \in N$, which implies $W$ restricted to each coset $Ns$ is constant up to a phase dependent on the elements of $Ns$. This motivates us to introduce
\begin{equation}
W^{\prime}(s):=W(c(s)),\ \ \forall s\in G_{[L]}.
\end{equation}
That is, $W^{\prime}(s)$ is constant on each coset $Ns$, and is determined by the chosen representative $c(s) \in Ns$. In fact, $W^{\prime}$ is also a $\omega_L''$-corep of $G_{[L]}$, which can be seen from the straightforward derivation:
\begin{equation}
\begin{aligned}
&\ \ \ \ W_{s_{1}}^{\prime} W_{s_{2}}^{\prime}  =W_{c(s_{1})} W_{c(s_{2})}=W_{c(s_{1}) s_{1}^{-1} s_{1}} W_{c(s_{2}) s_{2}^{-1} s_{2}} \\
& =\frac{1}{g(s_{1}) c_{s_1}(g(s_{2}))} W_{s_{1}} W_{s_{2}}=\frac{\omega_L'(s_{1}, s_{2})}{g(s_{1}) c_{s_1}(g(s_{2}))} W_{s_{1} s_{2}} \\
& =\frac{\omega_L'(s_{1}, s_{2})}{g(s_{1}) c_{s_1}(g(s_{2}))} W_{s_{1} s_{2}\left[c(s_{1} s_{2})\right]^{-1} c(s_{1} s_{2})} \\
& =\frac{\omega_L'(s_{1}, s_{2})}{g(s_{1}) c_{s_1}(g(s_{2})) \omega_L'(s_{1} s_{2}\left[c(s_{1} s_{2})\right]^{-1}, c(s_{1} s_{2}))} W_{c(s_{1} s_{2})} \\
& =\omega_L'(s_{1}, s_{2}) \frac{g(s_{1},s_2)}{g(s_{1}) c_{s_1}(g(s_{2}))} W_{s_{1} s_{2}}^{\prime}\\
&=\omega_L''(s_1,s_2) W_{s_{1} s_{2}}^{\prime}.
\end{aligned}
\end{equation}
In the second-to-last step, we have used the result:
\begin{equation}
\begin{aligned}
& \omega_L'(s_{1} s_{2}\left[c(s_{1} s_{2})\right]^{-1}, c(s_{1} s_{2})) \omega_L'(c(s_{1} s_{2})(s_{1} s_{2})^{-1}, s_{1} s_{2}) \\
= & \omega_L'(1, s_{1} s_{2})\omega_L'(s_{1} s_{2}\left[c(s_{1} s_{2})\right]^{-1}, c(s_{1} s_{2})(s_{1} s_{2})^{-1})=1.
\end{aligned}
\end{equation}
Since $W^{\prime}$ is constant on each coset, $\omega_L''$ is also constant on $ Ns_{1} \times Ns_{2} $ for all $s_{1}, s_{2} \in G_{[L]}$. Hence, $\omega_L''$ can be cast in the form $\omega_L''=\omega_{L} \circ(\pi \times \pi)$, where $\omega_{L} \in Z^{2,c}\left(G_{[L]} / N, U(1)\right)$ is uniquely determined by $\nu_{L}^{\prime}$. Furthermore, one can show that only the cohomological class of $\omega_{L}$ is uniquely determined by $L$ and $\nu$.

\textit{The construction of irreducible $\nu$-corepresentations}

We are now ready to state the way to construct all equivalence classes of irreducible corepresentations of $G$, from orbits of $G$-action on $\widehat{N}$ and ${\widehat{G_{[L]}/N}}^{\omega_{L}}$. Here, ${\widehat{G_{[L]}/N}}^{\omega_{L}}$ is the set of all $\omega_L$-coirreps of $G_{[L]}/N$.

The $G$-action $\mathrm{Act}$ on $\widehat{N}$, specified by
\begin{equation}
\begin{aligned}
&\mathrm{Act}_g(L)(x)=c_g[\xi(g| x) L(g^{-1}x g)], \\ 
&\xi(g |x)=\frac{\sigma\left(g^{-1}, x\right) \sigma(g^{-1} x, g)}{\sigma(g^{-1}, g)},
\end{aligned}
\end{equation}
partitions $\widehat{N}$ into orbits $\mathcal{O}_{i}$ indexed by $i$. For each orbit $\mathcal{O}_{i}$, we choose a representative $\left[L_{i}\right]$, which has a little group $G_{\left[L_{i}\right]}$. Then, following the procedure presented above, the $G_{\left[L_{i}\right]}$-action on equivalent elements of $\left[L_{i}\right]$ leads to a unitary $\tau_{L_{i}}$-corep of $G_{\left[L_{i}\right]}$ with factor $\tau_{L_{i}}$ :
\begin{equation}
\begin{aligned}
&\mathrm{Act}_s(L_{i})(x)=U_{L_{i}}^{\dagger}(s) L(x) U_{L_{i}}(s), \\
& U_{L_{i}}(s_{1}) c_{s_1}(U_{L_{i}}(s_{2}))=\tau_{L_{i}}(s_{1}, s_{2}) U_{L_{i}}(s_{1} s_{2}),
\end{aligned}
\end{equation}
where $x \in N$ and $s_1, s_{2} \in G_{\left[L_{i}\right]}$.

Since $N \subset G_{\left[L_{i}\right]}$, we consider the little cogroup $G_{\left[L_{i}\right]} / N$.  We can choose the phase of $U_{L_i}$ such that $\nu/\tau_{L_i}=\omega_{L_{i}} \circ(\pi \times \pi)$, where $\omega_{L_{i}}$ is a factor of $G_{\left[L_{i}\right]} / N$.

Then, we consider the set $\widehat{G_{\left[L_{i}\right]} / N}^{\omega_{L_{i}}}$ of all matrix unitary $\omega_{L_{i}}$-coirreps, and
index $\widehat{G_{\left[L_{i}\right]} / N}^{\omega_{L_{i}}}$ by $a_i$. Every $\tilde{D}_{a_i} \in \widehat{G_{\left[L_{i}\right]} / N}^{\omega_{L_{i}}}$ can be lifted to be a matrix $\omega_{L_{i}} \circ(\pi \times \pi)$-coirrep of $G_{\left[L_{i}\right]}$, which is given by
\begin{equation}
{D}_{a_i}:=\tilde{D}_{a_i} \circ \pi.
\end{equation}
Then, we can construct an irreducible $\nu$-corep of $G_{\left[L_{i}\right]}$ as
\begin{equation}
M_{i, a_{i}}(s):=U_{L_{i}}(s) \otimes {D}_{a_i}(s)c_s, \ \ \forall s\in G_{[L_i]},
\end{equation}
{{where} $c_s=\mathcal{K}$ if $s$ is an antiunitary element. }
Note that the factor of $M_{i, a_{i}}$ is the product of $\tau_{L_{i}}$ and $\omega_{L_{i}} \circ(\pi \times \pi)$, and therefore is $\nu$.

Over the orbit $\mathcal{O}_{i}, M_{i, a_{i}}$ induces a $\nu$-coirrep $\mathrm{Ind}_{G_{[L_i]}}^{G} M_{i, a_{i}}$ of $G$. All such induced $\nu$-coirreps exhaustively represent all equivalence classes of $\nu$-coirreps of $G$. For an introduction to induced representations, see Appendix \ref{Induce}.

\section{$\sigma$-irreps of the translational subgroup}\label{irreptrans}

Here we first show that when $\sigma$ is nontrivial, the quotient algebra $\mathrm{T}^\sigma/Z(\mathrm{T}^\sigma)$ is always a Heisenberg algebra, then we explicitly construct $\sigma$-irreps of the translational subgroup $\mathrm{T}$ .

First, we consider the case that only one $\sigma_{ij}=-1$. Without loss of generality, we assume $\sigma_{ab}=-1, \sigma_{bc}=\sigma_{ca}=1$. $Z(\mathrm{T}^\sigma)$ can be generated by $t_a'=t_a^2, t_b'=t_b^2$ and $t_c'=t_c$. The Brillouin zone is defined with respect to the new basis $\boldsymbol{t}_a',\boldsymbol{t}_b',\boldsymbol{t}_c'$. The quotient algebra $\mathrm{T}^\sigma/Z(\mathrm{T}^\sigma)$ is a Heisenberg algebra $\mathrm{Heis}(\mathbb{Z}_2\times \mathbb{Z}_2)$, which can be written as 
\begin{equation}
\mathrm{Heis}(\mathbb{Z}_2\times \mathbb{Z}_2)=\langle \tilde{t}_a,\tilde{t}_b|\tilde{t}_a^2=1, \tilde{t}_b^2=1,\tilde{t}_a\tilde{t}_b=-\tilde{t}_b\tilde{t}_a\rangle,
\end{equation}
where $\tilde{t}_a,\tilde{t}_b$ is the corresponding equivalence class of $t_a,t_b$ respectively, while $t_a^2, t_b^2$ and $t_c$ correspond to the trivial equivalence class.

Next, we consider the case when two $\sigma_{ij}=-1$. Without loss of generality, we assume $\sigma_{ab}=\sigma_{ca}=-1$ and $\sigma_{bc}=1$. $Z(\mathrm{T}^\sigma)$ is generated by $t_a'=t_a^2, t_b'=t_bt_c, t_c'=t_bt_c^{-1}$. The Brillouin zone is defined with respect to the new basis $\boldsymbol{t}_a',\boldsymbol{t}_b',\boldsymbol{t}_c'$. The quotient algebra $\mathrm{T}^\sigma/Z(\mathrm{T}^\sigma)$ is also a Heisenberg algebra $\mathrm{Heis}(\mathbb{Z}_2\times \mathbb{Z}_2)$, which can be written as 
\begin{equation}
\mathrm{Heis}(\mathbb{Z}_2\times \mathbb{Z}_2)=\langle \tilde{t}_a,\tilde{t}_b|\tilde{t}_a^2=1, \tilde{t}_b^2=1,\tilde{t}_a\tilde{t}_b=-\tilde{t}_b\tilde{t}_a\rangle,
\end{equation}
where $\tilde{t}_a,\tilde{t}_b$ are the corresponding equivalence class of $t_a,t_b$ respectively.

Finally, we consider the case when three $\sigma_{ij}=-1$, i.e., $\sigma_{ab}=\sigma_{bc}=\sigma_{ca}=-1$.  $Z(\mathrm{T}^\sigma)$ is generated by $t_a'=t_bt_ct_a^{-1}, t_b'=t_ct_at_b^{-1}, t_c'=t_at_bt_c^{-1}$. The Brillouin zone is defined with respect to the new basis $\boldsymbol{t}_a',\boldsymbol{t}_b',\boldsymbol{t}_c'$. The quotient algebra is still a Heisenberg algebra $\mathrm{Heis}(\mathbb{Z}_2\times \mathbb{Z}_2)$, which can be written as
\begin{equation}
\mathrm{Heis}(\mathbb{Z}_2\times \mathbb{Z}_2)=\langle \tilde{t}_a,\tilde{t}_c|\tilde{t}_a^2=1, \tilde{t}_c^2=1,\tilde{t}_a\tilde{t}_c=-\tilde{t}_c\tilde{t}_a\rangle,
\end{equation}
where $\tilde{t}_a,\tilde{t}_c$ is the corresponding equivalence class of $t_a, t_c$ respectively, while the corresponding equivalence class of $t_b$ is $\tilde{t}_a\tilde{t}_c$.

To construct $\sigma$-irreps of $\mathrm{T}$, we cast the short exact sequence
\begin{equation}
1 \to Z(\mathrm{T}^\sigma) \to  \mathrm{T}^\sigma \to \mathrm{Heis}(\mathbb{Z}_2\times \mathbb{Z}_2) \to 1
\end{equation}
back to a short exact sequence of groups 
\begin{equation}
1\to Z(\mathrm{T}^\sigma)\to \mathrm{T} \to \mathbb{Z}_2\times \mathbb{Z}_2\to 1.
\end{equation}
In the sense of equivalence class, the factor system of $\mathrm{T}$ can be decomposed into three parts according to the canonical decomposition of factor systems in Appendix \ref{appfactor}. Since $Z(\mathrm{T}^\sigma)$ is the center and $\mathbb{Z}_2\times \mathbb{Z}_2$ acts trivially on $Z(\mathrm{T}^\sigma)$, the factor system of $Z(\mathrm{T}^\sigma)$ and the mixing factor between $Z(\mathrm{T}^\sigma)$ and $\mathbb{Z}_2\times \mathbb{Z}_2$ are trivial, so the only factor of $\mathbb{Z}_2\times \mathbb{Z}_2$ is nontrivial, which is determined by the Heisenberg algebra.

Now we can construct $\sigma$-irreps of $\mathrm{T}$ via Mackey machine. The $\sigma$-irreps of $Z(\mathrm{T}^\sigma)$ are given by Eq. (\ref{tsirrep}). Since the action of $\mathbb{Z}_2\times \mathbb{Z}_2$ on $Z(\mathrm{T}^\sigma)$ is trivial, at every $\boldsymbol{k}$, the little group is the whole translation group $\mathrm{T}$, and the little cogroup is $\mathbb{Z}_2\times \mathbb{Z}_2$. The factor system of $\mathbb{Z}_2\times \mathbb{Z}_2$ is given by 
\begin{equation}
\tilde{\sigma}_{\boldsymbol{k}}(\tilde{t}_1,\tilde{t}_2)=e^{i\boldsymbol{k}\cdot \boldsymbol{\omega}(\tilde{t}_1,\tilde{t}_2)}\alpha_H(\tilde{t}_1,\tilde{t}_2), \ \ \forall \tilde{t}_1,\tilde{t}_2\in \mathbb{Z}_2\times \mathbb{Z}_2,
\end{equation}
where $\alpha_H$ is the factor system contributed by the Heisenberg algebra, and $e^{i\boldsymbol{k}\cdot \boldsymbol{\omega}(\tilde{t}_1,\tilde{t}_2)}$ is contributed by ``nonsymmorphicity". $\boldsymbol{\omega}(\tilde{t}_1,\tilde{t}_2)\in \mathcal{Z}_{\mathcal{L}}$ is obtained according to
\begin{equation}
s(\tilde{t}_1)s(\tilde{t}_2)=\omega(\tilde{t}_1,\tilde{t}_2)s(\tilde{t}_1\tilde{t}_2),\ \ \forall  \tilde{t}_1,\tilde{t}_2\in \mathbb{Z}_2\times \mathbb{Z}_2,
\end{equation}
with $s(\tilde{t})$ being the section of $\tilde{t}\in \mathbb{Z}_2\times \mathbb{Z}_2$ in $\mathrm{T}$. Suppose the $\tilde{\sigma}_{\boldsymbol{k}}$-irrep of $\mathbb{Z}_2\times \mathbb{Z}_2$ is $S_{\boldsymbol{k}}$ (which is unique up to equivalence), then the $\sigma$-irrep of $\mathrm{T}$ at $\boldsymbol{k}$ is given by
\begin{equation}
T^{\boldsymbol{k}}(s(\tilde{t})t_0)=S_{\boldsymbol{k}}(\tilde{t})e^{i\boldsymbol{k}\cdot \boldsymbol{t}_0}, \ \forall \tilde{t} \in \mathbb{Z}_2\times \mathbb{Z}_2, \ t_0\in Z(\mathrm{T}^\sigma).
\end{equation}
Note every element in $\mathrm{T}$ has a unique decomposition $s(\tilde{t})t_0$.

For the case with only one $\sigma_{ij}=-1$. The section of $\tilde{t}_a,\tilde{t}_b$ in $\mathrm{T}^\sigma$ can just be taken to be $s(\tilde{t}_a)=t_a,s(\tilde{t}_b)=t_b$. Thus $\omega(\tilde{t}_a,\tilde{t}_a)=t_a^2$, $\omega(\tilde{t}_b,\tilde{t}_b)=t_b^2$, $\omega(\tilde{t}_a,\tilde{t}_b)=\omega(\tilde{t}_b,\tilde{t}_a)=0$. Then the factor system $\tilde{\sigma}_{\boldsymbol{k}}$ of the little cogroup $\mathbb{Z}_2\times \mathbb{Z}_2$ at momentum $\boldsymbol{k}$ is given by
\begin{equation}\label{1sigmaf1}
\tilde{\sigma}_{\boldsymbol{k}}(\tilde{t}_a,\tilde{t}_a)=e^{i\boldsymbol{k}\cdot 2\boldsymbol{t}_a}, \ \tilde{\sigma}_{\boldsymbol{k}}(\tilde{t}_b,\tilde{t}_b)=e^{i\boldsymbol{k}\cdot 2\boldsymbol{t}_b},
\end{equation}
and
\begin{equation}\label{1sigmaf2}
\frac{\tilde{\sigma}_{\boldsymbol{k}}(\tilde{t}_a,\tilde{t}_b)}{\tilde{\sigma}_{\boldsymbol{k}}(\tilde{t}_b,\tilde{t}_a)}=-1.
\end{equation} 
The $\tilde{\sigma}_{\boldsymbol{k}}$-irrep $S_{\boldsymbol{k}}$ of $\mathbb{Z}_2\times \mathbb{Z}_2$ can be written as 
\begin{equation}\label{sk1}
\begin{aligned}
S_{\boldsymbol{k}}(\tilde{t}_a)=e^{i\boldsymbol{k}\cdot \boldsymbol{t}_a}\sigma_1,\ \ \ S_{\boldsymbol{k}}(\tilde{t}_b)=e^{i\boldsymbol{k}\cdot \boldsymbol{t}_b}\sigma_3.
\end{aligned}
\end{equation}
And the $\sigma$-irrep of $\mathrm{T}$ at $\boldsymbol{k}$ is just given by 
\begin{equation}
\begin{aligned}
&T^{\boldsymbol{k}}(t_a)=e^{i\boldsymbol{k}\cdot \boldsymbol{t}_a}\sigma_1,\ \ \ T^{\boldsymbol{k}}(t_b)=e^{i\boldsymbol{k}\cdot \boldsymbol{t}_b}\sigma_3,\\
&T^{\boldsymbol{k}}(t_c)=e^{i\boldsymbol{k}\cdot \boldsymbol{t}_c}\sigma_0.
\end{aligned}
\end{equation}
Note here we only give the representation matrices for generators of $\mathrm{T}$.

For the case with two $\sigma_{ij}=-1$, we can take the section to be $s(\tilde{t}_a)=t_a,s(\tilde{t}_b)=t_b$. Thus, we still have $\omega(\tilde{t}_a,\tilde{t}_a)=t_a^2$, $\omega(\tilde{t}_b,\tilde{t}_b)=t_b^2$, $\omega(\tilde{t}_a,\tilde{t}_b)=\omega(\tilde{t}_b,\tilde{t}_a)=0$. Note $t_b^2$ is an element in $Z(\mathrm{T}^\sigma)$.  Then the factor system $\tilde{\sigma}_{\boldsymbol{k}}$ is also given by Eq. (\ref{1sigmaf1}) and Eq. (\ref{1sigmaf2}), and the $\tilde{\sigma}_{\boldsymbol{k}}$-irrep $S_{\boldsymbol{k}}$ of $\mathbb{Z}_2\times \mathbb{Z}_2$ is given by Eq. (\ref{sk1}). Finally, the $\sigma$-irrep of $\mathrm{T}$ at $\boldsymbol{k}$ is given by 
\begin{equation}
\begin{aligned}
&T^{\boldsymbol{k}}(t_a)=S_{\boldsymbol{k}}(\tilde{t}_a)=e^{i\boldsymbol{k}\cdot \boldsymbol{t}_a}\sigma_1,\\ 
&T^{\boldsymbol{k}}(t_b)=S_{\boldsymbol{k}}(\tilde{t}_b)=e^{i\boldsymbol{k}\cdot \boldsymbol{t}_b}\sigma_3,\\
&T^{\boldsymbol{k}}(t_c)=T^{\boldsymbol{k}}(s(\tilde{t}_b^{-1})t_b')=S_{\boldsymbol{k}}(\tilde{t}_b^{-1})e^{i\boldsymbol{k}\cdot \boldsymbol{t}_b'}=e^{i\boldsymbol{k}\cdot \boldsymbol{t}_c}\sigma_3.
\end{aligned}
\end{equation}

For the case with all three $\sigma_{ij}=-1$, we can take the section to be $s(\tilde{t}_a)=t_a,s(\tilde{t}_c)=t_c$. Thus,  the factor system $\tilde{\sigma}_{\boldsymbol{k}}$ of $\mathbb{Z}_2\times \mathbb{Z}_2$ is given by 
\begin{equation}
\tilde{\sigma}_{\boldsymbol{k}}(\tilde{t}_a,\tilde{t}_a)=e^{i\boldsymbol{k}\cdot 2\boldsymbol{t}_a}, \ \tilde{\sigma}_{\boldsymbol{k}}(\tilde{t}_c,\tilde{t}_c)=e^{i\boldsymbol{k}\cdot 2\boldsymbol{t}_c},
\end{equation}
and
\begin{equation}
\frac{\tilde{\sigma}_{\boldsymbol{k}}(\tilde{t}_a,\tilde{t}_c)}{\tilde{\sigma}_{\boldsymbol{k}}(\tilde{t}_c,\tilde{t}_a)}=-1.
\end{equation} 
 The $\tilde{\sigma}_{\boldsymbol{k}}$-irrep $S_{\boldsymbol{k}}$ of $\mathbb{Z}_2\times \mathbb{Z}_2$ can be chosen as
\begin{equation}
S_{\boldsymbol{k}}(\tilde{t}_a)=e^{i\boldsymbol{k}\cdot \boldsymbol{t}_a}\sigma_1,\ \ \ S_{\boldsymbol{k}}(\tilde{t}_c)=e^{i\boldsymbol{k}\cdot \boldsymbol{t}_c}\sigma_3.
\end{equation}
 Finally, the $\sigma$-irrep of $\mathrm{T}$ at $\boldsymbol{k}$ is given by 
\begin{equation}\label{3sigma}
\begin{aligned}
&T^{\boldsymbol{k}}(t_a)=S_{\boldsymbol{k}}(\tilde{t}_a)=e^{i\boldsymbol{k}\cdot \boldsymbol{t}_a}\sigma_1,\\ 
&T^{\boldsymbol{k}}(t_b)=T^{\boldsymbol{k}}(s(\tilde{t}_a\tilde{t}_c)t_b'^{-1})=S_{\boldsymbol{k}}(\tilde{t}_a\tilde{t}_c)e^{-i\boldsymbol{k}\cdot \boldsymbol{t}_b'}=-ie^{i\boldsymbol{k}\cdot \boldsymbol{t}_b}\sigma_2,\\
&T^{\boldsymbol{k}}(t_c)=S_{\boldsymbol{k}}(\tilde{t}_c)=e^{i\boldsymbol{k}\cdot \boldsymbol{t}_c}\sigma_3.
\end{aligned}
\end{equation}
Since multiplying $T^{\boldsymbol{k}}(t)$ with $U(1)$ factor $f(t)$ results in an equivalent irrep in the same equivalence class of the factor system, we can take a more symmetric form of Eq. (\ref{3sigma}) as
\begin{equation}
\begin{aligned}
&T^{\boldsymbol{k}}(t_a)=-ie^{i\boldsymbol{k}\cdot \boldsymbol{t}_a}\sigma_1,\\ 
&T^{\boldsymbol{k}}(t_b)=-ie^{i\boldsymbol{k}\cdot \boldsymbol{t}_b}\sigma_2,\\
&T^{\boldsymbol{k}}(t_c)=-ie^{i\boldsymbol{k}\cdot \boldsymbol{t}_c}\sigma_3.
\end{aligned}
\end{equation}

Here, we note that the method we adopt above is also applicable for the translational subgroups with general rational factor systems, the only difference is that the Heisenberg algebra is in general $\mathrm{Heis}(\mathbb{Z}_q\times \mathbb{Z}_q)$ \cite{FangSymmetry2023}.

\section{{$k$-space nonsymmorphic actions realizable by SSGs}}\label{nonsymapp}

Here we prove two results in Sec. \ref{actonBZ}. One is that if the $k$-orbits of a $k$-space nonsymmorphic MSG all have dimensions larger than two, then this $k$-space MSG cannot be realized by SSGs. Another is that there is no essential $k$-space nonsymmorphic action when the factor system $\sigma$ is {nontrivial} for an MSG. To prove them, the key observation is that the factor system $\nu$ of $G$ is obtained from a two-dimensional unitary corep of $G$ (see Eq. (\ref{defact})). This means if a general factor system of $G$ does not have coirreps in dimension one or two, then it cannot be realized by lifting $SO(3)$ to $SU(2)$ in spin space groups.

Following the construction in Sec. \ref{construct}, we know the dimension of a coirrep of $G$ at $\boldsymbol{k}$ is given by
\begin{equation}\label{dimen}
d=d_{\mathrm{o}}\times d_{\mathrm{t}}\times d_{\mathrm{p}},
\end{equation}
where $d_{\mathrm{o}}$ is the dimension of the orbit of $\boldsymbol{k}$, $d_{\mathrm{t}}$ is the dimension of the translational subgroup, and $d_{\mathrm{p}}$ is the dimension of the coirrep of the little cogroup.

For a factor system $\nu$ of $G$, if $d_\mathrm{o}>2$ at all $\boldsymbol{k}$, then all $\nu$-coirreps of $G$ have dimensions larger than two, thus $\nu$ cannot be realized by lifting $SO(3)$ to $SU(2)$ in spin space groups.

For a factor system $\nu$ of $G$, if it is nontrivial on the translational subgroup, $d_{\mathrm{t}}=2$. Furthermore, if it leads to essential nonsymmorphic actions on BZ, there is no fixed point in BZ, and $d_{\mathrm{o}}\geq 2$ for all orbits. Then according to Eq. (\ref{dimen}), every $\nu$-coirrep of $G$ is at least four dimensional. Thus, $\nu$ cannot be realized by lifting $SO(3)$ to $SU(2)$ in spin space groups.

\section{Induced projective representation}\label{Induce}

$H$ is a subgroup of $G$; given a projective corepresentation $D$ of $H$ with factor system $\nu$, we can induce a $\nu$-corep of $G$ in the following.

We denote the basis of representation space of $D$ as $\{|\psi_{i}\rangle\}$, i.e., we have
\begin{equation}
\hat{h}|\psi_i\rangle =D(h)_{ji}|\psi_j\rangle,\ \ h\in H.
\end{equation}
Here $\hat{h}$ is the representation operator on the representation space, and repeated indices are summed. Since $H\subset G$,  we can decompose $G$ into cosets of $H$:
\begin{equation}
G=\cup_n g_n H,
\end{equation}
where $g_n$ is the representative element in coset $g_nH$. Any $g\in G$ can be decomposed as $g=g_\alpha h$ with $h\in H$. Then we define a set of new basis vectors$\{|\psi_{n,i}\rangle\}$:
\begin{equation}
|\psi_{n,i}\rangle=\hat{g}_n|\psi_i\rangle,\ \ n=1,2,..., i=1,2,....
\end{equation}
Then the vector space spanned by $\{|\psi_{n,i}\rangle\}$ is just a $\nu$-corep of $G$, which we call the induced $\nu$-corep of $G$ from $D$. The representation matrices can be obtained by acting with an element of $G$ on the basis. For $\forall g\in G$, we have
\begin{equation}
\begin{aligned}
\hat{g}|\psi_{n,i}\rangle&=\hat{g}\hat{g}_{n}|\psi_i\rangle\\
&=\nu(g,g_n)\widehat{gg_n}|\psi_i\rangle\\
&=\frac{\nu(g,g_n)}{\nu(g_m,h)}\hat{g}_m\hat{h}|\psi_i\rangle\\
&=\frac{\nu(g,g_n)}{\nu(g_m,h)}c_{g_m}(D(h)_{ji})\hat{g}_m|\psi_j\rangle\\
&=\frac{\nu(g,g_n)}{\nu(g_m,h)}c_{g_m}(D(h)_{ji})|\psi_{m,j}\rangle,
\end{aligned}
\end{equation}
in which $g_m h$ is the unique decomposition of $gg_n$, and $c_{g_m}=\mathcal{K}$ if $g_m$ is antiunitary. From this derivation, the matrix elements of the induced $\nu$-corep for $g$ are
\begin{equation}
\langle \psi_{m,j}|\hat{g}|\psi_{n,i}\rangle=\frac{\nu(g,g_n)}{\nu(g_m,h)}c_{g_m}(D(h)_{ji})\delta_{h,g_m^{-1}gg_n}.
\end{equation}

\section{Construct $\nu$-coirreps of antiunitary groups from $\nu$-irreps of its unitary subgroup}\label{Criteria}

Here, we introduce an important application of the extended Mackey's theory for antiunitary groups, i.e., constructing $\nu$-coirreps of antiunitary groups from $\nu$-irreps of their unitary subgroup. This important special case is discussed in \cite{ShawIrreducible1974}, which is a generalization of Wigner's classification of coirreps \cite{WignerGroup1959}.
For an antiunitary group $G$, its unitary elements form a subgroup $G_0$, and $G/G_0\cong Z_2$. So $G$ can be written as 
\begin{equation}
G=G_0+G_0 a,
\end{equation}
with $a$ an arbitrary antiunitary element. The $\nu$-coirreps of $G$ can be obtained by $\nu$-irreps of $G_0$ by Mackey machine. For a given $\nu$-irrep $D$ of $G_0$, we can construct a $\nu$-coirrep of $G$. Consider the action of $a$ on $D$,
\begin{equation}
\mathrm{Act}_a(D)(h)=\xi^*(a|h)D^*(a^{-1}ha),\ \ \forall h\in G_0,
\end{equation}
where
\begin{equation}
\xi(a|h)=\frac{\nu(a^{-1},h)\nu(a^{-1}h,a)}{\nu(a^{-1},a)}
\end{equation}
is caused by the factor system $\nu$. Referring to the relation between $\mathrm{Act}_a(D)$ and $D$, there are three cases:

Case (i): $\mathrm{Act}_a(D)$ is equivalent to $D$ by $\mathrm{Act}_a(D)(h)=U^\dagger(a)D(h)U(a),\ \forall h\in G_0$. And the transformation matrix $U(a)$ satisfies $U(a)U^*(a)=+\nu(a,a)D(a^2)$.

The $\nu$-coirrep of $G$ induced from $D$ is given by
\begin{equation}
\begin{aligned}
&\rho(h)=D(h),\ \ \forall h\in G_0,\\
&\rho(a)=\pm U(a)\mathcal{K},
\end{aligned}
\end{equation}
where $\pm$ give equivalent $\nu$-coirreps.

Case (ii): $\mathrm{Act}_a(D)$ is equivalent to $D$ by $\mathrm{Act}_a(D)(h)=U^\dagger(a)D(h)U(a),\ \forall h\in G_0$. And the transformation matrix $U(a)$ satisfies $U(a)U^*(a)=-\nu(a,a)D(a^2)$.

The $\nu$-coirrep of $G$ induced from $D$ is given by
\begin{equation}
\begin{aligned}
&\rho(h)= \sigma_0\otimes D(h),\ \ \forall h\in G_0,\\
&\rho(a)=i\sigma_2\otimes U(a)  \mathcal{K}.
\end{aligned}
\end{equation}

Case(iii): $\mathrm{Act}_a(D)$ is not equivalent to $D$.

The $\nu$-coirrep of $G$ induced from $D$ is given by
\begin{equation}
\begin{aligned}
&\rho(h)=\left(\begin{array}{cc} D(h)& 0 \\ 0 & \mathrm{Act}_a(D)(h) \end{array}\right),\ \ \forall h\in G_0,\\
&\rho(a)=\left(\begin{array}{cc} 0 & \nu(a,a)D(a^2) \\ I & 0 \end{array}\right)  \mathcal{K}.
\end{aligned}
\end{equation}

The three cases can be determined by the following criterion \cite{YangHamiltonian2021}:
\begin{equation}
\sum_{g\in aG_0}\nu(g,g)\chi(g^2)=
\left\{
\begin{aligned}
&+|G_0|,\ \ \ \mathrm{case (i)}\\
&-|G_0|,\ \ \ \mathrm{case (ii)}\\
&0,\ \ \ \ \ \  \ \ \ \ \mathrm{case (iii)}.
\end{aligned}
\right.
\end{equation}
\vspace{0.3cm}

\bibliography{spinref2} 

\end{document}